\documentclass[aps,reprint,amssymb,amsfonts,amsmath,prl,superscriptaddress,longbibliography]{revtex4-2}
\usepackage{amsfonts}
\usepackage{textcomp,url}
\usepackage{braket,chngcntr,comment,dsfont}
\usepackage{times}
\usepackage{graphicx}
\usepackage{float}
\usepackage{latexsym,amsmath,amssymb,bm,euscript}
\usepackage{color}
\usepackage{subfigure}
\usepackage{epstopdf}
\usepackage[colorlinks=true,linkcolor=blue,citecolor=blue]{hyperref}
\usepackage{soul}
\usepackage[normalem]{ulem}
\usepackage{mathrsfs}
\usepackage{lipsum}
\usepackage{xspace}
\usepackage{natbib}
\usepackage{etoolbox}
\usepackage{tikz,listings}
\usepackage{booktabs,makecell}
\usepackage{pifont}

\newcommand{\Rmnum}[1]{\expandafter\@slowromancap\romannumeral #1@}

\newcommand{\LNO}{La$_3$Ni$_2$O$_7$}
\newcommand{\LPNO}{La$_2$PrNi$_2$O$_7$}
\newcommand{\XO}{\ensuremath{d_{x^2-y^2}}~}
\newcommand{\ZO}{\ensuremath{d_{z^2}}~}

\begin{document}

\title{Hund's Rule, Interorbital Hybridization, and High-$T_c$ Superconductivity in the Bilayer Nickelate}
\author{Xing-Zhou Qu}
\thanks{These authors contributed equally to this work.}
\affiliation{Kavli Institute for Theoretical Sciences, University of Chinese Academy of Sciences, Beijing 100190, China}
\affiliation{Institute of Theoretical Physics, Chinese Academy of Sciences, Beijing 100190, China}

\author{Dai-Wei Qu}
\thanks{These authors contributed equally to this work.}
\affiliation{Kavli Institute for Theoretical Sciences, University of Chinese Academy of Sciences, Beijing 100190, China}
\affiliation{Institute of Theoretical Physics, Chinese Academy of Sciences, Beijing 100190, China}

\author{Xin-Wei Yi}
\affiliation{School of Physical Sciences, University of Chinese Academy of Sciences, Beijing 100049, China}

\author{Wei Li}
\email{w.li@itp.ac.cn}
\affiliation{Institute of Theoretical Physics, Chinese Academy of Sciences, Beijing 100190, China}

\author{Gang Su}
\email{sugang@itp.ac.cn}
\affiliation{Institute of Theoretical Physics, Chinese Academy of Sciences, Beijing 100190, China}
\affiliation{Kavli Institute for Theoretical Sciences, University of Chinese Academy of Sciences, Beijing 100190, China}
\affiliation{School of Physical Sciences, University of Chinese Academy of Sciences, Beijing 100049, China}

\begin{abstract}
Understanding the pairing mechanism in bilayer nickelate superconductors constitutes a fascinating quest. Here we investigate the intriguing interplay between Hund's rule coupling and interorbital hybridization in a two-orbital model for bilayer nickelates, using a comprehensive tensor network approach: density matrix renormalization group for finite-size systems, infinite projected entangled-pair states in the thermodynamic limit, and thermal tensor networks for finite-temperature properties. We explain the pressure-dependent high-$T_c$ superconductivity observed in experiment, by identifying three distinct superconductive (SC) regimes: hybridization dominant, Hund's rule dominant, and the hybrid-Hund synergistic SC regimes. In these SC regimes, both $d_{x^2-y^2}$ and $d_{z^2}$ orbitals exhibit algebraic pairing correlations with similar Luttinger parameters $K_{\rm SC}$. However, the former exhibits a much stronger amplitude than the latter, with a distinctly higher SC characteristic temperature $T_c^*$, below which the pairing susceptibility diverges as $\chi_{\rm SC}(T) \sim 1/T^{2-K_{\rm SC}}$. With realistic model parameters, we find the pressurized La$_3$Ni$_2$O$_7$ falls into the Hund's rule dominated SC regime. As hybridization further enhances with pressure, it leads to significant interorbital frustration and in turn suppresses the SC correlations, explaining the rise and fall of high-$T_c$ superconductivity under high pressure. Our results offer a comprehensive understanding of the interlayer pairing in superconducting La$_3$Ni$_2$O$_7$.
\end{abstract}

\maketitle

\textit{Introduction.---}
The recent discovery of nearly 80~K superconductive (SC) transition in the pressurized Ruddlesden-Popper 
perovskite \LNO~\cite{Nickelate80K} has sparked significant research activities in both experiment
\cite{Wang2024Bulk,Ko2024ambient,Liu2023correlation,Hou2023emergence,Zhang2023hightemperature,
yang2023orbitaldependent,zhang2023effects,fwang2023pressureinduced,wang2023observation,
cui2023strain,chen2023evidence,kumar2023softening,Dong2024,li2024pressuredriven,liu2024growth,
wen2024probing,chen2024electronic,xie2024neutron} and theory~\cite{Luo2023Model, zhang2023electronic, yang2023possible, lechermann2023electronic, sakakibara2023possible, gu2023effective, shen2023effective, christiansson2023correlated, Shilenko2023Correlated, wu2023charge, cao2023flat, chen2023critical, liu2023spmwave, lu2023interlayer, qu2023bilayer, oh2023type, zhang2023structural, liao2023electron, yang2023minimal, jiang2023high, zhang2023trends, huang2023impurity, qin2023hightc, tian2023correlation, lu2023superconductivity, jiang2023pressure, kitamine2023theoretical, luo2023hightc, zhang2023strong, pan2023effect, sakakibara2023theoretical, lange2023pairing, geisler2023structural, yang2023strong, rhodes2023structural, lange2023feshbach, labollita2023electronic, kaneko2023pair, lu2023interplay, ryee2023critical, schlmer2023superconductivity, chen2023orbitalselective, liu2023role, ouyang2023hund, chang2023fermi, sui2023electronic, zheng2023superconductivity,xue2024magnetism,bhatta2025structural}. 
The bilayer structure and orbital selectivity are believed to be key factors in the formation of SC order, and the interlayer antiferromagnetic (AF) coupling is considered as the pairing driving
force~\cite{lu2023interlayer,qu2023bilayer,zhang2023structural,oh2023type,liao2023electron,yang2023minimal,qin2023hightc,chen2023orbitalselective}. Nonetheless, debate persists regarding the SC pairing mechanism, particularly with respect to the intriguing roles of Hund's rule coupling and the hybridization between the two $e_g$ orbitals.

\begin{figure}[!htbp]
\includegraphics[width=1\linewidth]{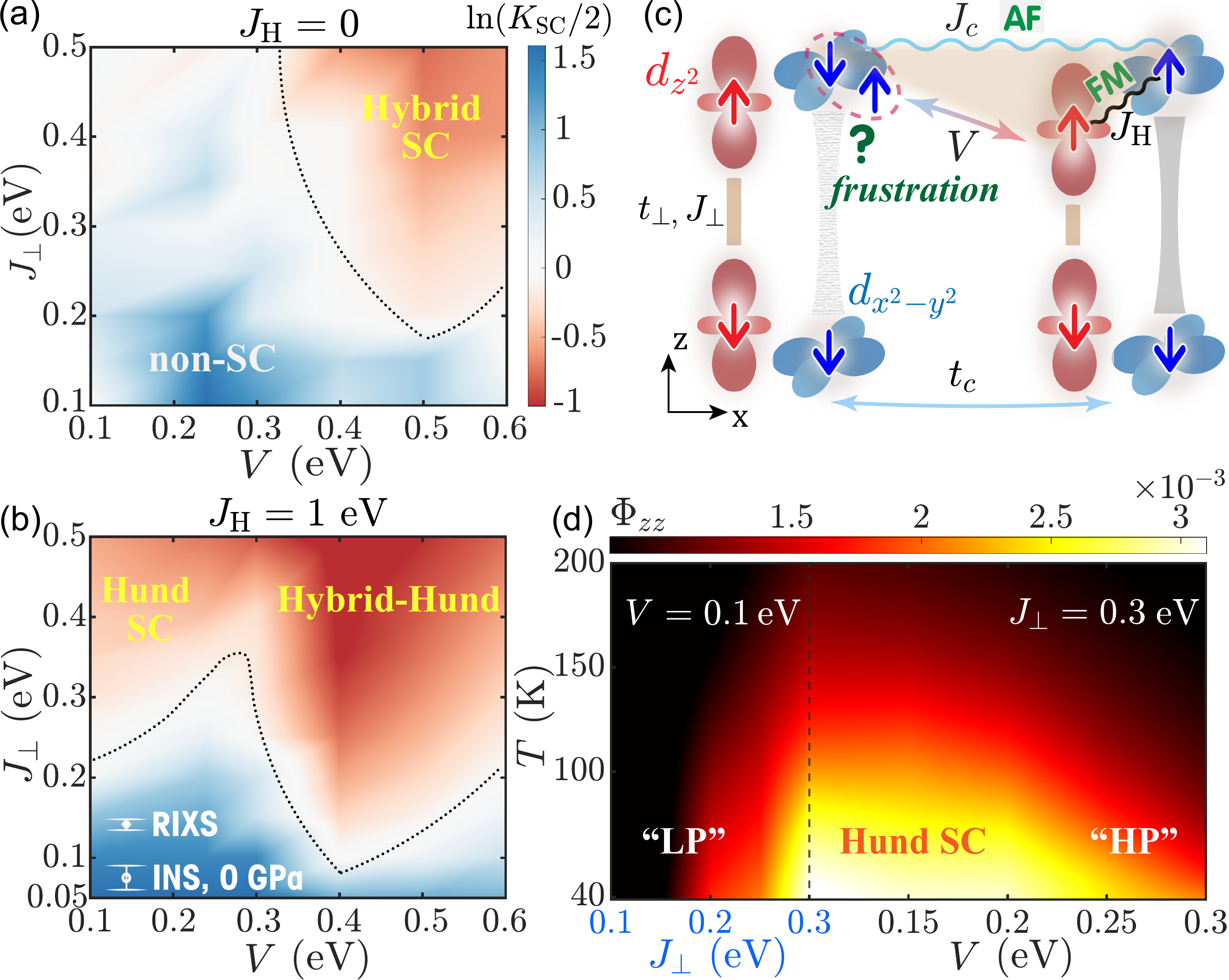}
\caption{The ground state $J_\perp$-$V$ phase diagrams are depicted with (a) $J_{\rm H} = 0$ and (b) $J_{\rm H} =1$~eV, 
respectively. Other model parameters are fixed according to DFT calculations under about 30~GPa~\cite{Luo2023Model}. 
The dotted lines separating 
the SC and non-SC regimes are guides for the eye. Three distinct SC regimes include ``Hybrid" dominated by $V$, ``Hund" by 
$J_{\rm H}$, and ``Hybrid-Hund" by both. Measured $J_\perp$ values from ambient resonant inelastic X-ray scattering (RIXS)~\cite{chen2024electronic} 
and inelastic neutron scattering (INS)~\cite{xie2024neutron} experiments are marked in (b). (c) The $d_{x^2-y^2}$ orbital exhibits intralayer couplings ($t_c$, $J_c$), 
while the $d_{z^2}$ orbital features interlayer couplings ($t_\perp$, $J_\perp$) and intralayer hopping ($t_d$). $J_\mathrm{H}$ 
denotes the on-site FM Hund's rule coupling and $V$ represents the interorbital hybridization. The \XO-\XO super-exchange and 
the \XO-\ZO double-exchange-like couplings compete and give rise to interorbital frustration highlighted by the shaded triangle. 
(d) The contour plot of the interlayer pairing correlation $\Phi_{zz}(r)$ of \XO orbital, averaged over $r=2$ to $L/4$.}
\label{Fig1}
\end{figure}

The hybridization theory~\cite{yang2023minimal,qin2023hightc,shen2023effective,sakakibara2023possible,
kaneko2023pair,luo2023hightc,Masataka2024pair,zheng2023superconductivity,shen2024numerical,
wang2024highly} considers the interlayer \ZO pairs correlated via the strong interlayer AF coupling. 
There are few holes in the \ZO orbital, and due to the limited intralayer hopping of \ZO electrons, 
the preformed pairs necessitate hybridization with itinerant \XO orbitals to attain phase coherence 
[cf., Appendix Fig.~\ref{FigE1}(a)]. Nevertheless, it is debated whether such SC order in the \ZO orbitals could render high $T_c$ through this mechanism~\cite{lu2023interplay, tian2023correlation, chen2023orbitalselective, shen2024numerical, bhatta2025structural}. The significant interlayer $t_\perp$ between \ZO orbitals could result in pronounced Pauli blocking~\cite{Hilker2023pairing}. The relatively low hole density tends to suppress potential SC order in the \ZO orbitals~\cite{chen2023orbitalselective, shen2024numerical}, suggesting this scenario warrants further investigation.

On the other hand, a different scenario suggests that the Hund's rule coupling plays a critical role in forming the high-$T_c$ SC order~\cite{lu2023interlayer, qu2023bilayer, oh2023type, chen2023orbitalselective, lu2023interplay, tian2023correlation, ouyang2023hund}. Although the interlayer spin exchange is quite small in the \XO orbitals, the substantial ferromagnetic (FM) Hund's rule coupling $J_{\rm H} \sim$ 1~eV can effectively transfer the interlayer AF coupling from the \ZO to \XO orbitals, passing a strong pairing force to the latter~\cite{lu2023interlayer, qu2023bilayer, oh2023type} [cf., Fig.~\ref{FigE1}(b)]. Moreover, as the quarter-filling \XO orbital possesses ample hole density and adequate intralayer hopping amplitude, the hole pairs can thus move coherently within each layer. By integrating out the \ZO orbitals in the large $J_{\rm H}$ limit and neglecting the interorbital hybridization, a single-orbital $t$-$J$-$J_\perp$ model has been proposed, which demonstrates high-$T_c$ superconductivity~\cite{lu2023interlayer, qu2023bilayer}. This bilayer pairing mechanism is quite robust and has been discussed before in the context of fermion ladders~\cite{Zhu2018pairing, Jiang2020critical} and bilayer square lattices~\cite{Ma2022doping}. Nonetheless, recent experiments find that the SC order is suppressed under higher pressure, up to 104~GPa, leading to a right-angled triangular SC phase~\cite{li2024pressuredriven}. Such an observation remains to be explained within the Hund's coupling dominant scenario, and underscores the need for a more comprehensive study incorporating both $e_g$ orbitals of Ni$^{2.5+}$ in \LNO.

In this study, we consider a two-orbital bilayer model with both Hund's rule coupling and interorbital hybridization, and conduct a comprehensive numerical study. We find that the \XO orbital consistently serves as the primary host for robust SC order, exhibiting stronger algebraic pairing correlations and a higher $T_c^*$ determined from pairing susceptibility compared to those of the \ZO orbitals. We highlight that there are three distinct SC regimes: one dominated by hybridization, another by Hund's rule, and a third by both mechanisms. With realistic parameters, we find the pressurized {\LNO} resides in the Hund SC regime and obtain its finite-temperature phase diagram. 
Our results explain the non-monotonic pressure-dependent behaviors of SC order observed in recent experiments.

\textit{Model and method.---}
Below we consider the following two-orbital bilayer $t$-$J$ model
\begin{widetext}
\begin{eqnarray*}
  H =&-&t_{c} \sum_{\langle i, j\rangle, \mu, \sigma}\left(c_{i, \mu, \sigma}^{\dagger} c_{j, \mu, \sigma}+ \mathrm{H.c.}\right)
  +J_{c} \sum_{\langle i, j\rangle, \mu}\left(\mathbf{S}_{i, \mu}^{c} \cdot \mathbf{S}_{j, \mu}^{c}-\frac{1}{4} n_{i, \mu}^{c} n_{j, \mu}^{c}\right) \notag \\
  &-&t_{\perp} \sum_{i, \sigma}\left(d_{i, \mu = 1, \sigma}^{\dagger} d_{j, \mu = -1, \sigma}+ \mathrm{H.c.}\right)
  +J_{\perp} \sum_{i}\left(\mathbf{S}_{i, \mu= 1}^{d} \cdot \mathbf{S}_{i, \mu = -1}^{d}-\frac{1}{4} n_{i, \mu = 1}^{d} n_{i, \mu = -1}^{d}\right) \\
  &-&t_{d} \sum_{\langle i, j\rangle, \mu, \sigma}\left(d_{i, \mu, \sigma}^{\dagger} d_{j, \mu, \sigma}+ \mathrm{H.c.}\right)
  -V  \sum_{i, \mu, \sigma}\left(c_{i, \mu, \sigma}^{\dagger} \tilde{d}_{i, \mu, \sigma}+ \mathrm{H.c.}\right)
  -J_{\mathrm{H}} \sum_{i, \mu} \mathbf{S}_{i, \mu}^{c} \cdot \mathbf{S}_{i, \mu}^{d}+\varepsilon_{c} \sum_{i, \mu} n_{i, \mu}^{c}
  +\varepsilon_{d} \sum_{i, \mu} n_{i, \mu}^{d} \notag,
\label{Eq:2orb-t-J}
\end{eqnarray*}
\end{widetext}
where $c_{i, \mu, \sigma}$ ($d_{i, \mu, \sigma}$) denotes the $d_{x^2-y^2}$ ($d_{z^2}$) electron at site $i$, layer $\mu = \pm 1$, and spin $\sigma = \{\uparrow, \downarrow\}$. Similarly, spin operators $\mathbf{S}^c_{i,\mu}$ ($\mathbf{S}^d_{i,\mu}$) and density operators $n_{i, \mu}^{c}$ ($n_{i, \mu}^{d}$) are defined for the two $e_g$ orbitals. We denote $\tilde{d}_{i, \mu, \sigma} = (d_{i+x, \mu, \sigma} + d_{i-x, \mu, \sigma} - d_{i+y, \mu, \sigma} - d_{i-y, \mu, \sigma})/2$ in the hybridization term according to the orbital symmetry. Parameter $t_c$ ($t_d$) labels the intralayer hopping of the \XO ($d_{z^2}$) orbital. The intralayer spin coupling between the \XO electrons is denoted as $J_c$, while that of \ZO is negligibly small. The interlayer hopping and coupling of the \ZO orbital are labeled as $t_\perp$ and $J_\perp$, respectively. The interorbital hybridization $V$ and the on-site Hund's rule coupling $J_{\mathrm{H}}$ between the two $e_g$ orbitals are considered, with $\varepsilon_{c}$ and $\varepsilon_{d}$ representing the respective site energies. In the present study, the electron filling is fixed as $n_e = 1.5$ per site for the two $e_g$ orbitals unless otherwise specified.

\begin{figure}[h!]
\includegraphics[width=1\linewidth]{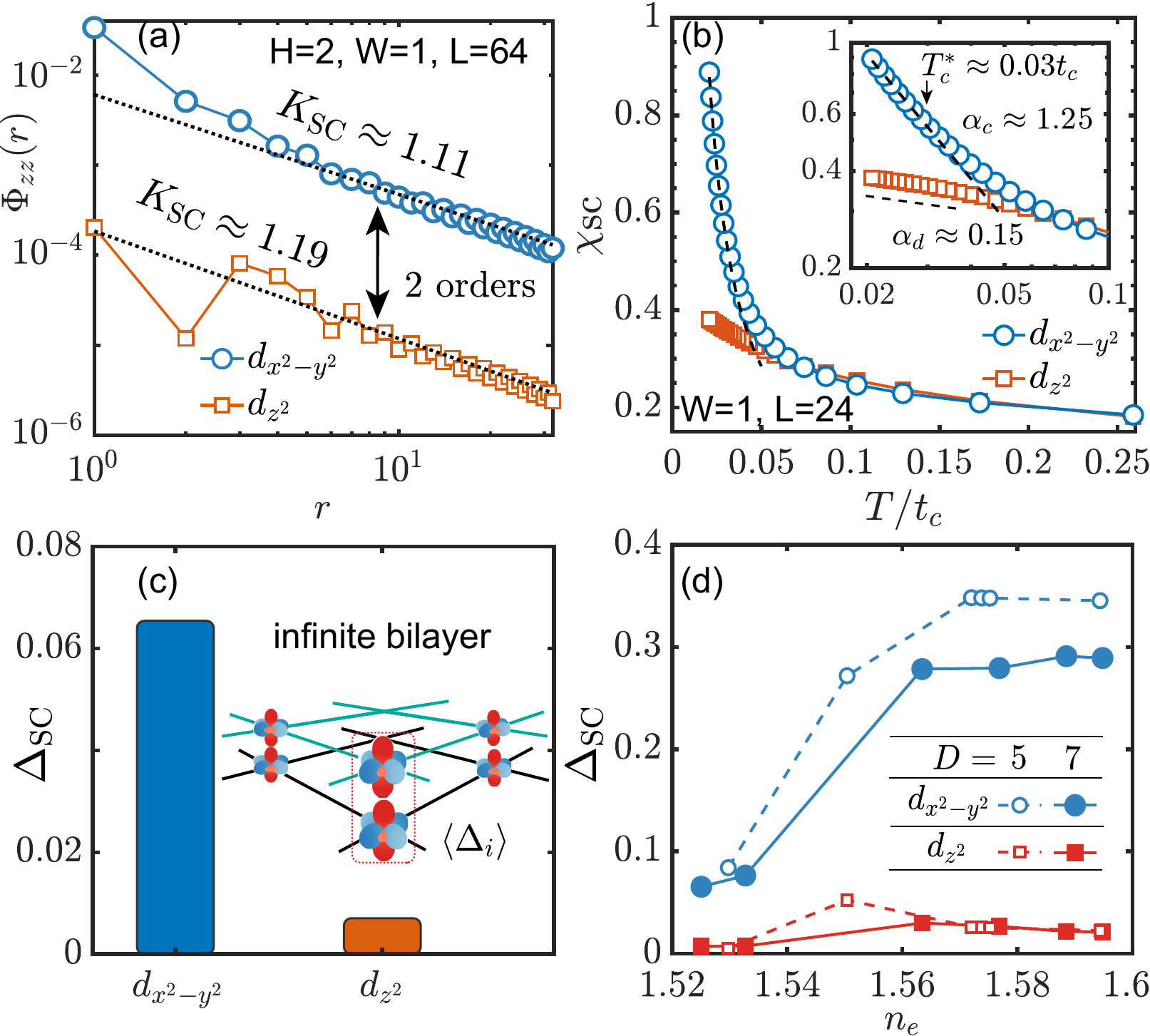}
\caption{(a) Pairing correlation $\Phi_{zz}(r)$ of the \XO and \ZO orbitals obtained from DMRG on a $W=1$ ladder, with $r$ the distance between two rung pairs. (b) The pairing susceptibility calculated with the finite-temperature tensor networks, which diverges algebraically as $\chi_{\rm SC} \sim 1/T^{\alpha_c}$ (with $\alpha_c \simeq 1.25$) below $T_c^* \approx 0.03 t_c$ for the \XO orbital, while exhibits weak divergence in the \ZO orbital. (c) Interlayer SC pairing order parameter $\Delta_{\rm SC} = |\langle \Delta_i \rangle|$, averaged within the $2\times2$ unit cell (see the inset). The iPEPS calculations on an infinite bilayer lattice and at filling $n_e \simeq 1.525$. (d) Order parameter $\Delta_{\rm SC}$ increases systematically with electron filling $n_e$. Solid and dashed lines correspond to results with different kept states ($D=5, 7$). Realistic model parameters for {\LNO} (see main text) are adopted in the calculations.
}
\label{Fig2}
\end{figure}

To simulate the bilayer nickelate, we mainly consider realistic parameters for \LNO~under about 30~GPa, i.e., $t_c = 0.483$~eV, $t_\perp = 0.635$~eV, $t_d = 0.110$~eV, $V = 0.239$~eV$, \varepsilon_c = 0.776$~eV, and $\varepsilon_d = 0.409$~eV~\cite{Luo2023Model}. With properly chosen Hubbard $U = 4$~eV~\cite{christiansson2023correlated,tian2023correlation}, the spin couplings are set as $J_c \simeq 4t_c^2 / U = 0.233$~eV and $J_\perp \simeq 4t_\perp^2 / U = 0.403$~eV, and the Hund's rule coupling is determined as $J_{\mathrm{H}}=1$~eV ($\sim 2.5J_\perp$)~\cite{christiansson2023correlated, jiang2023pressure,cao2023flat}. In this work, we employ multiple tensor-network methods: density matrix renormalization grou (DMRG)~\cite{White1992, Schollwock2011MPS,Andreas2023CBE} for the finite-size system at $T=0$, and infinite projected entangled-pair state (iPEPS)~\cite{Verstraete2004renorm, Jordan2008Classical, Cirac2021RMP, Corboz2010Simulation} in the thermodynamic limit, together with tangent-space tensor renormalization group (tanTRG) methods~\cite{Li2011a, Chen2018XTRG, tanTRG2023} for finite-temperature calculations. Our DMRG and tanTRG calculations are performed on a lattice of size $H \times W\times L$ (with height $H=2$ for bilayer, length $L$ up to 64, and width $W$ up to 2), while the iPEPS calculations are on infinite lattice with $2\times 2$ unit cell. The DFT simulations are employed to determine the coupling parameters, particularly the hybridization $V$, under different pressures. 

\textit{Orbital selectivity and interlayer pairing.---}
In both DMRG and iPEPS calculations, we find the \ZO orbitals are nearly half-filled while the \XO orbitals are approximately quarter-filled (see Appendix). Such clear orbital selectivity can lead to distinct SC pairing behaviors. In Fig.~\ref{Fig2}(a), we show the interlayer correlation $\Phi_{zz}(r) =  \langle \Delta_{i}^\dagger \Delta_{j} \rangle$, where $\Delta_i^{\dagger}$ is the pairing operator. For the $d_{x^2-y^2}$ orbital, $\Delta_i^{\dagger} = \frac{1}{\sqrt{2}} \sum_{\mu=\pm1} \, c_{i,\mu,\uparrow}^{\dagger} c_{i,-\mu,\downarrow}^{\dagger}$, where $r \equiv |j-i|$ represents distance along the length direction. When switched from \XO to the \ZO orbital, we replace $c_{i,\mu,\sigma}$ with $d_{i,\mu,\sigma}$. The pairing susceptibility is defined as $\chi_{\rm SC}(T) = \frac{1}{WL} \partial \langle \Delta_\text{tot} \rangle_T/\partial h_p$, computed with a small pairing field $h_p=0.002$ coupled to $\Delta_\text{tot} = \frac{1}{2} \sum_{i}[\Delta_{i}+ (\Delta_{i})^\dagger]$. 

Figure~\ref{Fig2}(a) presents the pairing correlations $\Phi_{zz}(r)$ for both $e_g$ orbitals, where the \XO orbital exhibits much stronger pairing correlations, about two orders of magnitude greater than that of the \ZO orbitals. From the power-law fitting, we find the Luttinger parameter $K_{\rm SC} \simeq 1.11$ for the \XO orbital, indicating the presence of (quasi-long range) SC order. Pairing correlations of the \ZO orbital also follow an algebraic scaling with $K_{\rm SC}\simeq 1.19$. Despite a significant difference --- more than an order of magnitude --- in the pairing correlation strength between two $e_g$ orbitals, the proximity effect~\cite{lu2023interplay} renders them comparable $K_{\rm SC}$ values. Beside the two-orbital ladder with $W=1$, in the Appendix we show DMRG results on $W=2$ lattice ($2\times2\times32$) and draw consistent conclusions.

The small $K_{\rm SC} \lesssim 1$ implies the divergence of pairing susceptibility. In Fig.~\ref{Fig2}(b), we show the calculated results of $\chi_{\rm SC}(T)$ in the two $e_g$ orbitals. In the \ZO orbital, $\chi_{\text{SC}}(T)$ exhibits an increase at lower temperature and shows significantly weaker divergence, thus being secondary in determining the critical temperature $T_c$. Conversely, the $\chi_{\text{SC}}(T)$ behaviors of the \XO orbital suggest $T_c^* \simeq 0.03 t_c$, corresponding to a high transition temperature in the order of 100 K. Moreover, we find the $\chi_{\rm SC}$ for the \XO orbital exhibits a algebraic divergence behavior $\sim 1/T^{\alpha}$ with an exponent $\alpha \approx 2-K_{\rm SC}$, consistent with $K_{\rm SC}$ determined from the ground-state pairing correlations. 

Figures~\ref{Fig2}(c,d) show the iPEPS results on an infinite 2D lattice. We compute the interlayer SC pairing order parameter $\Delta_{\rm SC}$, which is the mean absolute value of $\langle \Delta_i \rangle$ within the ($2\times 2$) unit cell. In Fig~\ref{Fig2}(c) we fix $n_e \simeq 1.525$ and confirm that \XO orbital exhibits significantly stronger SC order than the \ZO orbital. As \ZO orbital is nearly half-filled and behaves like a local moment, in the large $J_{\rm H}$ limit we can integrate out the \ZO electrons, thereby recovering the single-band $t$-$J$-$J_\perp$ model~\cite{lu2023interlayer, qu2023bilayer} with strong interlayer pairing. In Fig~\ref{Fig2}(d) we tune the chemical potential and dope electrons into the system, which increases the carrier density of \XO orbital and enhances the SC order.

\begin{figure}[t]
\centering
\includegraphics[width=1\linewidth]{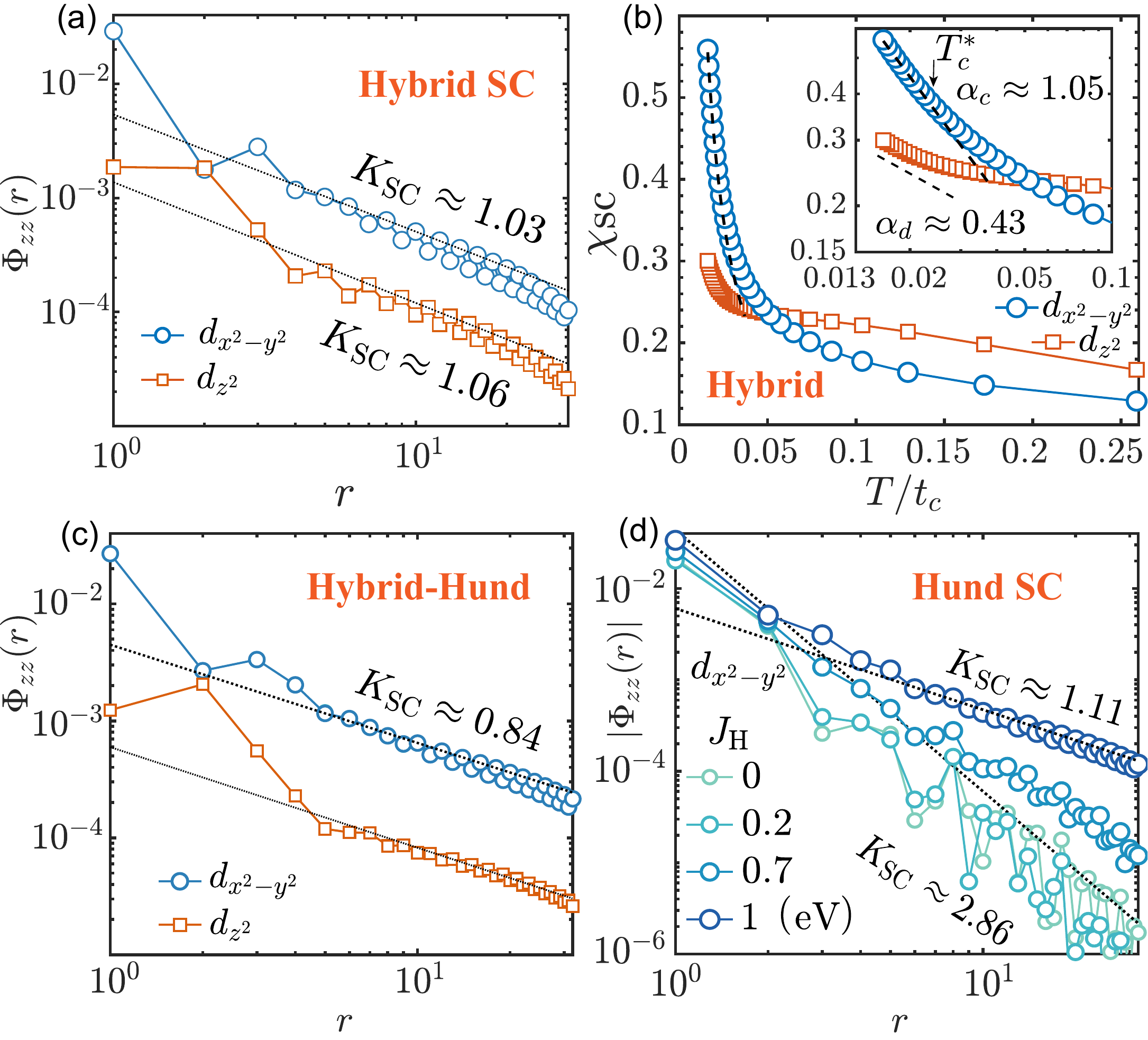}
\caption{(a) Pairing correlation $\Phi_{zz}$ and (b) finite-temperature pairing susceptibilities 
of the \XO and \ZO orbitals within the hybridization SC regime ($J_{\rm H} = 0$, $V = 0.5$~eV). 
Same correlations are shown for the (c) Hybrid-Hund ($J_{\rm H} = 1$~eV, $V = 0.5$~eV) and 
(d) Hund SC regime ($V = 0.239$~eV), where the Hund's rule coupling $J_{\mathrm{H}}$
varies from zero to 1~eV.}
\label{Fig3}
\end{figure}

\textit{Hund SC regime in the two-orbital bilayer model.---}
In Figs.~\ref{Fig1}(a,b), we distinguish between SC and non-SC regimes using the fitted Luttinger exponent $K_{\rm SC}$ of the \XO orbital. We find a strong interlayer $J_\perp$ is crucial for forming a robust SC order in pressurized nickelates. Recent experiments~\cite{chen2024electronic, xie2024neutron} reveal the magnitude of interlayer spin coupling $J_\perp$ around $0.1$~eV in ambient bulk \LNO, which is located in the non-SC regime [cf., Fig.~\ref{Fig1}(b)]. The influence of other 
parameters like the intralayer hopping $t_c$ are relatively small, as demonstrated in the Appendix.

As shown in Figs.~\ref{Fig1}(a,b), we identify three distinct SC regimes, namely, the ``Hybrid SC'' regime dominated by hybridization $V$, ``Hund SC'' regime with strong coupling $J_{\rm H}$, and the integrated ``Hybrid-Hund SC'' regime where both couplings are essential and synergistic. In Fig.~\ref{Fig3}(a), we show that $\Phi_{zz}(r) \sim r^{-K_{\rm SC}}$ with $K_{\rm SC} \approx 1.03$ for the \XO orbital at $J_{\rm H} = 0$ and $V=0.5$~eV. In Fig.~\ref{Fig3}(b), we uncover that the SC pairing susceptibility $\chi_{\text{SC}}$ of the \XO orbital adheres to a scaling law of $1/T^{\alpha_c}$, where $\alpha_c \approx 2 - K_{\text{SC}} \approx 1$. Notably, the \ZO orbital also demonstrates a comparable Luttinger parameter of $K_{\text{SC}} \approx 1$ [cf., Fig.~\ref{Fig3}(a)]; however, the pairing correlation and susceptibility $\chi_{\rm SC}$ are substantially weaker in intensity. Consequently, even within the hybrid SC scenario, the \ZO orbital pairing arises primarily through a proximity-induced mechanism rather than being the dominant contributor.

In Fig.~\ref{Fig1}(b), there exists a bybrid-Hund SC regime where both $V$ and $J_{\rm H}$ cooperate to render a robust SC order. In Fig.~\ref{Fig3}(c), we set $J_{\rm H} = 1$~eV and introduce large $V = 0.5$~eV, finding that the interlayer pairing correlations for both orbitals exhibit power-law scaling with $K_{\rm SC} \approx 0.84$. Nevertheless, the much larger amplitude ensures again the dominance of SC pairing with the \XO electrons. Although the hybrid and hybrid-Hund SC regimes can have $K_{\rm SC} \lesssim 2$, the required hybridization $V \gtrsim 0.4$~eV exceeds realistic value in the compound \LNO. 

Within a realistic range of $V$, we demonstrate in Fig.~\ref{Fig1}(b) a SC regime where Hund's rule coupling $J_{\rm H}$ plays a dominant role. In Fig.~\ref{Fig3}(d), with fixed $V = 0.239$~eV and increasing $J_{\rm H}$, we find the SC pairing correlation gradually enhances and a quasi-long-range order develops. The results with $J_{\rm H} = 1$~eV are plotted again here as a reference. Therefore, we conclude that the pressurized {\LNO} is located within the Hund SC regime, and $J_{\rm H}$ is essential for driving the SC pairing.

\textit{Pressure evolution of the SC order.---}
High-$T_c$ SC emerges in \LNO~\cite{Nickelate80K,li2024pressuredriven} under moderately high pressure above 15~GPa; however, further increasing the pressure suppresses rather than enhances the SC order~\cite{li2024pressuredriven}. Such a non-monotonic behavior can be captured by our two-orbital model. 

\begin{figure}[]
\centering
\includegraphics[width=1\linewidth]{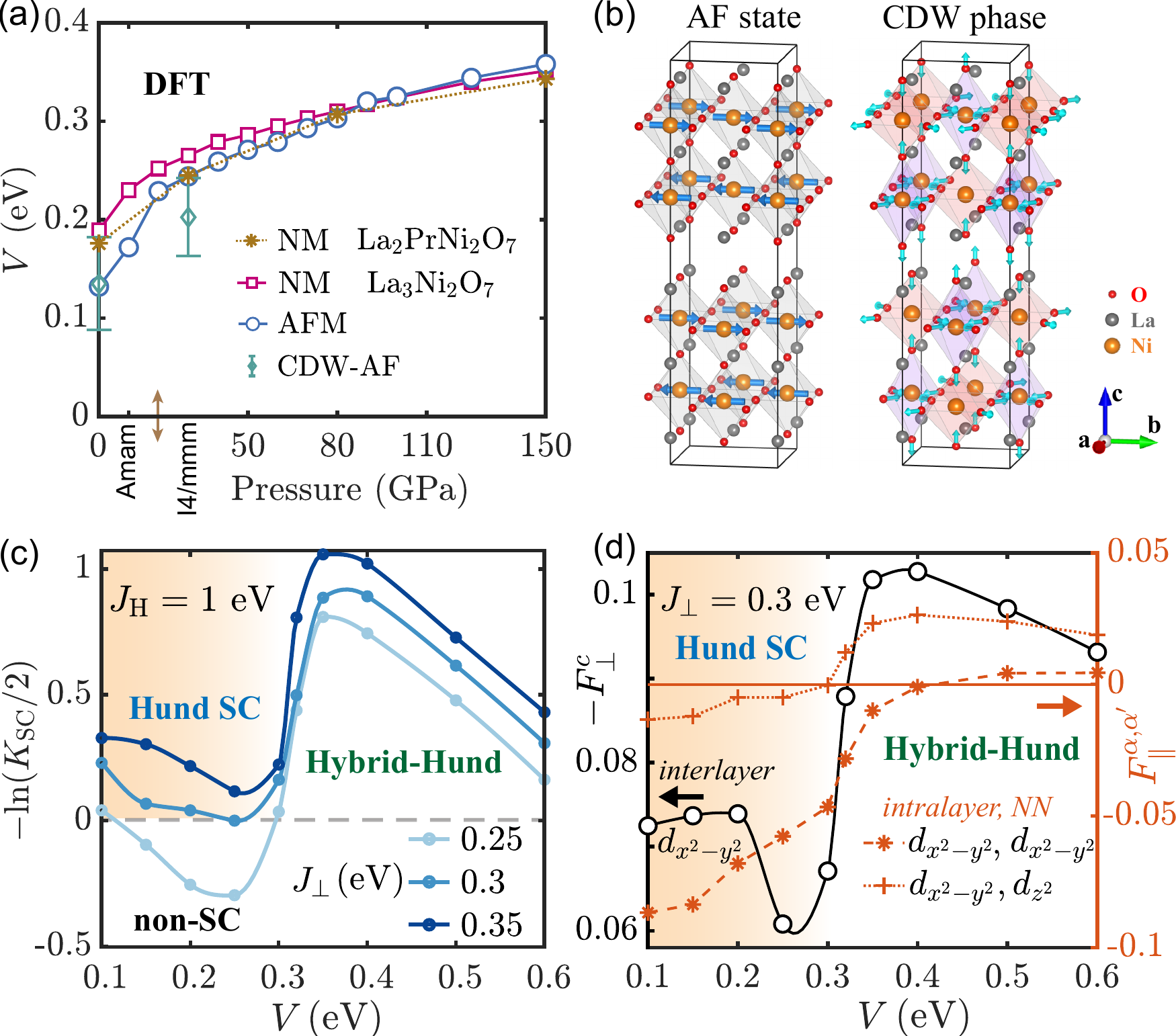}
\caption{(a) DFT results for the pristine and CDW phases with both NM and AF configurations as illustrated in (b). 
For the CDW-AF phase, error bars in the hybridization $V$ reflect its non-uniform distribution due to oxygen 
octahedra distortion. The ambient and 10 GPa data are obtained from \textit{Amam} phase while others are 
from \textit{I4/mmm} phase. (c) The SC Luttinger parameters $K_{\rm SC}$ and (d) the interlayer spin correlation 
$-F_\perp^c$ and intralayer $F^{\alpha,\alpha'}_{\parallel}$ ($\alpha,\alpha'\in \{c, d\}$) between nearest-neighboring sites are plotted. 
$J_{\rm H}$ is fixed as 1~eV in the calculations. }
\label{Fig4}
\end{figure}

Following a significant enhancement in $J_\perp$ due to the pressure-induced structural transition~\cite{jiang2023pressure, ryee2023critical, yang2023strong, wu2023charge}, the system moves upwards in the phase diagram, thereby entering the Hund SC regime (see Fig.~\ref{Fig1}(b)). However, further increases in pressure can lead to substantial hybridization ($V$) (see Fig.~\ref{Fig4}(a)), 
causing deviation from the Hund SC regime and suppression of the SC order. Within such a scenario, the overall two-step pressure evolution is depicted in Fig.~\ref{Fig1}(d), where we show the interlayer \XO pairing correlations $\Phi_{zz}(r)$ averaged form $r=2$ to $L/4$. From low-pressure (``LP") to high-pressure (``HP") regime, the increased interlayer spin coupling $J_\perp$ renders the emergence of high-$T_c$ SC. Nevertheless, further increasing pressure would enhance $V$, leading to the suppression of SC order gradually in the over-pressurized regime. Our results provide a possible 
explanation of the reported pressure-temperature phase diagram in recent experiments~\cite{li2024pressuredriven}. 

\textit{Suppression of SC order and interorbital frustration.---}
To uncover the coupling parameters at varying pressure levels, we employ DFT calculations. These calculations reveal the presence of an AF order in the magnetic ground state of \LNO~\cite{yi2024antiferromagnetic}, as opposed to a nonmagnetic (NM) configuration. In addition, strong Fermi surface nesting can also induce charge density wave (CDW) instability~\cite{yi2024antiferromagnetic}. For each configuration depicted in Fig.~\ref{Fig4}(b), we perform Wannier downfolding of the DFT band structure to obtain the hopping parameters of the two-orbital bilayer model (see Appendix for details). Similar analysis is also performed for Pr-doped \LPNO~with the NM configuration, with the obtained hybridization $V$ shown in Fig.~\ref{Fig4}(a). These results show that $V$ increases from $\sim 0.15$~eV to $\sim 0.33$ eV with pressure in both compounds. 

In Fig.~\ref{Fig4}(c), we illustrate the influence of $V$ on the SC order by examining the behaviors of $K_{\rm SC}$. For $J_\perp = 0.25$~eV and 0.3 eV, $K_{\rm SC}$ becomes greater than 2 as $V \gtrsim 0.3$~eV, which corresponds to about 100 GPa pressure in experiment [cf., Fig.~\ref{Fig4}(a)]. The system thus leaves the Hund SC regime due to the enhancement of $V$ as shown in Fig.~\ref{Fig1}(b). We attribute the suppression of SC order to the magnetic frustration effect illustrated in Fig.~\ref{Fig1}(c). This effect intensifies as pressure — and particularly $V$ — increases, leading to weakened interlayer AF correlations as well as SC pairing between \XO orbitals.

To better quantify the interorbital frustration, we compute the interlayer \XO spin correlation $F_\perp^c$ and intralayer correlations $F^{\alpha,\alpha'}_{\parallel} \equiv \frac{1}{2(L-1)} \sum_{i, \mu} \langle \mathbf{S}_{i, \mu}^\alpha \cdot \mathbf{S}_{i+1, \mu}^{\alpha'} \rangle$, where $\alpha, \alpha' \in \{c, d\}$ denote the two orbitals. As depicted in Fig.~\ref{Fig4}(d), within the Hund SC regime, the intralayer spin correlation between the \XO orbitals is dominated by the intralayer super-exchange $J_c>0$. As $V$ increases, the FM correlation strengthens due to the differing electron fillings between the two orbitals and the presence of Hund's rule coupling, which together facilitate a double-exchange-like FM coupling. Their competition introduces interorbital frustration to the system, which leads to a switch in intralayer spin correlations from AF to FM. As shown in Fig.~\ref{Fig4}(d), the strength of the interlayer AF correlation $|F_\perp^c|$ also decreases, weakening the interlayer pairing. Upon further enhancing the hybridization $V$ to overcome the AF $J_c > 0$, spin frustration becomes alleviated in the hybrid-Hund regime. Consequently in the hybrid-Hund regime, $V$ and $J_{\rm H}$ cooperate and stabilize a robust SC order. These results predict that a reentrant SC phase may appear at higher pressures, presenting an important target for future experimental studies.

\textit{Discussion and outlook.---}
In this work, we perform a comprehensive numerical study of the two-orbital bilayer $t$-$J$ model, with both interorbital hybridization and Hund's rule coupling included and treated on equal footing. With realistic model parameters, we find the pressurized \LNO~resides in the Hund's rule dominated regime, in which the \XO orbital is mainly responsible for forming the SC order, while the \ZO orbital becomes superconducting via the proximity effect. Furthermore, in the finite-temperature phase diagram, we find that increasing pressure can suppress the SC order, which explains the recently observed right-triangle SC phase in \LNO~\cite{li2024pressuredriven}. Besides {\LNO}, we expect similar SC behaviors in {\LPNO} under high pressure~\cite{Wang2024Bulk,wen2024probing}. The intriguing interplay between the Hund's rule coupling and hybridization accounts for two possible SC phases, namely the Hund and hybrid-Hund regimes, which have different signs of intralayer spin correlations.

Our two-orbital model not only explains the emergent SC order in the pressurized bulk but can also be suggestive for the ultrathin films~\cite{Ko2024ambient, zhou2024ambient}. In bulk materials, a structural transition occurs from orthorhombic to tetragonal crystal structure~\cite{Nickelate80K, li2024pressuredriven}. This transition is accompanied by a significant increase in interlayer AF coupling, which stabilizes the SC order. In thin films, such high-symmetry tetragonal phase is proposed to be stabilized by strain and exists even at ambient pressure~\cite{zhou2024ambient}. According to our two-orbital scenario dominated by Hund's rule, the enhanced interlayer AF coupling likely accounts for the observed high-$T_c$ superconductivity. When the \ZO orbital approaches half filling, this would be beneficial rather than detrimental (see Appendix Fig.~\ref{FigE4}) to the robust SC order.

\begin{acknowledgements}
\textit{Acknowledgments.---}
The authors are indebted to Jialin Chen, Qiaoyi Li, Xian-Xin Wu, Rong Yu, Fan Yang, and Congjun Wu for stimulating 
discussions. This work was supported by the National Natural Science Foundation of China (Grant Nos. 
12222412 and 12047503), Innovation Program for Quantum Science and Technology (Nos.~2021ZD0301800 
and 2021ZD0301900), and CAS Project for Young Scientists in Basic Research (Grant No.~YSBR-057). 
We thank the HPC-ITP for the technical support and generous allocation of CPU time.
\end{acknowledgements}

\bibliography{nickelate}

\begin{thebibliography}{101}%
\makeatletter
\providecommand \@ifxundefined [1]{%
 \@ifx{#1\undefined}
}%
\providecommand \@ifnum [1]{%
 \ifnum #1\expandafter \@firstoftwo
 \else \expandafter \@secondoftwo
 \fi
}%
\providecommand \@ifx [1]{%
 \ifx #1\expandafter \@firstoftwo
 \else \expandafter \@secondoftwo
 \fi
}%
\providecommand \natexlab [1]{#1}%
\providecommand \enquote  [1]{``#1''}%
\providecommand \bibnamefont  [1]{#1}%
\providecommand \bibfnamefont [1]{#1}%
\providecommand \citenamefont [1]{#1}%
\providecommand \href@noop [0]{\@secondoftwo}%
\providecommand \href [0]{\begingroup \@sanitize@url \@href}%
\providecommand \@href[1]{\@@startlink{#1}\@@href}%
\providecommand \@@href[1]{\endgroup#1\@@endlink}%
\providecommand \@sanitize@url [0]{\catcode `\\12\catcode `\$12\catcode
  `\&12\catcode `\#12\catcode `\^12\catcode `\_12\catcode `\%12\relax}%
\providecommand \@@startlink[1]{}%
\providecommand \@@endlink[0]{}%
\providecommand \url  [0]{\begingroup\@sanitize@url \@url }%
\providecommand \@url [1]{\endgroup\@href {#1}{\urlprefix }}%
\providecommand \urlprefix  [0]{URL }%
\providecommand \Eprint [0]{\href }%
\providecommand \doibase [0]{https://doi.org/}%
\providecommand \selectlanguage [0]{\@gobble}%
\providecommand \bibinfo  [0]{\@secondoftwo}%
\providecommand \bibfield  [0]{\@secondoftwo}%
\providecommand \translation [1]{[#1]}%
\providecommand \BibitemOpen [0]{}%
\providecommand \bibitemStop [0]{}%
\providecommand \bibitemNoStop [0]{.\EOS\space}%
\providecommand \EOS [0]{\spacefactor3000\relax}%
\providecommand \BibitemShut  [1]{\csname bibitem#1\endcsname}%
\let\auto@bib@innerbib\@empty
\bibitem [{\citenamefont {Sun}\ \emph {et~al.}(2023)\citenamefont {Sun},
  \citenamefont {Huo}, \citenamefont {Hu}, \citenamefont {Li}, \citenamefont
  {Liu}, \citenamefont {Han}, \citenamefont {Tang}, \citenamefont {Mao},
  \citenamefont {Yang}, \citenamefont {Wang}, \citenamefont {Cheng},
  \citenamefont {Yao}, \citenamefont {Zhang},\ and\ \citenamefont
  {Wang}}]{Nickelate80K}%
  \BibitemOpen
  \bibfield  {author} {\bibinfo {author} {\bibfnamefont {H.}~\bibnamefont
  {Sun}}, \bibinfo {author} {\bibfnamefont {M.}~\bibnamefont {Huo}}, \bibinfo
  {author} {\bibfnamefont {X.}~\bibnamefont {Hu}}, \bibinfo {author}
  {\bibfnamefont {J.}~\bibnamefont {Li}}, \bibinfo {author} {\bibfnamefont
  {Z.}~\bibnamefont {Liu}}, \bibinfo {author} {\bibfnamefont {Y.}~\bibnamefont
  {Han}}, \bibinfo {author} {\bibfnamefont {L.}~\bibnamefont {Tang}}, \bibinfo
  {author} {\bibfnamefont {Z.}~\bibnamefont {Mao}}, \bibinfo {author}
  {\bibfnamefont {P.}~\bibnamefont {Yang}}, \bibinfo {author} {\bibfnamefont
  {B.}~\bibnamefont {Wang}}, \bibinfo {author} {\bibfnamefont {J.}~\bibnamefont
  {Cheng}}, \bibinfo {author} {\bibfnamefont {D.-X.}\ \bibnamefont {Yao}},
  \bibinfo {author} {\bibfnamefont {G.-M.}\ \bibnamefont {Zhang}},\ and\
  \bibinfo {author} {\bibfnamefont {M.}~\bibnamefont {Wang}},\ }\bibfield
  {title} {\bibinfo {title} {Signatures of superconductivity near {80 K} in a
  nickelate under high pressure},\ }\href
  {https://doi.org/10.1038/s41586-023-06408-7} {\bibfield  {journal} {\bibinfo
  {journal} {Nature}\ }\textbf {\bibinfo {volume} {621}},\ \bibinfo {pages}
  {493} (\bibinfo {year} {2023})}\BibitemShut {NoStop}%
\bibitem [{\citenamefont {Wang}\ \emph
  {et~al.}(2024{\natexlab{a}})\citenamefont {Wang}, \citenamefont {Wang},
  \citenamefont {Shen}, \citenamefont {Hou}, \citenamefont {Luo}, \citenamefont
  {Ma}, \citenamefont {Yang}, \citenamefont {Shi}, \citenamefont {Dou},
  \citenamefont {Feng}, \citenamefont {Yang}, \citenamefont {Shi},
  \citenamefont {Ren}, \citenamefont {Ma}, \citenamefont {Yang}, \citenamefont
  {Liu}, \citenamefont {Liu}, \citenamefont {Zhang}, \citenamefont {Dong},
  \citenamefont {Wang}, \citenamefont {Jiang}, \citenamefont {Hu},
  \citenamefont {Nagasaki}, \citenamefont {Kitagawa}, \citenamefont {Calder},
  \citenamefont {Yan}, \citenamefont {Sun}, \citenamefont {Wang}, \citenamefont
  {Zhou}, \citenamefont {Uwatoko},\ and\ \citenamefont {Cheng}}]{Wang2024Bulk}%
  \BibitemOpen
  \bibfield  {author} {\bibinfo {author} {\bibfnamefont {N.}~\bibnamefont
  {Wang}}, \bibinfo {author} {\bibfnamefont {G.}~\bibnamefont {Wang}}, \bibinfo
  {author} {\bibfnamefont {X.}~\bibnamefont {Shen}}, \bibinfo {author}
  {\bibfnamefont {J.}~\bibnamefont {Hou}}, \bibinfo {author} {\bibfnamefont
  {J.}~\bibnamefont {Luo}}, \bibinfo {author} {\bibfnamefont {X.}~\bibnamefont
  {Ma}}, \bibinfo {author} {\bibfnamefont {H.}~\bibnamefont {Yang}}, \bibinfo
  {author} {\bibfnamefont {L.}~\bibnamefont {Shi}}, \bibinfo {author}
  {\bibfnamefont {J.}~\bibnamefont {Dou}}, \bibinfo {author} {\bibfnamefont
  {J.}~\bibnamefont {Feng}}, \bibinfo {author} {\bibfnamefont {J.}~\bibnamefont
  {Yang}}, \bibinfo {author} {\bibfnamefont {Y.}~\bibnamefont {Shi}}, \bibinfo
  {author} {\bibfnamefont {Z.}~\bibnamefont {Ren}}, \bibinfo {author}
  {\bibfnamefont {H.}~\bibnamefont {Ma}}, \bibinfo {author} {\bibfnamefont
  {P.}~\bibnamefont {Yang}}, \bibinfo {author} {\bibfnamefont {Z.}~\bibnamefont
  {Liu}}, \bibinfo {author} {\bibfnamefont {Y.}~\bibnamefont {Liu}}, \bibinfo
  {author} {\bibfnamefont {H.}~\bibnamefont {Zhang}}, \bibinfo {author}
  {\bibfnamefont {X.}~\bibnamefont {Dong}}, \bibinfo {author} {\bibfnamefont
  {Y.}~\bibnamefont {Wang}}, \bibinfo {author} {\bibfnamefont {K.}~\bibnamefont
  {Jiang}}, \bibinfo {author} {\bibfnamefont {J.}~\bibnamefont {Hu}}, \bibinfo
  {author} {\bibfnamefont {S.}~\bibnamefont {Nagasaki}}, \bibinfo {author}
  {\bibfnamefont {K.}~\bibnamefont {Kitagawa}}, \bibinfo {author}
  {\bibfnamefont {S.}~\bibnamefont {Calder}}, \bibinfo {author} {\bibfnamefont
  {J.}~\bibnamefont {Yan}}, \bibinfo {author} {\bibfnamefont {J.}~\bibnamefont
  {Sun}}, \bibinfo {author} {\bibfnamefont {B.}~\bibnamefont {Wang}}, \bibinfo
  {author} {\bibfnamefont {R.}~\bibnamefont {Zhou}}, \bibinfo {author}
  {\bibfnamefont {Y.}~\bibnamefont {Uwatoko}},\ and\ \bibinfo {author}
  {\bibfnamefont {J.}~\bibnamefont {Cheng}},\ }\bibfield  {title} {\bibinfo
  {title} {Bulk high-temperature superconductivity in pressurized tetragonal
  {La$_{2}$PrNi$_{2}$O$_7$}},\ }\href
  {https://doi.org/10.1038/s41586-024-07996-8} {\bibfield  {journal} {\bibinfo
  {journal} {Nature}\ }\textbf {\bibinfo {volume} {634}},\ \bibinfo {pages}
  {579} (\bibinfo {year} {2024}{\natexlab{a}})}\BibitemShut {NoStop}%
\bibitem [{\citenamefont {Ko}\ \emph {et~al.}(2024)\citenamefont {Ko},
  \citenamefont {Yu}, \citenamefont {Liu}, \citenamefont {Bhatt}, \citenamefont
  {Li}, \citenamefont {Thampy}, \citenamefont {Kuo}, \citenamefont {Wang},
  \citenamefont {Lee}, \citenamefont {Lee}, \citenamefont {Lee}, \citenamefont
  {Goodge}, \citenamefont {Muller},\ and\ \citenamefont
  {Hwang}}]{Ko2024ambient}%
  \BibitemOpen
  \bibfield  {author} {\bibinfo {author} {\bibfnamefont {E.~K.}\ \bibnamefont
  {Ko}}, \bibinfo {author} {\bibfnamefont {Y.}~\bibnamefont {Yu}}, \bibinfo
  {author} {\bibfnamefont {Y.}~\bibnamefont {Liu}}, \bibinfo {author}
  {\bibfnamefont {L.}~\bibnamefont {Bhatt}}, \bibinfo {author} {\bibfnamefont
  {J.}~\bibnamefont {Li}}, \bibinfo {author} {\bibfnamefont {V.}~\bibnamefont
  {Thampy}}, \bibinfo {author} {\bibfnamefont {C.-T.}\ \bibnamefont {Kuo}},
  \bibinfo {author} {\bibfnamefont {B.~Y.}\ \bibnamefont {Wang}}, \bibinfo
  {author} {\bibfnamefont {Y.}~\bibnamefont {Lee}}, \bibinfo {author}
  {\bibfnamefont {K.}~\bibnamefont {Lee}}, \bibinfo {author} {\bibfnamefont
  {J.-S.}\ \bibnamefont {Lee}}, \bibinfo {author} {\bibfnamefont {B.~H.}\
  \bibnamefont {Goodge}}, \bibinfo {author} {\bibfnamefont {D.~A.}\
  \bibnamefont {Muller}},\ and\ \bibinfo {author} {\bibfnamefont {H.~Y.}\
  \bibnamefont {Hwang}},\ }\bibfield  {title} {\bibinfo {title} {Signatures of
  ambient pressure superconductivity in thin film {La$_{3}$Ni$_{2}$O$_7$}},\
  }\bibfield  {journal} {\bibinfo  {journal} {Nature}\ }\href
  {https://doi.org/10.1038/s41586-024-08525-3} {10.1038/s41586-024-08525-3}
  (\bibinfo {year} {2024})\BibitemShut {NoStop}%
\bibitem [{\citenamefont {Liu}\ \emph {et~al.}(2024{\natexlab{a}})\citenamefont
  {Liu}, \citenamefont {Huo}, \citenamefont {Li}, \citenamefont {Li},
  \citenamefont {Liu}, \citenamefont {Dai}, \citenamefont {Zhou}, \citenamefont
  {Hao}, \citenamefont {Lu}, \citenamefont {Wang},\ and\ \citenamefont
  {Wen}}]{Liu2023correlation}%
  \BibitemOpen
  \bibfield  {author} {\bibinfo {author} {\bibfnamefont {Z.}~\bibnamefont
  {Liu}}, \bibinfo {author} {\bibfnamefont {M.}~\bibnamefont {Huo}}, \bibinfo
  {author} {\bibfnamefont {J.}~\bibnamefont {Li}}, \bibinfo {author}
  {\bibfnamefont {Q.}~\bibnamefont {Li}}, \bibinfo {author} {\bibfnamefont
  {Y.}~\bibnamefont {Liu}}, \bibinfo {author} {\bibfnamefont {Y.}~\bibnamefont
  {Dai}}, \bibinfo {author} {\bibfnamefont {X.}~\bibnamefont {Zhou}}, \bibinfo
  {author} {\bibfnamefont {J.}~\bibnamefont {Hao}}, \bibinfo {author}
  {\bibfnamefont {Y.}~\bibnamefont {Lu}}, \bibinfo {author} {\bibfnamefont
  {M.}~\bibnamefont {Wang}},\ and\ \bibinfo {author} {\bibfnamefont {H.-H.}\
  \bibnamefont {Wen}},\ }\bibfield  {title} {\bibinfo {title} {Electronic
  correlations and partial gap in the bilayer nickelate
  {La$_{3}$Ni$_{2}$O$_7$}},\ }\href
  {https://doi.org/10.1038/s41467-024-52001-5} {\bibfield  {journal} {\bibinfo
  {journal} {Nature Communications}\ }\textbf {\bibinfo {volume} {15}},\
  \bibinfo {pages} {7570} (\bibinfo {year} {2024}{\natexlab{a}})}\BibitemShut
  {NoStop}%
\bibitem [{\citenamefont {Hou}\ \emph {et~al.}(2023)\citenamefont {Hou},
  \citenamefont {Yang}, \citenamefont {Liu}, \citenamefont {Li}, \citenamefont
  {Shan}, \citenamefont {Ma}, \citenamefont {Wang}, \citenamefont {Wang},
  \citenamefont {Guo}, \citenamefont {Sun}, \citenamefont {Uwatoko},
  \citenamefont {Wang}, \citenamefont {Zhang}, \citenamefont {Wang},\ and\
  \citenamefont {Cheng}}]{Hou2023emergence}%
  \BibitemOpen
  \bibfield  {author} {\bibinfo {author} {\bibfnamefont {J.}~\bibnamefont
  {Hou}}, \bibinfo {author} {\bibfnamefont {P.-T.}\ \bibnamefont {Yang}},
  \bibinfo {author} {\bibfnamefont {Z.-Y.}\ \bibnamefont {Liu}}, \bibinfo
  {author} {\bibfnamefont {J.-Y.}\ \bibnamefont {Li}}, \bibinfo {author}
  {\bibfnamefont {P.-F.}\ \bibnamefont {Shan}}, \bibinfo {author}
  {\bibfnamefont {L.}~\bibnamefont {Ma}}, \bibinfo {author} {\bibfnamefont
  {G.}~\bibnamefont {Wang}}, \bibinfo {author} {\bibfnamefont {N.-N.}\
  \bibnamefont {Wang}}, \bibinfo {author} {\bibfnamefont {H.-Z.}\ \bibnamefont
  {Guo}}, \bibinfo {author} {\bibfnamefont {J.-P.}\ \bibnamefont {Sun}},
  \bibinfo {author} {\bibfnamefont {Y.}~\bibnamefont {Uwatoko}}, \bibinfo
  {author} {\bibfnamefont {M.}~\bibnamefont {Wang}}, \bibinfo {author}
  {\bibfnamefont {G.-M.}\ \bibnamefont {Zhang}}, \bibinfo {author}
  {\bibfnamefont {B.-S.}\ \bibnamefont {Wang}},\ and\ \bibinfo {author}
  {\bibfnamefont {J.-G.}\ \bibnamefont {Cheng}},\ }\bibfield  {title} {\bibinfo
  {title} {{Emergence of High-Temperature Superconducting Phase in Pressurized
  {La$_{3}$Ni$_{2}$O$_7$} Crystals}},\ }\href
  {https://doi.org/10.1088/0256-307X/40/11/117302} {\bibfield  {journal}
  {\bibinfo  {journal} {Chinese Physics Letters}\ }\textbf {\bibinfo {volume}
  {40}},\ \bibinfo {eid} {117302} (\bibinfo {year} {2023})}\BibitemShut
  {NoStop}%
\bibitem [{\citenamefont {Zhang}\ \emph
  {et~al.}(2024{\natexlab{a}})\citenamefont {Zhang}, \citenamefont {Su},
  \citenamefont {Huang}, \citenamefont {Shan}, \citenamefont {Sun},
  \citenamefont {Huo}, \citenamefont {Ye}, \citenamefont {Zhang}, \citenamefont
  {Yang}, \citenamefont {Xu}, \citenamefont {Su}, \citenamefont {Li},
  \citenamefont {Smidman}, \citenamefont {Wang}, \citenamefont {Jiao},\ and\
  \citenamefont {Yuan}}]{Zhang2023hightemperature}%
  \BibitemOpen
  \bibfield  {author} {\bibinfo {author} {\bibfnamefont {Y.}~\bibnamefont
  {Zhang}}, \bibinfo {author} {\bibfnamefont {D.}~\bibnamefont {Su}}, \bibinfo
  {author} {\bibfnamefont {Y.}~\bibnamefont {Huang}}, \bibinfo {author}
  {\bibfnamefont {Z.}~\bibnamefont {Shan}}, \bibinfo {author} {\bibfnamefont
  {H.}~\bibnamefont {Sun}}, \bibinfo {author} {\bibfnamefont {M.}~\bibnamefont
  {Huo}}, \bibinfo {author} {\bibfnamefont {K.}~\bibnamefont {Ye}}, \bibinfo
  {author} {\bibfnamefont {J.}~\bibnamefont {Zhang}}, \bibinfo {author}
  {\bibfnamefont {Z.}~\bibnamefont {Yang}}, \bibinfo {author} {\bibfnamefont
  {Y.}~\bibnamefont {Xu}}, \bibinfo {author} {\bibfnamefont {Y.}~\bibnamefont
  {Su}}, \bibinfo {author} {\bibfnamefont {R.}~\bibnamefont {Li}}, \bibinfo
  {author} {\bibfnamefont {M.}~\bibnamefont {Smidman}}, \bibinfo {author}
  {\bibfnamefont {M.}~\bibnamefont {Wang}}, \bibinfo {author} {\bibfnamefont
  {L.}~\bibnamefont {Jiao}},\ and\ \bibinfo {author} {\bibfnamefont
  {H.}~\bibnamefont {Yuan}},\ }\bibfield  {title} {\bibinfo {title}
  {High-temperature superconductivity with zero resistance and strange-metal
  behaviour in {La$_3$Ni$_2$O$_{7-\delta}$}},\ }\href
  {https://doi.org/10.1038/s41567-024-02515-y} {\bibfield  {journal} {\bibinfo
  {journal} {Nature Physics}\ }\textbf {\bibinfo {volume} {20}},\ \bibinfo
  {pages} {1269} (\bibinfo {year} {2024}{\natexlab{a}})}\BibitemShut {NoStop}%
\bibitem [{\citenamefont {Yang}\ \emph
  {et~al.}(2024{\natexlab{a}})\citenamefont {Yang}, \citenamefont {Sun},
  \citenamefont {Hu}, \citenamefont {Xie}, \citenamefont {Miao}, \citenamefont
  {Luo}, \citenamefont {Chen}, \citenamefont {Liang}, \citenamefont {Zhu},
  \citenamefont {Qu}, \citenamefont {Chen}, \citenamefont {Huo}, \citenamefont
  {Huang}, \citenamefont {Zhang}, \citenamefont {Zhang}, \citenamefont {Yang},
  \citenamefont {Wang}, \citenamefont {Peng}, \citenamefont {Mao},
  \citenamefont {Liu}, \citenamefont {Xu}, \citenamefont {Qian}, \citenamefont
  {Yao}, \citenamefont {Wang}, \citenamefont {Zhao},\ and\ \citenamefont
  {Zhou}}]{yang2023orbitaldependent}%
  \BibitemOpen
  \bibfield  {author} {\bibinfo {author} {\bibfnamefont {J.}~\bibnamefont
  {Yang}}, \bibinfo {author} {\bibfnamefont {H.}~\bibnamefont {Sun}}, \bibinfo
  {author} {\bibfnamefont {X.}~\bibnamefont {Hu}}, \bibinfo {author}
  {\bibfnamefont {Y.}~\bibnamefont {Xie}}, \bibinfo {author} {\bibfnamefont
  {T.}~\bibnamefont {Miao}}, \bibinfo {author} {\bibfnamefont {H.}~\bibnamefont
  {Luo}}, \bibinfo {author} {\bibfnamefont {H.}~\bibnamefont {Chen}}, \bibinfo
  {author} {\bibfnamefont {B.}~\bibnamefont {Liang}}, \bibinfo {author}
  {\bibfnamefont {W.}~\bibnamefont {Zhu}}, \bibinfo {author} {\bibfnamefont
  {G.}~\bibnamefont {Qu}}, \bibinfo {author} {\bibfnamefont {C.-Q.}\
  \bibnamefont {Chen}}, \bibinfo {author} {\bibfnamefont {M.}~\bibnamefont
  {Huo}}, \bibinfo {author} {\bibfnamefont {Y.}~\bibnamefont {Huang}}, \bibinfo
  {author} {\bibfnamefont {S.}~\bibnamefont {Zhang}}, \bibinfo {author}
  {\bibfnamefont {F.}~\bibnamefont {Zhang}}, \bibinfo {author} {\bibfnamefont
  {F.}~\bibnamefont {Yang}}, \bibinfo {author} {\bibfnamefont {Z.}~\bibnamefont
  {Wang}}, \bibinfo {author} {\bibfnamefont {Q.}~\bibnamefont {Peng}}, \bibinfo
  {author} {\bibfnamefont {H.}~\bibnamefont {Mao}}, \bibinfo {author}
  {\bibfnamefont {G.}~\bibnamefont {Liu}}, \bibinfo {author} {\bibfnamefont
  {Z.}~\bibnamefont {Xu}}, \bibinfo {author} {\bibfnamefont {T.}~\bibnamefont
  {Qian}}, \bibinfo {author} {\bibfnamefont {D.-X.}\ \bibnamefont {Yao}},
  \bibinfo {author} {\bibfnamefont {M.}~\bibnamefont {Wang}}, \bibinfo {author}
  {\bibfnamefont {L.}~\bibnamefont {Zhao}},\ and\ \bibinfo {author}
  {\bibfnamefont {X.~J.}\ \bibnamefont {Zhou}},\ }\bibfield  {title} {\bibinfo
  {title} {Orbital-dependent electron correlation in double-layer nickelate
  {La$_3$Ni$_2$O$_7$}},\ }\href {https://doi.org/10.1038/s41467-024-48701-7}
  {\bibfield  {journal} {\bibinfo  {journal} {Nature Communications}\ }\textbf
  {\bibinfo {volume} {15}},\ \bibinfo {pages} {4373} (\bibinfo {year}
  {2024}{\natexlab{a}})}\BibitemShut {NoStop}%
\bibitem [{\citenamefont {Zhang}\ \emph
  {et~al.}(2024{\natexlab{b}})\citenamefont {Zhang}, \citenamefont {Pei},
  \citenamefont {Wang}, \citenamefont {Zhao}, \citenamefont {Li}, \citenamefont
  {Cao}, \citenamefont {Zhu}, \citenamefont {Wu},\ and\ \citenamefont
  {Qi}}]{zhang2023effects}%
  \BibitemOpen
  \bibfield  {author} {\bibinfo {author} {\bibfnamefont {M.}~\bibnamefont
  {Zhang}}, \bibinfo {author} {\bibfnamefont {C.}~\bibnamefont {Pei}}, \bibinfo
  {author} {\bibfnamefont {Q.}~\bibnamefont {Wang}}, \bibinfo {author}
  {\bibfnamefont {Y.}~\bibnamefont {Zhao}}, \bibinfo {author} {\bibfnamefont
  {C.}~\bibnamefont {Li}}, \bibinfo {author} {\bibfnamefont {W.}~\bibnamefont
  {Cao}}, \bibinfo {author} {\bibfnamefont {S.}~\bibnamefont {Zhu}}, \bibinfo
  {author} {\bibfnamefont {J.}~\bibnamefont {Wu}},\ and\ \bibinfo {author}
  {\bibfnamefont {Y.}~\bibnamefont {Qi}},\ }\bibfield  {title} {\bibinfo
  {title} {Effects of pressure and doping on {Ruddlesden-Popper} phases
  {La$_{n+1}$Ni$_{n}$O$_{3n+1}$}},\ }\href
  {https://doi.org/https://doi.org/10.1016/j.jmst.2023.11.011} {\bibfield
  {journal} {\bibinfo  {journal} {Journal of Materials Science \& Technology}\
  }\textbf {\bibinfo {volume} {185}},\ \bibinfo {pages} {147} (\bibinfo {year}
  {2024}{\natexlab{b}})}\BibitemShut {NoStop}%
\bibitem [{\citenamefont {Wang}\ \emph
  {et~al.}(2024{\natexlab{b}})\citenamefont {Wang}, \citenamefont {Wang},
  \citenamefont {Shen}, \citenamefont {Hou}, \citenamefont {Ma}, \citenamefont
  {Shi}, \citenamefont {Ren}, \citenamefont {Gu}, \citenamefont {Ma},
  \citenamefont {Yang}, \citenamefont {Liu}, \citenamefont {Guo}, \citenamefont
  {Sun}, \citenamefont {Zhang}, \citenamefont {Calder}, \citenamefont {Yan},
  \citenamefont {Wang}, \citenamefont {Uwatoko},\ and\ \citenamefont
  {Cheng}}]{fwang2023pressureinduced}%
  \BibitemOpen
  \bibfield  {author} {\bibinfo {author} {\bibfnamefont {G.}~\bibnamefont
  {Wang}}, \bibinfo {author} {\bibfnamefont {N.~N.}\ \bibnamefont {Wang}},
  \bibinfo {author} {\bibfnamefont {X.~L.}\ \bibnamefont {Shen}}, \bibinfo
  {author} {\bibfnamefont {J.}~\bibnamefont {Hou}}, \bibinfo {author}
  {\bibfnamefont {L.}~\bibnamefont {Ma}}, \bibinfo {author} {\bibfnamefont
  {L.~F.}\ \bibnamefont {Shi}}, \bibinfo {author} {\bibfnamefont {Z.~A.}\
  \bibnamefont {Ren}}, \bibinfo {author} {\bibfnamefont {Y.~D.}\ \bibnamefont
  {Gu}}, \bibinfo {author} {\bibfnamefont {H.~M.}\ \bibnamefont {Ma}}, \bibinfo
  {author} {\bibfnamefont {P.~T.}\ \bibnamefont {Yang}}, \bibinfo {author}
  {\bibfnamefont {Z.~Y.}\ \bibnamefont {Liu}}, \bibinfo {author} {\bibfnamefont
  {H.~Z.}\ \bibnamefont {Guo}}, \bibinfo {author} {\bibfnamefont {J.~P.}\
  \bibnamefont {Sun}}, \bibinfo {author} {\bibfnamefont {G.~M.}\ \bibnamefont
  {Zhang}}, \bibinfo {author} {\bibfnamefont {S.}~\bibnamefont {Calder}},
  \bibinfo {author} {\bibfnamefont {J.-Q.}\ \bibnamefont {Yan}}, \bibinfo
  {author} {\bibfnamefont {B.~S.}\ \bibnamefont {Wang}}, \bibinfo {author}
  {\bibfnamefont {Y.}~\bibnamefont {Uwatoko}},\ and\ \bibinfo {author}
  {\bibfnamefont {J.-G.}\ \bibnamefont {Cheng}},\ }\bibfield  {title} {\bibinfo
  {title} {{Pressure-Induced Superconductivity In Polycrystalline
  {${\mathrm{La}}_{3}{\mathrm{Ni}}_{2}{\mathrm{O}}_{7-\delta}$}}},\ }\href
  {https://doi.org/10.1103/PhysRevX.14.011040} {\bibfield  {journal} {\bibinfo
  {journal} {Phys. Rev. X}\ }\textbf {\bibinfo {volume} {14}},\ \bibinfo
  {pages} {011040} (\bibinfo {year} {2024}{\natexlab{b}})}\BibitemShut
  {NoStop}%
\bibitem [{\citenamefont {Wang}\ \emph {et~al.}(2023)\citenamefont {Wang},
  \citenamefont {Wang}, \citenamefont {Wang}, \citenamefont {Shi},
  \citenamefont {Shen}, \citenamefont {Hou}, \citenamefont {Ma}, \citenamefont
  {Yang}, \citenamefont {Liu}, \citenamefont {Zhang}, \citenamefont {Dong},
  \citenamefont {Sun}, \citenamefont {Wang}, \citenamefont {Jiang},
  \citenamefont {Hu}, \citenamefont {Uwatoko},\ and\ \citenamefont
  {Cheng}}]{wang2023observation}%
  \BibitemOpen
  \bibfield  {author} {\bibinfo {author} {\bibfnamefont {G.}~\bibnamefont
  {Wang}}, \bibinfo {author} {\bibfnamefont {N.}~\bibnamefont {Wang}}, \bibinfo
  {author} {\bibfnamefont {Y.}~\bibnamefont {Wang}}, \bibinfo {author}
  {\bibfnamefont {L.}~\bibnamefont {Shi}}, \bibinfo {author} {\bibfnamefont
  {X.}~\bibnamefont {Shen}}, \bibinfo {author} {\bibfnamefont {J.}~\bibnamefont
  {Hou}}, \bibinfo {author} {\bibfnamefont {H.}~\bibnamefont {Ma}}, \bibinfo
  {author} {\bibfnamefont {P.}~\bibnamefont {Yang}}, \bibinfo {author}
  {\bibfnamefont {Z.}~\bibnamefont {Liu}}, \bibinfo {author} {\bibfnamefont
  {H.}~\bibnamefont {Zhang}}, \bibinfo {author} {\bibfnamefont
  {X.}~\bibnamefont {Dong}}, \bibinfo {author} {\bibfnamefont {J.}~\bibnamefont
  {Sun}}, \bibinfo {author} {\bibfnamefont {B.}~\bibnamefont {Wang}}, \bibinfo
  {author} {\bibfnamefont {K.}~\bibnamefont {Jiang}}, \bibinfo {author}
  {\bibfnamefont {J.}~\bibnamefont {Hu}}, \bibinfo {author} {\bibfnamefont
  {Y.}~\bibnamefont {Uwatoko}},\ and\ \bibinfo {author} {\bibfnamefont
  {J.}~\bibnamefont {Cheng}},\ }\href@noop {} {\bibinfo {title} {Observation of
  high-temperature superconductivity in the high-pressure tetragonal phase of
  {La$_2$PrNi$_2$O$_{7-\delta}$}}} (\bibinfo {year} {2023}),\ \Eprint
  {https://arxiv.org/abs/2311.08212} {arXiv:2311.08212 [cond-mat.supr-con]}
  \BibitemShut {NoStop}%
\bibitem [{\citenamefont {Cui}\ \emph {et~al.}(2024)\citenamefont {Cui},
  \citenamefont {Choi}, \citenamefont {Lin}, \citenamefont {Liu}, \citenamefont
  {Wang}, \citenamefont {Wang}, \citenamefont {Chen}, \citenamefont {Hong},
  \citenamefont {Rong}, \citenamefont {Wang}, \citenamefont {Jin},
  \citenamefont {Wang}, \citenamefont {Gu}, \citenamefont {Ge}, \citenamefont
  {Wang}, \citenamefont {Cheng}, \citenamefont {Zhang}, \citenamefont {Si},
  \citenamefont {Jin},\ and\ \citenamefont {Guo}}]{cui2023strain}%
  \BibitemOpen
  \bibfield  {author} {\bibinfo {author} {\bibfnamefont {T.}~\bibnamefont
  {Cui}}, \bibinfo {author} {\bibfnamefont {S.}~\bibnamefont {Choi}}, \bibinfo
  {author} {\bibfnamefont {T.}~\bibnamefont {Lin}}, \bibinfo {author}
  {\bibfnamefont {C.}~\bibnamefont {Liu}}, \bibinfo {author} {\bibfnamefont
  {G.}~\bibnamefont {Wang}}, \bibinfo {author} {\bibfnamefont {N.}~\bibnamefont
  {Wang}}, \bibinfo {author} {\bibfnamefont {S.}~\bibnamefont {Chen}}, \bibinfo
  {author} {\bibfnamefont {H.}~\bibnamefont {Hong}}, \bibinfo {author}
  {\bibfnamefont {D.}~\bibnamefont {Rong}}, \bibinfo {author} {\bibfnamefont
  {Q.}~\bibnamefont {Wang}}, \bibinfo {author} {\bibfnamefont {Q.}~\bibnamefont
  {Jin}}, \bibinfo {author} {\bibfnamefont {J.-O.}\ \bibnamefont {Wang}},
  \bibinfo {author} {\bibfnamefont {L.}~\bibnamefont {Gu}}, \bibinfo {author}
  {\bibfnamefont {C.}~\bibnamefont {Ge}}, \bibinfo {author} {\bibfnamefont
  {C.}~\bibnamefont {Wang}}, \bibinfo {author} {\bibfnamefont {J.-G.}\
  \bibnamefont {Cheng}}, \bibinfo {author} {\bibfnamefont {Q.}~\bibnamefont
  {Zhang}}, \bibinfo {author} {\bibfnamefont {L.}~\bibnamefont {Si}}, \bibinfo
  {author} {\bibfnamefont {K.-j.}\ \bibnamefont {Jin}},\ and\ \bibinfo {author}
  {\bibfnamefont {E.-J.}\ \bibnamefont {Guo}},\ }\bibfield  {title} {\bibinfo
  {title} {Strain-mediated phase crossover in {Ruddlesden--Popper}
  nickelates},\ }\href {https://doi.org/10.1038/s43246-024-00478-4} {\bibfield
  {journal} {\bibinfo  {journal} {Communications Materials}\ }\textbf {\bibinfo
  {volume} {5}},\ \bibinfo {pages} {32} (\bibinfo {year} {2024})}\BibitemShut
  {NoStop}%
\bibitem [{\citenamefont {Chen}\ \emph
  {et~al.}(2024{\natexlab{a}})\citenamefont {Chen}, \citenamefont {Liu},
  \citenamefont {Jiao}, \citenamefont {Zou}, \citenamefont {Jiang},
  \citenamefont {Li}, \citenamefont {Luo}, \citenamefont {Wu}, \citenamefont
  {Zhang}, \citenamefont {Guo},\ and\ \citenamefont {Shu}}]{chen2023evidence}%
  \BibitemOpen
  \bibfield  {author} {\bibinfo {author} {\bibfnamefont {K.}~\bibnamefont
  {Chen}}, \bibinfo {author} {\bibfnamefont {X.}~\bibnamefont {Liu}}, \bibinfo
  {author} {\bibfnamefont {J.}~\bibnamefont {Jiao}}, \bibinfo {author}
  {\bibfnamefont {M.}~\bibnamefont {Zou}}, \bibinfo {author} {\bibfnamefont
  {C.}~\bibnamefont {Jiang}}, \bibinfo {author} {\bibfnamefont
  {X.}~\bibnamefont {Li}}, \bibinfo {author} {\bibfnamefont {Y.}~\bibnamefont
  {Luo}}, \bibinfo {author} {\bibfnamefont {Q.}~\bibnamefont {Wu}}, \bibinfo
  {author} {\bibfnamefont {N.}~\bibnamefont {Zhang}}, \bibinfo {author}
  {\bibfnamefont {Y.}~\bibnamefont {Guo}},\ and\ \bibinfo {author}
  {\bibfnamefont {L.}~\bibnamefont {Shu}},\ }\bibfield  {title} {\bibinfo
  {title} {{Evidence of Spin Density Waves in
  {${\mathrm{La}}_{3}{\mathrm{Ni}}_{2}{\mathrm{O}}_{7-\delta}$}}},\ }\href
  {https://doi.org/10.1103/PhysRevLett.132.256503} {\bibfield  {journal}
  {\bibinfo  {journal} {Phys. Rev. Lett.}\ }\textbf {\bibinfo {volume} {132}},\
  \bibinfo {pages} {256503} (\bibinfo {year} {2024}{\natexlab{a}})}\BibitemShut
  {NoStop}%
\bibitem [{\citenamefont {Kumar}\ \emph {et~al.}(2025)\citenamefont {Kumar},
  \citenamefont {Melnick},\ and\ \citenamefont {Kotliar}}]{kumar2023softening}%
  \BibitemOpen
  \bibfield  {author} {\bibinfo {author} {\bibfnamefont {U.}~\bibnamefont
  {Kumar}}, \bibinfo {author} {\bibfnamefont {C.}~\bibnamefont {Melnick}},\
  and\ \bibinfo {author} {\bibfnamefont {G.}~\bibnamefont {Kotliar}},\
  }\bibfield  {title} {\bibinfo {title} {{Softening of $\mathit{dd}$ excitation
  in the resonant inelastic x-ray scattering spectra as a signature of Hund's
  coupling in nickelates}},\ }\href
  {https://doi.org/10.1103/PhysRevResearch.7.L012066} {\bibfield  {journal}
  {\bibinfo  {journal} {Phys. Rev. Res.}\ }\textbf {\bibinfo {volume} {7}},\
  \bibinfo {pages} {L012066} (\bibinfo {year} {2025})}\BibitemShut {NoStop}%
\bibitem [{\citenamefont {Dong}\ \emph {et~al.}(2024)\citenamefont {Dong},
  \citenamefont {Huo}, \citenamefont {Li}, \citenamefont {Li}, \citenamefont
  {Li}, \citenamefont {Sun}, \citenamefont {Gu}, \citenamefont {Lu},
  \citenamefont {Wang}, \citenamefont {Wang},\ and\ \citenamefont
  {Chen}}]{Dong2024}%
  \BibitemOpen
  \bibfield  {author} {\bibinfo {author} {\bibfnamefont {Z.}~\bibnamefont
  {Dong}}, \bibinfo {author} {\bibfnamefont {M.}~\bibnamefont {Huo}}, \bibinfo
  {author} {\bibfnamefont {J.}~\bibnamefont {Li}}, \bibinfo {author}
  {\bibfnamefont {J.}~\bibnamefont {Li}}, \bibinfo {author} {\bibfnamefont
  {P.}~\bibnamefont {Li}}, \bibinfo {author} {\bibfnamefont {H.}~\bibnamefont
  {Sun}}, \bibinfo {author} {\bibfnamefont {L.}~\bibnamefont {Gu}}, \bibinfo
  {author} {\bibfnamefont {Y.}~\bibnamefont {Lu}}, \bibinfo {author}
  {\bibfnamefont {M.}~\bibnamefont {Wang}}, \bibinfo {author} {\bibfnamefont
  {Y.}~\bibnamefont {Wang}},\ and\ \bibinfo {author} {\bibfnamefont
  {Z.}~\bibnamefont {Chen}},\ }\bibfield  {title} {\bibinfo {title}
  {Visualization of oxygen vacancies and self-doped ligand holes in
  {La$_3$Ni$_2$O$_{7-\delta}$}},\ }\href
  {https://doi.org/10.1038/s41586-024-07482-1} {\bibfield  {journal} {\bibinfo
  {journal} {Nature}\ }\textbf {\bibinfo {volume} {630}},\ \bibinfo {pages}
  {847} (\bibinfo {year} {2024})}\BibitemShut {NoStop}%
\bibitem [{\citenamefont {Li}\ \emph {et~al.}(2025)\citenamefont {Li},
  \citenamefont {Peng}, \citenamefont {Ma}, \citenamefont {Zhang},
  \citenamefont {Xing}, \citenamefont {Huang}, \citenamefont {Huang},
  \citenamefont {Huo}, \citenamefont {Hu}, \citenamefont {Dong}, \citenamefont
  {Chen}, \citenamefont {Xie}, \citenamefont {Dong}, \citenamefont {Sun},
  \citenamefont {Zeng}, \citenamefont {Mao},\ and\ \citenamefont
  {Wang}}]{li2024pressuredriven}%
  \BibitemOpen
  \bibfield  {author} {\bibinfo {author} {\bibfnamefont {J.}~\bibnamefont
  {Li}}, \bibinfo {author} {\bibfnamefont {D.}~\bibnamefont {Peng}}, \bibinfo
  {author} {\bibfnamefont {P.}~\bibnamefont {Ma}}, \bibinfo {author}
  {\bibfnamefont {H.}~\bibnamefont {Zhang}}, \bibinfo {author} {\bibfnamefont
  {Z.}~\bibnamefont {Xing}}, \bibinfo {author} {\bibfnamefont {X.}~\bibnamefont
  {Huang}}, \bibinfo {author} {\bibfnamefont {C.}~\bibnamefont {Huang}},
  \bibinfo {author} {\bibfnamefont {M.}~\bibnamefont {Huo}}, \bibinfo {author}
  {\bibfnamefont {D.}~\bibnamefont {Hu}}, \bibinfo {author} {\bibfnamefont
  {Z.}~\bibnamefont {Dong}}, \bibinfo {author} {\bibfnamefont {X.}~\bibnamefont
  {Chen}}, \bibinfo {author} {\bibfnamefont {T.}~\bibnamefont {Xie}}, \bibinfo
  {author} {\bibfnamefont {H.}~\bibnamefont {Dong}}, \bibinfo {author}
  {\bibfnamefont {H.}~\bibnamefont {Sun}}, \bibinfo {author} {\bibfnamefont
  {Q.}~\bibnamefont {Zeng}}, \bibinfo {author} {\bibfnamefont {H.-k.}\
  \bibnamefont {Mao}},\ and\ \bibinfo {author} {\bibfnamefont {M.}~\bibnamefont
  {Wang}},\ }\bibfield  {title} {\bibinfo {title} {{Identification of
  Superconductivity in Bilayer Nickelate La$_3$Ni$_2$O$_7$ under High Pressure
  up to 100 GPa}},\ }\href {https://doi.org/10.1093/nsr/nwaf220} {\bibfield
  {journal} {\bibinfo  {journal} {National Science Review}\ ,\ \bibinfo {pages}
  {nwaf220}} (\bibinfo {year} {2025})}\BibitemShut {NoStop}%
\bibitem [{\citenamefont {Liu}\ \emph {et~al.}(2024{\natexlab{b}})\citenamefont
  {Liu}, \citenamefont {Ou}, \citenamefont {Chu}, \citenamefont {Yang},
  \citenamefont {Li}, \citenamefont {Zhang},\ and\ \citenamefont
  {Wen}}]{liu2024growth}%
  \BibitemOpen
  \bibfield  {author} {\bibinfo {author} {\bibfnamefont {Y.}~\bibnamefont
  {Liu}}, \bibinfo {author} {\bibfnamefont {M.}~\bibnamefont {Ou}}, \bibinfo
  {author} {\bibfnamefont {H.}~\bibnamefont {Chu}}, \bibinfo {author}
  {\bibfnamefont {H.}~\bibnamefont {Yang}}, \bibinfo {author} {\bibfnamefont
  {Q.}~\bibnamefont {Li}}, \bibinfo {author} {\bibfnamefont {Y.-J.}\
  \bibnamefont {Zhang}},\ and\ \bibinfo {author} {\bibfnamefont {H.-H.}\
  \bibnamefont {Wen}},\ }\bibfield  {title} {\bibinfo {title} {{Growth and
  characterization of the
  ${\mathrm{La}}_{3}{\mathrm{Ni}}_{2}{\mathrm{O}}_{7\ensuremath{-}\ensuremath{\delta}}$
  thin films: Dominant contribution of the ${d}_{{x}^{2}\ensuremath{-}{y}^{2}}$
  orbital at ambient pressure}},\ }\href
  {https://doi.org/10.1103/PhysRevMaterials.8.124801} {\bibfield  {journal}
  {\bibinfo  {journal} {Phys. Rev. Mater.}\ }\textbf {\bibinfo {volume} {8}},\
  \bibinfo {pages} {124801} (\bibinfo {year} {2024}{\natexlab{b}})}\BibitemShut
  {NoStop}%
\bibitem [{\citenamefont {Wen}\ \emph {et~al.}(2024)\citenamefont {Wen},
  \citenamefont {Xu}, \citenamefont {Wang}, \citenamefont {He}, \citenamefont
  {Chen}, \citenamefont {Wang}, \citenamefont {Lu}, \citenamefont {Ma},
  \citenamefont {Jin}, \citenamefont {Chen}, \citenamefont {Liu}, \citenamefont
  {Fan}, \citenamefont {Liu}, \citenamefont {Pan}, \citenamefont {Liu},
  \citenamefont {Cheng},\ and\ \citenamefont {Yu}}]{wen2024probing}%
  \BibitemOpen
  \bibfield  {author} {\bibinfo {author} {\bibfnamefont {J.}~\bibnamefont
  {Wen}}, \bibinfo {author} {\bibfnamefont {Y.}~\bibnamefont {Xu}}, \bibinfo
  {author} {\bibfnamefont {G.}~\bibnamefont {Wang}}, \bibinfo {author}
  {\bibfnamefont {Z.-X.}\ \bibnamefont {He}}, \bibinfo {author} {\bibfnamefont
  {Y.}~\bibnamefont {Chen}}, \bibinfo {author} {\bibfnamefont {N.}~\bibnamefont
  {Wang}}, \bibinfo {author} {\bibfnamefont {T.}~\bibnamefont {Lu}}, \bibinfo
  {author} {\bibfnamefont {X.}~\bibnamefont {Ma}}, \bibinfo {author}
  {\bibfnamefont {F.}~\bibnamefont {Jin}}, \bibinfo {author} {\bibfnamefont
  {L.}~\bibnamefont {Chen}}, \bibinfo {author} {\bibfnamefont {M.}~\bibnamefont
  {Liu}}, \bibinfo {author} {\bibfnamefont {J.-W.}\ \bibnamefont {Fan}},
  \bibinfo {author} {\bibfnamefont {X.}~\bibnamefont {Liu}}, \bibinfo {author}
  {\bibfnamefont {X.-Y.}\ \bibnamefont {Pan}}, \bibinfo {author} {\bibfnamefont
  {G.-Q.}\ \bibnamefont {Liu}}, \bibinfo {author} {\bibfnamefont
  {J.}~\bibnamefont {Cheng}},\ and\ \bibinfo {author} {\bibfnamefont
  {X.}~\bibnamefont {Yu}},\ }\href@noop {} {\bibinfo {title} {Probing the
  {Meissner} effect in pressurized bilayer nickelate superconductors using
  diamond quantum sensors}} (\bibinfo {year} {2024}),\ \Eprint
  {https://arxiv.org/abs/2410.10275} {arXiv:2410.10275 [cond-mat.supr-con]}
  \BibitemShut {NoStop}%
\bibitem [{\citenamefont {Chen}\ \emph
  {et~al.}(2024{\natexlab{b}})\citenamefont {Chen}, \citenamefont {Choi},
  \citenamefont {Jiang}, \citenamefont {Mei}, \citenamefont {Jiang},
  \citenamefont {Li}, \citenamefont {Agrestini}, \citenamefont
  {Garcia-Fernandez}, \citenamefont {Sun}, \citenamefont {Huang}, \citenamefont
  {Shen}, \citenamefont {Wang}, \citenamefont {Hu}, \citenamefont {Lu},
  \citenamefont {Zhou},\ and\ \citenamefont {Feng}}]{chen2024electronic}%
  \BibitemOpen
  \bibfield  {author} {\bibinfo {author} {\bibfnamefont {X.}~\bibnamefont
  {Chen}}, \bibinfo {author} {\bibfnamefont {J.}~\bibnamefont {Choi}}, \bibinfo
  {author} {\bibfnamefont {Z.}~\bibnamefont {Jiang}}, \bibinfo {author}
  {\bibfnamefont {J.}~\bibnamefont {Mei}}, \bibinfo {author} {\bibfnamefont
  {K.}~\bibnamefont {Jiang}}, \bibinfo {author} {\bibfnamefont
  {J.}~\bibnamefont {Li}}, \bibinfo {author} {\bibfnamefont {S.}~\bibnamefont
  {Agrestini}}, \bibinfo {author} {\bibfnamefont {M.}~\bibnamefont
  {Garcia-Fernandez}}, \bibinfo {author} {\bibfnamefont {H.}~\bibnamefont
  {Sun}}, \bibinfo {author} {\bibfnamefont {X.}~\bibnamefont {Huang}}, \bibinfo
  {author} {\bibfnamefont {D.}~\bibnamefont {Shen}}, \bibinfo {author}
  {\bibfnamefont {M.}~\bibnamefont {Wang}}, \bibinfo {author} {\bibfnamefont
  {J.}~\bibnamefont {Hu}}, \bibinfo {author} {\bibfnamefont {Y.}~\bibnamefont
  {Lu}}, \bibinfo {author} {\bibfnamefont {K.-J.}\ \bibnamefont {Zhou}},\ and\
  \bibinfo {author} {\bibfnamefont {D.}~\bibnamefont {Feng}},\ }\bibfield
  {title} {\bibinfo {title} {Electronic and magnetic excitations in
  {La$_3$Ni$_2$O$_7$}},\ }\href {https://doi.org/10.1038/s41467-024-53863-5}
  {\bibfield  {journal} {\bibinfo  {journal} {Nature Communications}\ }\textbf
  {\bibinfo {volume} {15}},\ \bibinfo {pages} {9597} (\bibinfo {year}
  {2024}{\natexlab{b}})}\BibitemShut {NoStop}%
\bibitem [{\citenamefont {Xie}\ \emph {et~al.}(2024)\citenamefont {Xie},
  \citenamefont {Huo}, \citenamefont {Ni}, \citenamefont {Shen}, \citenamefont
  {Huang}, \citenamefont {Sun}, \citenamefont {Walker}, \citenamefont {Adroja},
  \citenamefont {Yu}, \citenamefont {Shen}, \citenamefont {He}, \citenamefont
  {Cao},\ and\ \citenamefont {Wang}}]{xie2024neutron}%
  \BibitemOpen
  \bibfield  {author} {\bibinfo {author} {\bibfnamefont {T.}~\bibnamefont
  {Xie}}, \bibinfo {author} {\bibfnamefont {M.}~\bibnamefont {Huo}}, \bibinfo
  {author} {\bibfnamefont {X.}~\bibnamefont {Ni}}, \bibinfo {author}
  {\bibfnamefont {F.}~\bibnamefont {Shen}}, \bibinfo {author} {\bibfnamefont
  {X.}~\bibnamefont {Huang}}, \bibinfo {author} {\bibfnamefont
  {H.}~\bibnamefont {Sun}}, \bibinfo {author} {\bibfnamefont {H.~C.}\
  \bibnamefont {Walker}}, \bibinfo {author} {\bibfnamefont {D.}~\bibnamefont
  {Adroja}}, \bibinfo {author} {\bibfnamefont {D.}~\bibnamefont {Yu}}, \bibinfo
  {author} {\bibfnamefont {B.}~\bibnamefont {Shen}}, \bibinfo {author}
  {\bibfnamefont {L.}~\bibnamefont {He}}, \bibinfo {author} {\bibfnamefont
  {K.}~\bibnamefont {Cao}},\ and\ \bibinfo {author} {\bibfnamefont
  {M.}~\bibnamefont {Wang}},\ }\bibfield  {title} {\bibinfo {title} {Strong
  interlayer magnetic exchange coupling in {La$_3$Ni$_2$O$_{7-\delta}$}
  revealed by inelastic neutron scattering},\ }\href
  {https://doi.org/https://doi.org/10.1016/j.scib.2024.07.030} {\bibfield
  {journal} {\bibinfo  {journal} {Science Bulletin}\ }\textbf {\bibinfo
  {volume} {69}},\ \bibinfo {pages} {3221} (\bibinfo {year}
  {2024})}\BibitemShut {NoStop}%
\bibitem [{\citenamefont {Luo}\ \emph {et~al.}(2023)\citenamefont {Luo},
  \citenamefont {Hu}, \citenamefont {Wang}, \citenamefont {W\'u},\ and\
  \citenamefont {Yao}}]{Luo2023Model}%
  \BibitemOpen
  \bibfield  {author} {\bibinfo {author} {\bibfnamefont {Z.}~\bibnamefont
  {Luo}}, \bibinfo {author} {\bibfnamefont {X.}~\bibnamefont {Hu}}, \bibinfo
  {author} {\bibfnamefont {M.}~\bibnamefont {Wang}}, \bibinfo {author}
  {\bibfnamefont {W.}~\bibnamefont {W\'u}},\ and\ \bibinfo {author}
  {\bibfnamefont {D.-X.}\ \bibnamefont {Yao}},\ }\bibfield  {title} {\bibinfo
  {title} {{Bilayer Two-Orbital Model of
  $\mathrm{L}{\mathrm{a}}_{3}\mathrm{N}{\mathrm{i}}_{2}{\mathrm{O}}_{7}$ under
  Pressure}},\ }\href {https://doi.org/10.1103/PhysRevLett.131.126001}
  {\bibfield  {journal} {\bibinfo  {journal} {Phys. Rev. Lett.}\ }\textbf
  {\bibinfo {volume} {131}},\ \bibinfo {pages} {126001} (\bibinfo {year}
  {2023})}\BibitemShut {NoStop}%
\bibitem [{\citenamefont {Zhang}\ \emph
  {et~al.}(2023{\natexlab{a}})\citenamefont {Zhang}, \citenamefont {Lin},
  \citenamefont {Moreo},\ and\ \citenamefont {Dagotto}}]{zhang2023electronic}%
  \BibitemOpen
  \bibfield  {author} {\bibinfo {author} {\bibfnamefont {Y.}~\bibnamefont
  {Zhang}}, \bibinfo {author} {\bibfnamefont {L.-F.}\ \bibnamefont {Lin}},
  \bibinfo {author} {\bibfnamefont {A.}~\bibnamefont {Moreo}},\ and\ \bibinfo
  {author} {\bibfnamefont {E.}~\bibnamefont {Dagotto}},\ }\bibfield  {title}
  {\bibinfo {title} {Electronic structure, dimer physics, orbital-selective
  behavior, and magnetic tendencies in the bilayer nickelate superconductor
  {${\mathrm{La}}_{3}{\mathrm{Ni}}_{2}{\mathrm{O}}_{7}$ under pressure}},\
  }\href {https://doi.org/10.1103/PhysRevB.108.L180510} {\bibfield  {journal}
  {\bibinfo  {journal} {Phys. Rev. B}\ }\textbf {\bibinfo {volume} {108}},\
  \bibinfo {pages} {L180510} (\bibinfo {year}
  {2023}{\natexlab{a}})}\BibitemShut {NoStop}%
\bibitem [{\citenamefont {Yang}\ \emph
  {et~al.}(2023{\natexlab{a}})\citenamefont {Yang}, \citenamefont {Wang},\ and\
  \citenamefont {Wang}}]{yang2023possible}%
  \BibitemOpen
  \bibfield  {author} {\bibinfo {author} {\bibfnamefont {Q.-G.}\ \bibnamefont
  {Yang}}, \bibinfo {author} {\bibfnamefont {D.}~\bibnamefont {Wang}},\ and\
  \bibinfo {author} {\bibfnamefont {Q.-H.}\ \bibnamefont {Wang}},\ }\bibfield
  {title} {\bibinfo {title} {Possible ${s}_{\ifmmode\pm\else\textpm\fi{}}$-wave
  superconductivity in
  {${\mathrm{La}}_{3}{\mathrm{Ni}}_{2}{\mathrm{O}}_{7}$}},\ }\href
  {https://doi.org/10.1103/PhysRevB.108.L140505} {\bibfield  {journal}
  {\bibinfo  {journal} {Phys. Rev. B}\ }\textbf {\bibinfo {volume} {108}},\
  \bibinfo {pages} {L140505} (\bibinfo {year}
  {2023}{\natexlab{a}})}\BibitemShut {NoStop}%
\bibitem [{\citenamefont {Lechermann}\ \emph {et~al.}(2023)\citenamefont
  {Lechermann}, \citenamefont {Gondolf}, \citenamefont {B\"otzel},\ and\
  \citenamefont {Eremin}}]{lechermann2023electronic}%
  \BibitemOpen
  \bibfield  {author} {\bibinfo {author} {\bibfnamefont {F.}~\bibnamefont
  {Lechermann}}, \bibinfo {author} {\bibfnamefont {J.}~\bibnamefont {Gondolf}},
  \bibinfo {author} {\bibfnamefont {S.}~\bibnamefont {B\"otzel}},\ and\
  \bibinfo {author} {\bibfnamefont {I.~M.}\ \bibnamefont {Eremin}},\ }\bibfield
   {title} {\bibinfo {title} {Electronic correlations and superconducting
  instability in {${\mathrm{La}}_{3}{\mathrm{Ni}}_{2}{\mathrm{O}}_{7}$} under
  high pressure},\ }\href {https://doi.org/10.1103/PhysRevB.108.L201121}
  {\bibfield  {journal} {\bibinfo  {journal} {Phys. Rev. B}\ }\textbf {\bibinfo
  {volume} {108}},\ \bibinfo {pages} {L201121} (\bibinfo {year}
  {2023})}\BibitemShut {NoStop}%
\bibitem [{\citenamefont {Sakakibara}\ \emph
  {et~al.}(2024{\natexlab{a}})\citenamefont {Sakakibara}, \citenamefont
  {Kitamine}, \citenamefont {Ochi},\ and\ \citenamefont
  {Kuroki}}]{sakakibara2023possible}%
  \BibitemOpen
  \bibfield  {author} {\bibinfo {author} {\bibfnamefont {H.}~\bibnamefont
  {Sakakibara}}, \bibinfo {author} {\bibfnamefont {N.}~\bibnamefont
  {Kitamine}}, \bibinfo {author} {\bibfnamefont {M.}~\bibnamefont {Ochi}},\
  and\ \bibinfo {author} {\bibfnamefont {K.}~\bibnamefont {Kuroki}},\
  }\bibfield  {title} {\bibinfo {title} {{Possible High {${T}_{c}$}
  Superconductivity in {${\mathrm{La}}_{3}{\mathrm{Ni}}_{2}{\mathrm{O}}_{7}$}
  under High Pressure through Manifestation of a Nearly Half-Filled Bilayer
  Hubbard Model}},\ }\href {https://doi.org/10.1103/PhysRevLett.132.106002}
  {\bibfield  {journal} {\bibinfo  {journal} {Phys. Rev. Lett.}\ }\textbf
  {\bibinfo {volume} {132}},\ \bibinfo {pages} {106002} (\bibinfo {year}
  {2024}{\natexlab{a}})}\BibitemShut {NoStop}%
\bibitem [{\citenamefont {Gu}\ \emph {et~al.}(2025)\citenamefont {Gu},
  \citenamefont {Le}, \citenamefont {Yang}, \citenamefont {Wu},\ and\
  \citenamefont {Hu}}]{gu2023effective}%
  \BibitemOpen
  \bibfield  {author} {\bibinfo {author} {\bibfnamefont {Y.}~\bibnamefont
  {Gu}}, \bibinfo {author} {\bibfnamefont {C.}~\bibnamefont {Le}}, \bibinfo
  {author} {\bibfnamefont {Z.}~\bibnamefont {Yang}}, \bibinfo {author}
  {\bibfnamefont {X.}~\bibnamefont {Wu}},\ and\ \bibinfo {author}
  {\bibfnamefont {J.}~\bibnamefont {Hu}},\ }\bibfield  {title} {\bibinfo
  {title} {{Effective model and pairing tendency in the bilayer Ni-based
  superconductor ${\mathrm{La}}_{3}{\mathrm{Ni}}_{2}{\mathrm{O}}_{7}$}},\
  }\href {https://doi.org/10.1103/PhysRevB.111.174506} {\bibfield  {journal}
  {\bibinfo  {journal} {Phys. Rev. B}\ }\textbf {\bibinfo {volume} {111}},\
  \bibinfo {pages} {174506} (\bibinfo {year} {2025})}\BibitemShut {NoStop}%
\bibitem [{\citenamefont {Shen}\ \emph {et~al.}(2023)\citenamefont {Shen},
  \citenamefont {Qin},\ and\ \citenamefont {Zhang}}]{shen2023effective}%
  \BibitemOpen
  \bibfield  {author} {\bibinfo {author} {\bibfnamefont {Y.}~\bibnamefont
  {Shen}}, \bibinfo {author} {\bibfnamefont {M.}~\bibnamefont {Qin}},\ and\
  \bibinfo {author} {\bibfnamefont {G.-M.}\ \bibnamefont {Zhang}},\ }\bibfield
  {title} {\bibinfo {title} {{Effective Bi-Layer Model Hamiltonian and
  Density-Matrix Renormalization Group Study for the High-{$T_{\rm c}$}
  Superconductivity in {La$_{3}$Ni$_{2}$O$_{7}$} under High Pressure}},\ }\href
  {https://doi.org/10.1088/0256-307X/40/12/127401} {\bibfield  {journal}
  {\bibinfo  {journal} {Chinese Physics Letters}\ }\textbf {\bibinfo {volume}
  {40}},\ \bibinfo {eid} {127401} (\bibinfo {year} {2023})}\BibitemShut
  {NoStop}%
\bibitem [{\citenamefont {Christiansson}\ \emph {et~al.}(2023)\citenamefont
  {Christiansson}, \citenamefont {Petocchi},\ and\ \citenamefont
  {Werner}}]{christiansson2023correlated}%
  \BibitemOpen
  \bibfield  {author} {\bibinfo {author} {\bibfnamefont {V.}~\bibnamefont
  {Christiansson}}, \bibinfo {author} {\bibfnamefont {F.}~\bibnamefont
  {Petocchi}},\ and\ \bibinfo {author} {\bibfnamefont {P.}~\bibnamefont
  {Werner}},\ }\bibfield  {title} {\bibinfo {title} {{Correlated Electronic
  Structure of {${\mathrm{La}}_{3}{\text{Ni}}_{2}{\mathrm{O}}_{7}$} under
  Pressure}},\ }\href {https://doi.org/10.1103/PhysRevLett.131.206501}
  {\bibfield  {journal} {\bibinfo  {journal} {Phys. Rev. Lett.}\ }\textbf
  {\bibinfo {volume} {131}},\ \bibinfo {pages} {206501} (\bibinfo {year}
  {2023})}\BibitemShut {NoStop}%
\bibitem [{\citenamefont {Shilenko}\ and\ \citenamefont
  {Leonov}(2023)}]{Shilenko2023Correlated}%
  \BibitemOpen
  \bibfield  {author} {\bibinfo {author} {\bibfnamefont {D.~A.}\ \bibnamefont
  {Shilenko}}\ and\ \bibinfo {author} {\bibfnamefont {I.~V.}\ \bibnamefont
  {Leonov}},\ }\bibfield  {title} {\bibinfo {title} {Correlated electronic
  structure, orbital-selective behavior, and magnetic correlations in
  double-layer {${\mathrm{La}}_{3}{\mathrm{Ni}}_{2}{\mathrm{O}}_{7}$} under
  pressure},\ }\href {https://doi.org/10.1103/PhysRevB.108.125105} {\bibfield
  {journal} {\bibinfo  {journal} {Phys. Rev. B}\ }\textbf {\bibinfo {volume}
  {108}},\ \bibinfo {pages} {125105} (\bibinfo {year} {2023})}\BibitemShut
  {NoStop}%
\bibitem [{\citenamefont {Wú}\ \emph {et~al.}(2024)\citenamefont {Wú},
  \citenamefont {Luo}, \citenamefont {Yao},\ and\ \citenamefont
  {Wang}}]{wu2023charge}%
  \BibitemOpen
  \bibfield  {author} {\bibinfo {author} {\bibfnamefont {W.}~\bibnamefont
  {Wú}}, \bibinfo {author} {\bibfnamefont {Z.}~\bibnamefont {Luo}}, \bibinfo
  {author} {\bibfnamefont {D.-X.}\ \bibnamefont {Yao}},\ and\ \bibinfo {author}
  {\bibfnamefont {M.}~\bibnamefont {Wang}},\ }\bibfield  {title} {\bibinfo
  {title} {Superexchange and charge transfer in the nickelate superconductor
  {La$_3$Ni$_2$O$_7$} under pressure},\ }\href
  {https://doi.org/10.1007/s11433-023-2300-4} {\bibfield  {journal} {\bibinfo
  {journal} {Science China Physics, Mechanics {\&} Astronomy}\ }\textbf
  {\bibinfo {volume} {67}},\ \bibinfo {pages} {117402} (\bibinfo {year}
  {2024})}\BibitemShut {NoStop}%
\bibitem [{\citenamefont {Cao}\ and\ \citenamefont {Yang}(2024)}]{cao2023flat}%
  \BibitemOpen
  \bibfield  {author} {\bibinfo {author} {\bibfnamefont {Y.}~\bibnamefont
  {Cao}}\ and\ \bibinfo {author} {\bibfnamefont {Y.-f.}\ \bibnamefont {Yang}},\
  }\bibfield  {title} {\bibinfo {title} {Flat bands promoted by {Hund's} rule
  coupling in the candidate double-layer high-temperature superconductor
  {${\mathrm{La}}_{3}{\mathrm{Ni}}_{2}{\mathrm{O}}_{7}$} under high pressure},\
  }\href {https://doi.org/10.1103/PhysRevB.109.L081105} {\bibfield  {journal}
  {\bibinfo  {journal} {Phys. Rev. B}\ }\textbf {\bibinfo {volume} {109}},\
  \bibinfo {pages} {L081105} (\bibinfo {year} {2024})}\BibitemShut {NoStop}%
\bibitem [{\citenamefont {Chen}\ \emph
  {et~al.}(2025{\natexlab{a}})\citenamefont {Chen}, \citenamefont {Jiang},
  \citenamefont {Li}, \citenamefont {Zhong},\ and\ \citenamefont
  {Lu}}]{chen2023critical}%
  \BibitemOpen
  \bibfield  {author} {\bibinfo {author} {\bibfnamefont {X.}~\bibnamefont
  {Chen}}, \bibinfo {author} {\bibfnamefont {P.}~\bibnamefont {Jiang}},
  \bibinfo {author} {\bibfnamefont {J.}~\bibnamefont {Li}}, \bibinfo {author}
  {\bibfnamefont {Z.}~\bibnamefont {Zhong}},\ and\ \bibinfo {author}
  {\bibfnamefont {Y.}~\bibnamefont {Lu}},\ }\bibfield  {title} {\bibinfo
  {title} {{Charge and spin instabilities in superconducting
  ${\mathrm{La}}_{3}{\mathrm{Ni}}_{2}{\mathrm{O}}_{7}$}},\ }\href
  {https://doi.org/10.1103/PhysRevB.111.014515} {\bibfield  {journal} {\bibinfo
   {journal} {Phys. Rev. B}\ }\textbf {\bibinfo {volume} {111}},\ \bibinfo
  {pages} {014515} (\bibinfo {year} {2025}{\natexlab{a}})}\BibitemShut
  {NoStop}%
\bibitem [{\citenamefont {Liu}\ \emph {et~al.}(2023)\citenamefont {Liu},
  \citenamefont {Mei}, \citenamefont {Ye}, \citenamefont {Chen},\ and\
  \citenamefont {Yang}}]{liu2023spmwave}%
  \BibitemOpen
  \bibfield  {author} {\bibinfo {author} {\bibfnamefont {Y.-B.}\ \bibnamefont
  {Liu}}, \bibinfo {author} {\bibfnamefont {J.-W.}\ \bibnamefont {Mei}},
  \bibinfo {author} {\bibfnamefont {F.}~\bibnamefont {Ye}}, \bibinfo {author}
  {\bibfnamefont {W.-Q.}\ \bibnamefont {Chen}},\ and\ \bibinfo {author}
  {\bibfnamefont {F.}~\bibnamefont {Yang}},\ }\bibfield  {title} {\bibinfo
  {title} {{${\mathrm{s}}^{\ifmmode\pm\else\textpm\fi{}}$-Wave Pairing and the
  Destructive Role of Apical-Oxygen Deficiencies in
  {${\mathrm{La}}_{3}{\mathrm{Ni}}_{2}{\mathrm{O}}_{7}$} under Pressure}},\
  }\href {https://doi.org/10.1103/PhysRevLett.131.236002} {\bibfield  {journal}
  {\bibinfo  {journal} {Phys. Rev. Lett.}\ }\textbf {\bibinfo {volume} {131}},\
  \bibinfo {pages} {236002} (\bibinfo {year} {2023})}\BibitemShut {NoStop}%
\bibitem [{\citenamefont {Lu}\ \emph {et~al.}(2024{\natexlab{a}})\citenamefont
  {Lu}, \citenamefont {Pan}, \citenamefont {Yang},\ and\ \citenamefont
  {Wu}}]{lu2023interlayer}%
  \BibitemOpen
  \bibfield  {author} {\bibinfo {author} {\bibfnamefont {C.}~\bibnamefont
  {Lu}}, \bibinfo {author} {\bibfnamefont {Z.}~\bibnamefont {Pan}}, \bibinfo
  {author} {\bibfnamefont {F.}~\bibnamefont {Yang}},\ and\ \bibinfo {author}
  {\bibfnamefont {C.}~\bibnamefont {Wu}},\ }\bibfield  {title} {\bibinfo
  {title} {{Interlayer-Coupling-Driven High-Temperature Superconductivity in
  {${\mathrm{La}}_{3}{\mathrm{Ni}}_{2}{\mathrm{O}}_{7}$} under Pressure}},\
  }\href {https://doi.org/10.1103/PhysRevLett.132.146002} {\bibfield  {journal}
  {\bibinfo  {journal} {Phys. Rev. Lett.}\ }\textbf {\bibinfo {volume} {132}},\
  \bibinfo {pages} {146002} (\bibinfo {year} {2024}{\natexlab{a}})}\BibitemShut
  {NoStop}%
\bibitem [{\citenamefont {Qu}\ \emph {et~al.}(2024)\citenamefont {Qu},
  \citenamefont {Qu}, \citenamefont {Chen}, \citenamefont {Wu}, \citenamefont
  {Yang}, \citenamefont {Li},\ and\ \citenamefont {Su}}]{qu2023bilayer}%
  \BibitemOpen
  \bibfield  {author} {\bibinfo {author} {\bibfnamefont {X.-Z.}\ \bibnamefont
  {Qu}}, \bibinfo {author} {\bibfnamefont {D.-W.}\ \bibnamefont {Qu}}, \bibinfo
  {author} {\bibfnamefont {J.}~\bibnamefont {Chen}}, \bibinfo {author}
  {\bibfnamefont {C.}~\bibnamefont {Wu}}, \bibinfo {author} {\bibfnamefont
  {F.}~\bibnamefont {Yang}}, \bibinfo {author} {\bibfnamefont {W.}~\bibnamefont
  {Li}},\ and\ \bibinfo {author} {\bibfnamefont {G.}~\bibnamefont {Su}},\
  }\bibfield  {title} {\bibinfo {title} {{Bilayer
  ${t\text{\ensuremath{-}}J\text{\ensuremath{-}}J}_{\ensuremath{\perp}}$ Model
  and Magnetically Mediated Pairing in the Pressurized Nickelate
  {${\mathrm{La}}_{3}{\mathrm{Ni}}_{2}{\mathrm{O}}_{7}$}}},\ }\href
  {https://doi.org/10.1103/PhysRevLett.132.036502} {\bibfield  {journal}
  {\bibinfo  {journal} {Phys. Rev. Lett.}\ }\textbf {\bibinfo {volume} {132}},\
  \bibinfo {pages} {036502} (\bibinfo {year} {2024})}\BibitemShut {NoStop}%
\bibitem [{\citenamefont {Oh}\ and\ \citenamefont {Zhang}(2023)}]{oh2023type}%
  \BibitemOpen
  \bibfield  {author} {\bibinfo {author} {\bibfnamefont {H.}~\bibnamefont
  {Oh}}\ and\ \bibinfo {author} {\bibfnamefont {Y.-H.}\ \bibnamefont {Zhang}},\
  }\bibfield  {title} {\bibinfo {title} {Type-{II} {$t\ensuremath{-}J$} model
  and shared superexchange coupling from {Hund's} rule in superconducting
  {${\mathrm{La}}_{3}{\mathrm{Ni}}_{2}{\mathrm{O}}_{7}$}},\ }\href
  {https://doi.org/10.1103/PhysRevB.108.174511} {\bibfield  {journal} {\bibinfo
   {journal} {Phys. Rev. B}\ }\textbf {\bibinfo {volume} {108}},\ \bibinfo
  {pages} {174511} (\bibinfo {year} {2023})}\BibitemShut {NoStop}%
\bibitem [{\citenamefont {Zhang}\ \emph
  {et~al.}(2024{\natexlab{c}})\citenamefont {Zhang}, \citenamefont {Lin},
  \citenamefont {Moreo}, \citenamefont {Maier},\ and\ \citenamefont
  {Dagotto}}]{zhang2023structural}%
  \BibitemOpen
  \bibfield  {author} {\bibinfo {author} {\bibfnamefont {Y.}~\bibnamefont
  {Zhang}}, \bibinfo {author} {\bibfnamefont {L.-F.}\ \bibnamefont {Lin}},
  \bibinfo {author} {\bibfnamefont {A.}~\bibnamefont {Moreo}}, \bibinfo
  {author} {\bibfnamefont {T.~A.}\ \bibnamefont {Maier}},\ and\ \bibinfo
  {author} {\bibfnamefont {E.}~\bibnamefont {Dagotto}},\ }\bibfield  {title}
  {\bibinfo {title} {Structural phase transition, s{\textpm}-wave pairing, and
  magnetic stripe order in bilayered superconductor {La$_3$Ni$_2$O$_7$} under
  pressure},\ }\href {https://doi.org/10.1038/s41467-024-46622-z} {\bibfield
  {journal} {\bibinfo  {journal} {Nature Communications}\ }\textbf {\bibinfo
  {volume} {15}},\ \bibinfo {pages} {2470} (\bibinfo {year}
  {2024}{\natexlab{c}})}\BibitemShut {NoStop}%
\bibitem [{\citenamefont {Liao}\ \emph {et~al.}(2023)\citenamefont {Liao},
  \citenamefont {Chen}, \citenamefont {Duan}, \citenamefont {Wang},
  \citenamefont {Liu}, \citenamefont {Yu},\ and\ \citenamefont
  {Si}}]{liao2023electron}%
  \BibitemOpen
  \bibfield  {author} {\bibinfo {author} {\bibfnamefont {Z.}~\bibnamefont
  {Liao}}, \bibinfo {author} {\bibfnamefont {L.}~\bibnamefont {Chen}}, \bibinfo
  {author} {\bibfnamefont {G.}~\bibnamefont {Duan}}, \bibinfo {author}
  {\bibfnamefont {Y.}~\bibnamefont {Wang}}, \bibinfo {author} {\bibfnamefont
  {C.}~\bibnamefont {Liu}}, \bibinfo {author} {\bibfnamefont {R.}~\bibnamefont
  {Yu}},\ and\ \bibinfo {author} {\bibfnamefont {Q.}~\bibnamefont {Si}},\
  }\bibfield  {title} {\bibinfo {title} {Electron correlations and
  superconductivity in {${\mathrm{La}}_{3}{\mathrm{Ni}}_{2}{\mathrm{O}}_{7}$}
  under pressure tuning},\ }\href {https://doi.org/10.1103/PhysRevB.108.214522}
  {\bibfield  {journal} {\bibinfo  {journal} {Phys. Rev. B}\ }\textbf {\bibinfo
  {volume} {108}},\ \bibinfo {pages} {214522} (\bibinfo {year}
  {2023})}\BibitemShut {NoStop}%
\bibitem [{\citenamefont {Yang}\ \emph
  {et~al.}(2023{\natexlab{b}})\citenamefont {Yang}, \citenamefont {Zhang},\
  and\ \citenamefont {Zhang}}]{yang2023minimal}%
  \BibitemOpen
  \bibfield  {author} {\bibinfo {author} {\bibfnamefont {Y.-f.}\ \bibnamefont
  {Yang}}, \bibinfo {author} {\bibfnamefont {G.-M.}\ \bibnamefont {Zhang}},\
  and\ \bibinfo {author} {\bibfnamefont {F.-C.}\ \bibnamefont {Zhang}},\
  }\bibfield  {title} {\bibinfo {title} {Interlayer valence bonds and
  two-component theory for high-{${T}_{c}$} superconductivity of
  {${\mathrm{La}}_{3}{\mathrm{Ni}}_{2}{\mathrm{O}}_{7}$} under pressure},\
  }\href {https://doi.org/10.1103/PhysRevB.108.L201108} {\bibfield  {journal}
  {\bibinfo  {journal} {Phys. Rev. B}\ }\textbf {\bibinfo {volume} {108}},\
  \bibinfo {pages} {L201108} (\bibinfo {year}
  {2023}{\natexlab{b}})}\BibitemShut {NoStop}%
\bibitem [{\citenamefont {Jiang}\ \emph
  {et~al.}(2024{\natexlab{a}})\citenamefont {Jiang}, \citenamefont {Wang},\
  and\ \citenamefont {Zhang}}]{jiang2023high}%
  \BibitemOpen
  \bibfield  {author} {\bibinfo {author} {\bibfnamefont {K.}~\bibnamefont
  {Jiang}}, \bibinfo {author} {\bibfnamefont {Z.}~\bibnamefont {Wang}},\ and\
  \bibinfo {author} {\bibfnamefont {F.-C.}\ \bibnamefont {Zhang}},\ }\bibfield
  {title} {\bibinfo {title} {{High-Temperature Superconductivity in
  {La$_3$Ni$_2$O$_7$}}},\ }\href
  {https://doi.org/10.1088/0256-307X/41/1/017402} {\bibfield  {journal}
  {\bibinfo  {journal} {Chin. Phys. Lett.}\ }\textbf {\bibinfo {volume} {41}},\
  \bibinfo {pages} {017402} (\bibinfo {year} {2024}{\natexlab{a}})}\BibitemShut
  {NoStop}%
\bibitem [{\citenamefont {Zhang}\ \emph
  {et~al.}(2023{\natexlab{b}})\citenamefont {Zhang}, \citenamefont {Lin},
  \citenamefont {Moreo}, \citenamefont {Maier},\ and\ \citenamefont
  {Dagotto}}]{zhang2023trends}%
  \BibitemOpen
  \bibfield  {author} {\bibinfo {author} {\bibfnamefont {Y.}~\bibnamefont
  {Zhang}}, \bibinfo {author} {\bibfnamefont {L.-F.}\ \bibnamefont {Lin}},
  \bibinfo {author} {\bibfnamefont {A.}~\bibnamefont {Moreo}}, \bibinfo
  {author} {\bibfnamefont {T.~A.}\ \bibnamefont {Maier}},\ and\ \bibinfo
  {author} {\bibfnamefont {E.}~\bibnamefont {Dagotto}},\ }\bibfield  {title}
  {\bibinfo {title} {Trends in electronic structures and
  ${s}_{\ifmmode\pm\else\textpm\fi{}}$-wave pairing for the rare-earth series
  in bilayer nickelate superconductor
  {${R}_{3}{\mathrm{Ni}}_{2}{\mathrm{O}}_{7}$}},\ }\href
  {https://doi.org/10.1103/PhysRevB.108.165141} {\bibfield  {journal} {\bibinfo
   {journal} {Phys. Rev. B}\ }\textbf {\bibinfo {volume} {108}},\ \bibinfo
  {pages} {165141} (\bibinfo {year} {2023}{\natexlab{b}})}\BibitemShut
  {NoStop}%
\bibitem [{\citenamefont {Huang}\ \emph {et~al.}(2023)\citenamefont {Huang},
  \citenamefont {Wang},\ and\ \citenamefont {Zhou}}]{huang2023impurity}%
  \BibitemOpen
  \bibfield  {author} {\bibinfo {author} {\bibfnamefont {J.}~\bibnamefont
  {Huang}}, \bibinfo {author} {\bibfnamefont {Z.~D.}\ \bibnamefont {Wang}},\
  and\ \bibinfo {author} {\bibfnamefont {T.}~\bibnamefont {Zhou}},\ }\bibfield
  {title} {\bibinfo {title} {Impurity and vortex states in the bilayer
  high-temperature superconductor
  {${\mathrm{La}}_{3}{\mathrm{Ni}}_{2}{\mathrm{O}}_{7}$}},\ }\href
  {https://doi.org/10.1103/PhysRevB.108.174501} {\bibfield  {journal} {\bibinfo
   {journal} {Phys. Rev. B}\ }\textbf {\bibinfo {volume} {108}},\ \bibinfo
  {pages} {174501} (\bibinfo {year} {2023})}\BibitemShut {NoStop}%
\bibitem [{\citenamefont {Qin}\ and\ \citenamefont
  {Yang}(2023)}]{qin2023hightc}%
  \BibitemOpen
  \bibfield  {author} {\bibinfo {author} {\bibfnamefont {Q.}~\bibnamefont
  {Qin}}\ and\ \bibinfo {author} {\bibfnamefont {Y.-f.}\ \bibnamefont {Yang}},\
  }\bibfield  {title} {\bibinfo {title} {High-${T}_{c}$ superconductivity by
  mobilizing local spin singlets and possible route to higher ${T}_{c}$ in
  pressurized {${\mathrm{La}}_{3}{\mathrm{Ni}}_{2}{\mathrm{O}}_{7}$}},\ }\href
  {https://doi.org/10.1103/PhysRevB.108.L140504} {\bibfield  {journal}
  {\bibinfo  {journal} {Phys. Rev. B}\ }\textbf {\bibinfo {volume} {108}},\
  \bibinfo {pages} {L140504} (\bibinfo {year} {2023})}\BibitemShut {NoStop}%
\bibitem [{\citenamefont {Tian}\ \emph {et~al.}(2024)\citenamefont {Tian},
  \citenamefont {Chen}, \citenamefont {Wang}, \citenamefont {He},\ and\
  \citenamefont {Lu}}]{tian2023correlation}%
  \BibitemOpen
  \bibfield  {author} {\bibinfo {author} {\bibfnamefont {Y.-H.}\ \bibnamefont
  {Tian}}, \bibinfo {author} {\bibfnamefont {Y.}~\bibnamefont {Chen}}, \bibinfo
  {author} {\bibfnamefont {J.-M.}\ \bibnamefont {Wang}}, \bibinfo {author}
  {\bibfnamefont {R.-Q.}\ \bibnamefont {He}},\ and\ \bibinfo {author}
  {\bibfnamefont {Z.-Y.}\ \bibnamefont {Lu}},\ }\bibfield  {title} {\bibinfo
  {title} {Correlation effects and concomitant two-orbital $s_{\pm}$-wave
  superconductivity in {${\mathrm{La}}_{3}{\mathrm{Ni}}_{2}{\mathrm{O}}_{7}$}
  under high pressure},\ }\href {https://doi.org/10.1103/PhysRevB.109.165154}
  {\bibfield  {journal} {\bibinfo  {journal} {Phys. Rev. B}\ }\textbf {\bibinfo
  {volume} {109}},\ \bibinfo {pages} {165154} (\bibinfo {year}
  {2024})}\BibitemShut {NoStop}%
\bibitem [{\citenamefont {Lu}\ \emph {et~al.}(2023)\citenamefont {Lu},
  \citenamefont {Li}, \citenamefont {Zeng}, \citenamefont {Hou}, \citenamefont
  {Wang}, \citenamefont {Yang},\ and\ \citenamefont
  {You}}]{lu2023superconductivity}%
  \BibitemOpen
  \bibfield  {author} {\bibinfo {author} {\bibfnamefont {D.-C.}\ \bibnamefont
  {Lu}}, \bibinfo {author} {\bibfnamefont {M.}~\bibnamefont {Li}}, \bibinfo
  {author} {\bibfnamefont {Z.-Y.}\ \bibnamefont {Zeng}}, \bibinfo {author}
  {\bibfnamefont {W.}~\bibnamefont {Hou}}, \bibinfo {author} {\bibfnamefont
  {J.}~\bibnamefont {Wang}}, \bibinfo {author} {\bibfnamefont {F.}~\bibnamefont
  {Yang}},\ and\ \bibinfo {author} {\bibfnamefont {Y.-Z.}\ \bibnamefont
  {You}},\ }\href@noop {} {\bibinfo {title} {{Superconductivity from Doping
  Symmetric Mass Generation Insulators: Application to {La$_3$Ni$_2$O$_7$}
  under Pressure}}} (\bibinfo {year} {2023}),\ \Eprint
  {https://arxiv.org/abs/2308.11195} {arXiv:2308.11195 [cond-mat.str-el]}
  \BibitemShut {NoStop}%
\bibitem [{\citenamefont {Jiang}\ \emph
  {et~al.}(2024{\natexlab{b}})\citenamefont {Jiang}, \citenamefont {Hou},
  \citenamefont {Fan}, \citenamefont {Lang},\ and\ \citenamefont
  {Ku}}]{jiang2023pressure}%
  \BibitemOpen
  \bibfield  {author} {\bibinfo {author} {\bibfnamefont {R.}~\bibnamefont
  {Jiang}}, \bibinfo {author} {\bibfnamefont {J.}~\bibnamefont {Hou}}, \bibinfo
  {author} {\bibfnamefont {Z.}~\bibnamefont {Fan}}, \bibinfo {author}
  {\bibfnamefont {Z.-J.}\ \bibnamefont {Lang}},\ and\ \bibinfo {author}
  {\bibfnamefont {W.}~\bibnamefont {Ku}},\ }\bibfield  {title} {\bibinfo
  {title} {{Pressure Driven Fractionalization of Ionic Spins Results in
  Cupratelike High-{${T}_{c}$} Superconductivity in
  {${\mathrm{La}}_{3}{\mathrm{Ni}}_{2}{\mathrm{O}}_{7}$}}},\ }\href
  {https://doi.org/10.1103/PhysRevLett.132.126503} {\bibfield  {journal}
  {\bibinfo  {journal} {Phys. Rev. Lett.}\ }\textbf {\bibinfo {volume} {132}},\
  \bibinfo {pages} {126503} (\bibinfo {year} {2024}{\natexlab{b}})}\BibitemShut
  {NoStop}%
\bibitem [{\citenamefont {Kitamine}\ \emph {et~al.}(2023)\citenamefont
  {Kitamine}, \citenamefont {Ochi},\ and\ \citenamefont
  {Kuroki}}]{kitamine2023theoretical}%
  \BibitemOpen
  \bibfield  {author} {\bibinfo {author} {\bibfnamefont {N.}~\bibnamefont
  {Kitamine}}, \bibinfo {author} {\bibfnamefont {M.}~\bibnamefont {Ochi}},\
  and\ \bibinfo {author} {\bibfnamefont {K.}~\bibnamefont {Kuroki}},\
  }\href@noop {} {\bibinfo {title} {{Theoretical designing of multiband
  Nickelate and Palladate superconductors with $d^{8+\delta}$ configuration}}}
  (\bibinfo {year} {2023}),\ \Eprint {https://arxiv.org/abs/2308.12750}
  {arXiv:2308.12750 [cond-mat.supr-con]} \BibitemShut {NoStop}%
\bibitem [{\citenamefont {Luo}\ \emph {et~al.}(2024)\citenamefont {Luo},
  \citenamefont {Lv}, \citenamefont {Wang}, \citenamefont {Wú},\ and\
  \citenamefont {Yao}}]{luo2023hightc}%
  \BibitemOpen
  \bibfield  {author} {\bibinfo {author} {\bibfnamefont {Z.}~\bibnamefont
  {Luo}}, \bibinfo {author} {\bibfnamefont {B.}~\bibnamefont {Lv}}, \bibinfo
  {author} {\bibfnamefont {M.}~\bibnamefont {Wang}}, \bibinfo {author}
  {\bibfnamefont {W.}~\bibnamefont {Wú}},\ and\ \bibinfo {author}
  {\bibfnamefont {D.-X.}\ \bibnamefont {Yao}},\ }\bibfield  {title} {\bibinfo
  {title} {High-{$T_C$} superconductivity in {La$_3$Ni$_2$O$_7$} based on the
  bilayer two-orbital {t-J} model},\ }\href
  {https://doi.org/10.1038/s41535-024-00668-w} {\bibfield  {journal} {\bibinfo
  {journal} {npj Quantum Materials}\ }\textbf {\bibinfo {volume} {9}},\
  \bibinfo {pages} {61} (\bibinfo {year} {2024})}\BibitemShut {NoStop}%
\bibitem [{\citenamefont {Zhang}\ \emph
  {et~al.}(2024{\natexlab{d}})\citenamefont {Zhang}, \citenamefont {Zhang},
  \citenamefont {You},\ and\ \citenamefont {Weng}}]{zhang2023strong}%
  \BibitemOpen
  \bibfield  {author} {\bibinfo {author} {\bibfnamefont {J.-X.}\ \bibnamefont
  {Zhang}}, \bibinfo {author} {\bibfnamefont {H.-K.}\ \bibnamefont {Zhang}},
  \bibinfo {author} {\bibfnamefont {Y.-Z.}\ \bibnamefont {You}},\ and\ \bibinfo
  {author} {\bibfnamefont {Z.-Y.}\ \bibnamefont {Weng}},\ }\bibfield  {title}
  {\bibinfo {title} {{Strong Pairing Originated from an Emergent
  {${\mathbb{Z}}_{2}$} Berry Phase in
  {${\mathrm{La}}_{3}{\mathrm{Ni}}_{2}{\mathrm{O}}_{7}$}}},\ }\href
  {https://doi.org/10.1103/PhysRevLett.133.126501} {\bibfield  {journal}
  {\bibinfo  {journal} {Phys. Rev. Lett.}\ }\textbf {\bibinfo {volume} {133}},\
  \bibinfo {pages} {126501} (\bibinfo {year} {2024}{\natexlab{d}})}\BibitemShut
  {NoStop}%
\bibitem [{\citenamefont {Pan}\ \emph {et~al.}(2024)\citenamefont {Pan},
  \citenamefont {Lu}, \citenamefont {Yang},\ and\ \citenamefont
  {Wu}}]{pan2023effect}%
  \BibitemOpen
  \bibfield  {author} {\bibinfo {author} {\bibfnamefont {Z.}~\bibnamefont
  {Pan}}, \bibinfo {author} {\bibfnamefont {C.}~\bibnamefont {Lu}}, \bibinfo
  {author} {\bibfnamefont {F.}~\bibnamefont {Yang}},\ and\ \bibinfo {author}
  {\bibfnamefont {C.}~\bibnamefont {Wu}},\ }\bibfield  {title} {\bibinfo
  {title} {{Effect of Rare-Earth Element Substitution in Superconducting
  {R$_3$Ni$_2$O$_7$} under Pressure}},\ }\href
  {https://doi.org/10.1088/0256-307X/41/8/087401} {\bibfield  {journal}
  {\bibinfo  {journal} {Chinese Physics Letters}\ }\textbf {\bibinfo {volume}
  {41}},\ \bibinfo {pages} {087401} (\bibinfo {year} {2024})}\BibitemShut
  {NoStop}%
\bibitem [{\citenamefont {Sakakibara}\ \emph
  {et~al.}(2024{\natexlab{b}})\citenamefont {Sakakibara}, \citenamefont {Ochi},
  \citenamefont {Nagata}, \citenamefont {Ueki}, \citenamefont {Sakurai},
  \citenamefont {Matsumoto}, \citenamefont {Terashima}, \citenamefont {Hirose},
  \citenamefont {Ohta}, \citenamefont {Kato}, \citenamefont {Takano},\ and\
  \citenamefont {Kuroki}}]{sakakibara2023theoretical}%
  \BibitemOpen
  \bibfield  {author} {\bibinfo {author} {\bibfnamefont {H.}~\bibnamefont
  {Sakakibara}}, \bibinfo {author} {\bibfnamefont {M.}~\bibnamefont {Ochi}},
  \bibinfo {author} {\bibfnamefont {H.}~\bibnamefont {Nagata}}, \bibinfo
  {author} {\bibfnamefont {Y.}~\bibnamefont {Ueki}}, \bibinfo {author}
  {\bibfnamefont {H.}~\bibnamefont {Sakurai}}, \bibinfo {author} {\bibfnamefont
  {R.}~\bibnamefont {Matsumoto}}, \bibinfo {author} {\bibfnamefont
  {K.}~\bibnamefont {Terashima}}, \bibinfo {author} {\bibfnamefont
  {K.}~\bibnamefont {Hirose}}, \bibinfo {author} {\bibfnamefont
  {H.}~\bibnamefont {Ohta}}, \bibinfo {author} {\bibfnamefont {M.}~\bibnamefont
  {Kato}}, \bibinfo {author} {\bibfnamefont {Y.}~\bibnamefont {Takano}},\ and\
  \bibinfo {author} {\bibfnamefont {K.}~\bibnamefont {Kuroki}},\ }\bibfield
  {title} {\bibinfo {title} {Theoretical analysis on the possibility of
  superconductivity in the trilayer {Ruddlesden-Popper} nickelate
  {${\mathrm{La}}_{4}{\mathrm{Ni}}_{3}{\mathrm{O}}_{10}$} under pressure and
  its experimental examination: Comparison with
  {${\mathrm{La}}_{3}{\mathrm{Ni}}_{2}{\mathrm{O}}_{7}$}},\ }\href
  {https://doi.org/10.1103/PhysRevB.109.144511} {\bibfield  {journal} {\bibinfo
   {journal} {Phys. Rev. B}\ }\textbf {\bibinfo {volume} {109}},\ \bibinfo
  {pages} {144511} (\bibinfo {year} {2024}{\natexlab{b}})}\BibitemShut
  {NoStop}%
\bibitem [{\citenamefont {Lange}\ \emph
  {et~al.}(2024{\natexlab{a}})\citenamefont {Lange}, \citenamefont {Homeier},
  \citenamefont {Demler}, \citenamefont {Schollw\"ock}, \citenamefont
  {Bohrdt},\ and\ \citenamefont {Grusdt}}]{lange2023pairing}%
  \BibitemOpen
  \bibfield  {author} {\bibinfo {author} {\bibfnamefont {H.}~\bibnamefont
  {Lange}}, \bibinfo {author} {\bibfnamefont {L.}~\bibnamefont {Homeier}},
  \bibinfo {author} {\bibfnamefont {E.}~\bibnamefont {Demler}}, \bibinfo
  {author} {\bibfnamefont {U.}~\bibnamefont {Schollw\"ock}}, \bibinfo {author}
  {\bibfnamefont {A.}~\bibnamefont {Bohrdt}},\ and\ \bibinfo {author}
  {\bibfnamefont {F.}~\bibnamefont {Grusdt}},\ }\bibfield  {title} {\bibinfo
  {title} {Pairing dome from an emergent {Feshbach} resonance in a strongly
  repulsive bilayer model},\ }\href
  {https://doi.org/10.1103/PhysRevB.110.L081113} {\bibfield  {journal}
  {\bibinfo  {journal} {Phys. Rev. B}\ }\textbf {\bibinfo {volume} {110}},\
  \bibinfo {pages} {L081113} (\bibinfo {year}
  {2024}{\natexlab{a}})}\BibitemShut {NoStop}%
\bibitem [{\citenamefont {Geisler}\ \emph {et~al.}(2024)\citenamefont
  {Geisler}, \citenamefont {Hamlin}, \citenamefont {Stewart}, \citenamefont
  {Hennig},\ and\ \citenamefont {Hirschfeld}}]{geisler2023structural}%
  \BibitemOpen
  \bibfield  {author} {\bibinfo {author} {\bibfnamefont {B.}~\bibnamefont
  {Geisler}}, \bibinfo {author} {\bibfnamefont {J.~J.}\ \bibnamefont {Hamlin}},
  \bibinfo {author} {\bibfnamefont {G.~R.}\ \bibnamefont {Stewart}}, \bibinfo
  {author} {\bibfnamefont {R.~G.}\ \bibnamefont {Hennig}},\ and\ \bibinfo
  {author} {\bibfnamefont {P.~J.}\ \bibnamefont {Hirschfeld}},\ }\bibfield
  {title} {\bibinfo {title} {Structural transitions, octahedral rotations, and
  electronic properties of {$A_3$Ni$_2$O$_7$} rare-earth nickelates under high
  pressure},\ }\href {https://doi.org/10.1038/s41535-024-00648-0} {\bibfield
  {journal} {\bibinfo  {journal} {npj Quantum Materials}\ }\textbf {\bibinfo
  {volume} {9}},\ \bibinfo {pages} {38} (\bibinfo {year} {2024})}\BibitemShut
  {NoStop}%
\bibitem [{\citenamefont {Yang}\ \emph
  {et~al.}(2024{\natexlab{b}})\citenamefont {Yang}, \citenamefont {Oh},\ and\
  \citenamefont {Zhang}}]{yang2023strong}%
  \BibitemOpen
  \bibfield  {author} {\bibinfo {author} {\bibfnamefont {H.}~\bibnamefont
  {Yang}}, \bibinfo {author} {\bibfnamefont {H.}~\bibnamefont {Oh}},\ and\
  \bibinfo {author} {\bibfnamefont {Y.-H.}\ \bibnamefont {Zhang}},\ }\bibfield
  {title} {\bibinfo {title} {Strong pairing from a small {Fermi} surface beyond
  weak coupling: Application to
  {${\mathrm{La}}_{3}{\mathrm{Ni}}_{2}{\mathrm{O}}_{7}$}},\ }\href
  {https://doi.org/10.1103/PhysRevB.110.104517} {\bibfield  {journal} {\bibinfo
   {journal} {Phys. Rev. B}\ }\textbf {\bibinfo {volume} {110}},\ \bibinfo
  {pages} {104517} (\bibinfo {year} {2024}{\natexlab{b}})}\BibitemShut
  {NoStop}%
\bibitem [{\citenamefont {Rhodes}\ and\ \citenamefont
  {Wahl}(2024)}]{rhodes2023structural}%
  \BibitemOpen
  \bibfield  {author} {\bibinfo {author} {\bibfnamefont {L.~C.}\ \bibnamefont
  {Rhodes}}\ and\ \bibinfo {author} {\bibfnamefont {P.}~\bibnamefont {Wahl}},\
  }\bibfield  {title} {\bibinfo {title} {Structural routes to stabilize
  superconducting {${\mathrm{La}}_{3}{\mathrm{Ni}}_{2}{\mathrm{O}}_{7}$} at
  ambient pressure},\ }\href
  {https://doi.org/10.1103/PhysRevMaterials.8.044801} {\bibfield  {journal}
  {\bibinfo  {journal} {Phys. Rev. Mater.}\ }\textbf {\bibinfo {volume} {8}},\
  \bibinfo {pages} {044801} (\bibinfo {year} {2024})}\BibitemShut {NoStop}%
\bibitem [{\citenamefont {Lange}\ \emph
  {et~al.}(2024{\natexlab{b}})\citenamefont {Lange}, \citenamefont {Homeier},
  \citenamefont {Demler}, \citenamefont {Schollw\"ock}, \citenamefont
  {Grusdt},\ and\ \citenamefont {Bohrdt}}]{lange2023feshbach}%
  \BibitemOpen
  \bibfield  {author} {\bibinfo {author} {\bibfnamefont {H.}~\bibnamefont
  {Lange}}, \bibinfo {author} {\bibfnamefont {L.}~\bibnamefont {Homeier}},
  \bibinfo {author} {\bibfnamefont {E.}~\bibnamefont {Demler}}, \bibinfo
  {author} {\bibfnamefont {U.}~\bibnamefont {Schollw\"ock}}, \bibinfo {author}
  {\bibfnamefont {F.}~\bibnamefont {Grusdt}},\ and\ \bibinfo {author}
  {\bibfnamefont {A.}~\bibnamefont {Bohrdt}},\ }\bibfield  {title} {\bibinfo
  {title} {Feshbach resonance in a strongly repulsive ladder of mixed
  dimensionality: A possible scenario for bilayer nickelate superconductors},\
  }\href {https://doi.org/10.1103/PhysRevB.109.045127} {\bibfield  {journal}
  {\bibinfo  {journal} {Phys. Rev. B}\ }\textbf {\bibinfo {volume} {109}},\
  \bibinfo {pages} {045127} (\bibinfo {year} {2024}{\natexlab{b}})}\BibitemShut
  {NoStop}%
\bibitem [{\citenamefont {LaBollita}\ \emph {et~al.}(2024)\citenamefont
  {LaBollita}, \citenamefont {Pardo}, \citenamefont {Norman},\ and\
  \citenamefont {Botana}}]{labollita2023electronic}%
  \BibitemOpen
  \bibfield  {author} {\bibinfo {author} {\bibfnamefont {H.}~\bibnamefont
  {LaBollita}}, \bibinfo {author} {\bibfnamefont {V.}~\bibnamefont {Pardo}},
  \bibinfo {author} {\bibfnamefont {M.~R.}\ \bibnamefont {Norman}},\ and\
  \bibinfo {author} {\bibfnamefont {A.~S.}\ \bibnamefont {Botana}},\ }\bibfield
   {title} {\bibinfo {title} {Assessing spin-density wave formation in
  {${\mathrm{La}}_{3}{\mathrm{Ni}}_{2}{\mathrm{O}}_{7}$} from electronic
  structure calculations},\ }\href
  {https://doi.org/10.1103/PhysRevMaterials.8.L111801} {\bibfield  {journal}
  {\bibinfo  {journal} {Phys. Rev. Mater.}\ }\textbf {\bibinfo {volume} {8}},\
  \bibinfo {pages} {L111801} (\bibinfo {year} {2024})}\BibitemShut {NoStop}%
\bibitem [{\citenamefont {Kaneko}\ \emph {et~al.}(2024)\citenamefont {Kaneko},
  \citenamefont {Sakakibara}, \citenamefont {Ochi},\ and\ \citenamefont
  {Kuroki}}]{kaneko2023pair}%
  \BibitemOpen
  \bibfield  {author} {\bibinfo {author} {\bibfnamefont {T.}~\bibnamefont
  {Kaneko}}, \bibinfo {author} {\bibfnamefont {H.}~\bibnamefont {Sakakibara}},
  \bibinfo {author} {\bibfnamefont {M.}~\bibnamefont {Ochi}},\ and\ \bibinfo
  {author} {\bibfnamefont {K.}~\bibnamefont {Kuroki}},\ }\bibfield  {title}
  {\bibinfo {title} {{Pair correlations in the two-orbital Hubbard ladder:
  Implications for superconductivity in the bilayer nickelate
  {${\mathrm{La}}_{3}{\mathrm{Ni}}_{2}{\mathrm{O}}_{7}$}}},\ }\href
  {https://doi.org/10.1103/PhysRevB.109.045154} {\bibfield  {journal} {\bibinfo
   {journal} {Phys. Rev. B}\ }\textbf {\bibinfo {volume} {109}},\ \bibinfo
  {pages} {045154} (\bibinfo {year} {2024})}\BibitemShut {NoStop}%
\bibitem [{\citenamefont {Lu}\ \emph {et~al.}(2024{\natexlab{b}})\citenamefont
  {Lu}, \citenamefont {Pan}, \citenamefont {Yang},\ and\ \citenamefont
  {Wu}}]{lu2023interplay}%
  \BibitemOpen
  \bibfield  {author} {\bibinfo {author} {\bibfnamefont {C.}~\bibnamefont
  {Lu}}, \bibinfo {author} {\bibfnamefont {Z.}~\bibnamefont {Pan}}, \bibinfo
  {author} {\bibfnamefont {F.}~\bibnamefont {Yang}},\ and\ \bibinfo {author}
  {\bibfnamefont {C.}~\bibnamefont {Wu}},\ }\bibfield  {title} {\bibinfo
  {title} {Interplay of two {${E}_{g}$} orbitals in superconducting
  {${\mathrm{La}}_{3}{\mathrm{Ni}}_{2}{\mathrm{O}}_{7}$} under pressure},\
  }\href {https://doi.org/10.1103/PhysRevB.110.094509} {\bibfield  {journal}
  {\bibinfo  {journal} {Phys. Rev. B}\ }\textbf {\bibinfo {volume} {110}},\
  \bibinfo {pages} {094509} (\bibinfo {year} {2024}{\natexlab{b}})}\BibitemShut
  {NoStop}%
\bibitem [{\citenamefont {Ryee}\ \emph {et~al.}(2024)\citenamefont {Ryee},
  \citenamefont {Witt},\ and\ \citenamefont {Wehling}}]{ryee2023critical}%
  \BibitemOpen
  \bibfield  {author} {\bibinfo {author} {\bibfnamefont {S.}~\bibnamefont
  {Ryee}}, \bibinfo {author} {\bibfnamefont {N.}~\bibnamefont {Witt}},\ and\
  \bibinfo {author} {\bibfnamefont {T.~O.}\ \bibnamefont {Wehling}},\
  }\bibfield  {title} {\bibinfo {title} {{Quenched Pair Breaking by Interlayer
  Correlations as a Key to Superconductivity in
  {${\mathrm{La}}_{3}{\mathrm{Ni}}_{2}{\mathrm{O}}_{7}$}}},\ }\href
  {https://doi.org/10.1103/PhysRevLett.133.096002} {\bibfield  {journal}
  {\bibinfo  {journal} {Phys. Rev. Lett.}\ }\textbf {\bibinfo {volume} {133}},\
  \bibinfo {pages} {096002} (\bibinfo {year} {2024})}\BibitemShut {NoStop}%
\bibitem [{\citenamefont {Schl{\"o}mer}\ \emph {et~al.}(2024)\citenamefont
  {Schl{\"o}mer}, \citenamefont {Schollw{\"o}ck}, \citenamefont {Grusdt},\ and\
  \citenamefont {Bohrdt}}]{schlmer2023superconductivity}%
  \BibitemOpen
  \bibfield  {author} {\bibinfo {author} {\bibfnamefont {H.}~\bibnamefont
  {Schl{\"o}mer}}, \bibinfo {author} {\bibfnamefont {U.}~\bibnamefont
  {Schollw{\"o}ck}}, \bibinfo {author} {\bibfnamefont {F.}~\bibnamefont
  {Grusdt}},\ and\ \bibinfo {author} {\bibfnamefont {A.}~\bibnamefont
  {Bohrdt}},\ }\bibfield  {title} {\bibinfo {title} {Superconductivity in the
  pressurized nickelate {La$_3$Ni$_2$O$_7$} in the vicinity of a {BEC--BCS}
  crossover},\ }\href {https://doi.org/10.1038/s42005-024-01854-9} {\bibfield
  {journal} {\bibinfo  {journal} {Communications Physics}\ }\textbf {\bibinfo
  {volume} {7}},\ \bibinfo {pages} {366} (\bibinfo {year} {2024})}\BibitemShut
  {NoStop}%
\bibitem [{\citenamefont {Chen}\ \emph
  {et~al.}(2024{\natexlab{c}})\citenamefont {Chen}, \citenamefont {Yang},\ and\
  \citenamefont {Li}}]{chen2023orbitalselective}%
  \BibitemOpen
  \bibfield  {author} {\bibinfo {author} {\bibfnamefont {J.}~\bibnamefont
  {Chen}}, \bibinfo {author} {\bibfnamefont {F.}~\bibnamefont {Yang}},\ and\
  \bibinfo {author} {\bibfnamefont {W.}~\bibnamefont {Li}},\ }\bibfield
  {title} {\bibinfo {title} {Orbital-selective superconductivity in the
  pressurized bilayer nickelate
  {${\mathrm{La}}_{3}{\mathrm{Ni}}_{2}{\mathrm{O}}_{7}$}: An infinite projected
  entangled-pair state study},\ }\href
  {https://doi.org/10.1103/PhysRevB.110.L041111} {\bibfield  {journal}
  {\bibinfo  {journal} {Phys. Rev. B}\ }\textbf {\bibinfo {volume} {110}},\
  \bibinfo {pages} {L041111} (\bibinfo {year}
  {2024}{\natexlab{c}})}\BibitemShut {NoStop}%
\bibitem [{\citenamefont {Xia}\ \emph {et~al.}(2025)\citenamefont {Xia},
  \citenamefont {Liu}, \citenamefont {Zhou},\ and\ \citenamefont
  {Chen}}]{liu2023role}%
  \BibitemOpen
  \bibfield  {author} {\bibinfo {author} {\bibfnamefont {C.}~\bibnamefont
  {Xia}}, \bibinfo {author} {\bibfnamefont {H.}~\bibnamefont {Liu}}, \bibinfo
  {author} {\bibfnamefont {S.}~\bibnamefont {Zhou}},\ and\ \bibinfo {author}
  {\bibfnamefont {H.}~\bibnamefont {Chen}},\ }\bibfield  {title} {\bibinfo
  {title} {{Sensitive dependence of pairing symmetry on Ni-$e_g$ crystal field
  splitting in the nickelate superconductor La$_3$Ni$_2$O$_7$}},\ }\href
  {https://doi.org/10.1038/s41467-025-56206-0} {\bibfield  {journal} {\bibinfo
  {journal} {Nature Communications}\ }\textbf {\bibinfo {volume} {16}},\
  \bibinfo {pages} {1054} (\bibinfo {year} {2025})}\BibitemShut {NoStop}%
\bibitem [{\citenamefont {Ouyang}\ \emph {et~al.}(2024)\citenamefont {Ouyang},
  \citenamefont {Wang}, \citenamefont {Wang}, \citenamefont {He}, \citenamefont
  {Huang},\ and\ \citenamefont {Lu}}]{ouyang2023hund}%
  \BibitemOpen
  \bibfield  {author} {\bibinfo {author} {\bibfnamefont {Z.}~\bibnamefont
  {Ouyang}}, \bibinfo {author} {\bibfnamefont {J.-M.}\ \bibnamefont {Wang}},
  \bibinfo {author} {\bibfnamefont {J.-X.}\ \bibnamefont {Wang}}, \bibinfo
  {author} {\bibfnamefont {R.-Q.}\ \bibnamefont {He}}, \bibinfo {author}
  {\bibfnamefont {L.}~\bibnamefont {Huang}},\ and\ \bibinfo {author}
  {\bibfnamefont {Z.-Y.}\ \bibnamefont {Lu}},\ }\bibfield  {title} {\bibinfo
  {title} {Hund electronic correlation in
  {${\mathrm{La}}_{3}{\mathrm{Ni}}_{2}{\mathrm{O}}_{7}$} under high pressure},\
  }\href {https://doi.org/10.1103/PhysRevB.109.115114} {\bibfield  {journal}
  {\bibinfo  {journal} {Phys. Rev. B}\ }\textbf {\bibinfo {volume} {109}},\
  \bibinfo {pages} {115114} (\bibinfo {year} {2024})}\BibitemShut {NoStop}%
\bibitem [{\citenamefont {Chang}\ \emph {et~al.}(2023)\citenamefont {Chang},
  \citenamefont {Guo}, \citenamefont {You},\ and\ \citenamefont
  {Li}}]{chang2023fermi}%
  \BibitemOpen
  \bibfield  {author} {\bibinfo {author} {\bibfnamefont {W.-X.}\ \bibnamefont
  {Chang}}, \bibinfo {author} {\bibfnamefont {S.}~\bibnamefont {Guo}}, \bibinfo
  {author} {\bibfnamefont {Y.-Z.}\ \bibnamefont {You}},\ and\ \bibinfo {author}
  {\bibfnamefont {Z.-X.}\ \bibnamefont {Li}},\ }\href@noop {} {\bibinfo {title}
  {Fermi surface symmetric mass generation: a quantum {Monte-Carlo} study}}
  (\bibinfo {year} {2023}),\ \Eprint {https://arxiv.org/abs/2311.09970}
  {arXiv:2311.09970 [cond-mat.str-el]} \BibitemShut {NoStop}%
\bibitem [{\citenamefont {Sui}\ \emph {et~al.}(2024)\citenamefont {Sui},
  \citenamefont {Han}, \citenamefont {Jin}, \citenamefont {Chen}, \citenamefont
  {Qiao}, \citenamefont {Shao},\ and\ \citenamefont
  {Huang}}]{sui2023electronic}%
  \BibitemOpen
  \bibfield  {author} {\bibinfo {author} {\bibfnamefont {X.}~\bibnamefont
  {Sui}}, \bibinfo {author} {\bibfnamefont {X.}~\bibnamefont {Han}}, \bibinfo
  {author} {\bibfnamefont {H.}~\bibnamefont {Jin}}, \bibinfo {author}
  {\bibfnamefont {X.}~\bibnamefont {Chen}}, \bibinfo {author} {\bibfnamefont
  {L.}~\bibnamefont {Qiao}}, \bibinfo {author} {\bibfnamefont {X.}~\bibnamefont
  {Shao}},\ and\ \bibinfo {author} {\bibfnamefont {B.}~\bibnamefont {Huang}},\
  }\bibfield  {title} {\bibinfo {title} {Electronic properties of the bilayer
  nickelates {${R}_{3}\mathrm{N}{\mathrm{i}}_{2}{\mathrm{O}}_{7}$} with oxygen
  vacancies {($R=\mathrm{La}$ or Ce)}},\ }\href
  {https://doi.org/10.1103/PhysRevB.109.205156} {\bibfield  {journal} {\bibinfo
   {journal} {Phys. Rev. B}\ }\textbf {\bibinfo {volume} {109}},\ \bibinfo
  {pages} {205156} (\bibinfo {year} {2024})}\BibitemShut {NoStop}%
\bibitem [{\citenamefont {Zheng}\ and\ \citenamefont
  {W\'u}(2025)}]{zheng2023superconductivity}%
  \BibitemOpen
  \bibfield  {author} {\bibinfo {author} {\bibfnamefont {Y.-Y.}\ \bibnamefont
  {Zheng}}\ and\ \bibinfo {author} {\bibfnamefont {W.}~\bibnamefont {W\'u}},\
  }\bibfield  {title} {\bibinfo {title}
  {{${s}_{\ifmmode\pm\else\textpm\fi{}}$-wave superconductivity in the bilayer
  two-orbital Hubbard model}},\ }\href
  {https://doi.org/10.1103/PhysRevB.111.035108} {\bibfield  {journal} {\bibinfo
   {journal} {Phys. Rev. B}\ }\textbf {\bibinfo {volume} {111}},\ \bibinfo
  {pages} {035108} (\bibinfo {year} {2025})}\BibitemShut {NoStop}%
\bibitem [{\citenamefont {Xue}\ and\ \citenamefont
  {Wang}(2024)}]{xue2024magnetism}%
  \BibitemOpen
  \bibfield  {author} {\bibinfo {author} {\bibfnamefont {J.-R.}\ \bibnamefont
  {Xue}}\ and\ \bibinfo {author} {\bibfnamefont {F.}~\bibnamefont {Wang}},\
  }\bibfield  {title} {\bibinfo {title} {{Magnetism and Superconductivity in
  the {$t$–$J$} Model of {La$_3$Ni$_2$O$_7$} Under Multiband {Gutzwiller}
  Approximation}},\ }\href {https://doi.org/10.1088/0256-307X/41/5/057403}
  {\bibfield  {journal} {\bibinfo  {journal} {Chinese Physics Letters}\
  }\textbf {\bibinfo {volume} {41}},\ \bibinfo {pages} {057403} (\bibinfo
  {year} {2024})}\BibitemShut {NoStop}%
\bibitem [{\citenamefont {Bhatta}\ \emph {et~al.}(2025)\citenamefont {Bhatta},
  \citenamefont {Zhang}, \citenamefont {Zhong},\ and\ \citenamefont
  {Jia}}]{bhatta2025structural}%
  \BibitemOpen
  \bibfield  {author} {\bibinfo {author} {\bibfnamefont {H.~C. R.~B.}\
  \bibnamefont {Bhatta}}, \bibinfo {author} {\bibfnamefont {X.}~\bibnamefont
  {Zhang}}, \bibinfo {author} {\bibfnamefont {Y.}~\bibnamefont {Zhong}},\ and\
  \bibinfo {author} {\bibfnamefont {C.}~\bibnamefont {Jia}},\ }\href
  {https://arxiv.org/abs/2502.01624} {\bibinfo {title} {{Structural and
  Electronic Evolution of Bilayer Nickelates Under Biaxial Strain}}} (\bibinfo
  {year} {2025}),\ \Eprint {https://arxiv.org/abs/2502.01624} {arXiv:2502.01624
  [cond-mat.supr-con]} \BibitemShut {NoStop}%
\bibitem [{\citenamefont {Kakoi}\ \emph {et~al.}(2024)\citenamefont {Kakoi},
  \citenamefont {Kaneko}, \citenamefont {Sakakibara}, \citenamefont {Ochi},\
  and\ \citenamefont {Kuroki}}]{Masataka2024pair}%
  \BibitemOpen
  \bibfield  {author} {\bibinfo {author} {\bibfnamefont {M.}~\bibnamefont
  {Kakoi}}, \bibinfo {author} {\bibfnamefont {T.}~\bibnamefont {Kaneko}},
  \bibinfo {author} {\bibfnamefont {H.}~\bibnamefont {Sakakibara}}, \bibinfo
  {author} {\bibfnamefont {M.}~\bibnamefont {Ochi}},\ and\ \bibinfo {author}
  {\bibfnamefont {K.}~\bibnamefont {Kuroki}},\ }\bibfield  {title} {\bibinfo
  {title} {Pair correlations of the hybridized orbitals in a ladder model for
  the bilayer nickelate
  {${\mathrm{La}}_{3}{\mathrm{Ni}}_{2}{\mathrm{O}}_{7}$}},\ }\href
  {https://doi.org/10.1103/PhysRevB.109.L201124} {\bibfield  {journal}
  {\bibinfo  {journal} {Phys. Rev. B}\ }\textbf {\bibinfo {volume} {109}},\
  \bibinfo {pages} {L201124} (\bibinfo {year} {2024})}\BibitemShut {NoStop}%
\bibitem [{\citenamefont {Shen}\ \emph {et~al.}(2025)\citenamefont {Shen},
  \citenamefont {Huang}, \citenamefont {Qian}, \citenamefont {Zhang},\ and\
  \citenamefont {Qin}}]{shen2024numerical}%
  \BibitemOpen
  \bibfield  {author} {\bibinfo {author} {\bibfnamefont {Y.}~\bibnamefont
  {Shen}}, \bibinfo {author} {\bibfnamefont {J.}~\bibnamefont {Huang}},
  \bibinfo {author} {\bibfnamefont {X.}~\bibnamefont {Qian}}, \bibinfo {author}
  {\bibfnamefont {G.-M.}\ \bibnamefont {Zhang}},\ and\ \bibinfo {author}
  {\bibfnamefont {M.}~\bibnamefont {Qin}},\ }\bibfield  {title} {\bibinfo
  {title} {{Numerical study of the bilayer two-orbital model for
  ${\mathrm{La}}_{3}{\mathrm{Ni}}_{2}{\mathrm{O}}_{7}$ on a plaquette
  ladder}},\ }\href {https://doi.org/10.1103/PhysRevB.111.L180508} {\bibfield
  {journal} {\bibinfo  {journal} {Phys. Rev. B}\ }\textbf {\bibinfo {volume}
  {111}},\ \bibinfo {pages} {L180508} (\bibinfo {year} {2025})}\BibitemShut
  {NoStop}%
\bibitem [{\citenamefont {Wang}\ and\ \citenamefont
  {Yang}(2025)}]{wang2024highly}%
  \BibitemOpen
  \bibfield  {author} {\bibinfo {author} {\bibfnamefont {J.}~\bibnamefont
  {Wang}}\ and\ \bibinfo {author} {\bibfnamefont {Y.-f.}\ \bibnamefont
  {Yang}},\ }\bibfield  {title} {\bibinfo {title} {{Highly asymmetric
  superconducting dome and strange metallicity in
  ${\mathrm{La}}_{3}{\mathrm{Ni}}_{2}{\mathrm{O}}_{7}$}},\ }\href
  {https://doi.org/10.1103/PhysRevB.111.014512} {\bibfield  {journal} {\bibinfo
   {journal} {Phys. Rev. B}\ }\textbf {\bibinfo {volume} {111}},\ \bibinfo
  {pages} {014512} (\bibinfo {year} {2025})}\BibitemShut {NoStop}%
\bibitem [{\citenamefont {Hirthe}\ \emph {et~al.}(2023)\citenamefont {Hirthe},
  \citenamefont {Chalopin}, \citenamefont {Bourgund}, \citenamefont
  {Bojovi{\'c}}, \citenamefont {Bohrdt}, \citenamefont {Demler}, \citenamefont
  {Grusdt}, \citenamefont {Bloch},\ and\ \citenamefont
  {Hilker}}]{Hilker2023pairing}%
  \BibitemOpen
  \bibfield  {author} {\bibinfo {author} {\bibfnamefont {S.}~\bibnamefont
  {Hirthe}}, \bibinfo {author} {\bibfnamefont {T.}~\bibnamefont {Chalopin}},
  \bibinfo {author} {\bibfnamefont {D.}~\bibnamefont {Bourgund}}, \bibinfo
  {author} {\bibfnamefont {P.}~\bibnamefont {Bojovi{\'c}}}, \bibinfo {author}
  {\bibfnamefont {A.}~\bibnamefont {Bohrdt}}, \bibinfo {author} {\bibfnamefont
  {E.}~\bibnamefont {Demler}}, \bibinfo {author} {\bibfnamefont
  {F.}~\bibnamefont {Grusdt}}, \bibinfo {author} {\bibfnamefont
  {I.}~\bibnamefont {Bloch}},\ and\ \bibinfo {author} {\bibfnamefont {T.~A.}\
  \bibnamefont {Hilker}},\ }\bibfield  {title} {\bibinfo {title} {Magnetically
  mediated hole pairing in fermionic ladders of ultracold atoms},\ }\href
  {https://doi.org/10.1038/s41586-022-05437-y} {\bibfield  {journal} {\bibinfo
  {journal} {Nature}\ }\textbf {\bibinfo {volume} {613}},\ \bibinfo {pages}
  {463} (\bibinfo {year} {2023})}\BibitemShut {NoStop}%
\bibitem [{\citenamefont {Zhu}\ \emph {et~al.}(2018)\citenamefont {Zhu},
  \citenamefont {Sheng},\ and\ \citenamefont {Weng}}]{Zhu2018pairing}%
  \BibitemOpen
  \bibfield  {author} {\bibinfo {author} {\bibfnamefont {Z.}~\bibnamefont
  {Zhu}}, \bibinfo {author} {\bibfnamefont {D.~N.}\ \bibnamefont {Sheng}},\
  and\ \bibinfo {author} {\bibfnamefont {Z.-Y.}\ \bibnamefont {Weng}},\
  }\bibfield  {title} {\bibinfo {title} {Pairing versus phase coherence of
  doped holes in distinct quantum spin backgrounds},\ }\href
  {https://doi.org/10.1103/PhysRevB.97.115144} {\bibfield  {journal} {\bibinfo
  {journal} {Phys. Rev. B}\ }\textbf {\bibinfo {volume} {97}},\ \bibinfo
  {pages} {115144} (\bibinfo {year} {2018})}\BibitemShut {NoStop}%
\bibitem [{\citenamefont {Jiang}\ \emph {et~al.}(2020)\citenamefont {Jiang},
  \citenamefont {Chen},\ and\ \citenamefont {Weng}}]{Jiang2020critical}%
  \BibitemOpen
  \bibfield  {author} {\bibinfo {author} {\bibfnamefont {H.-C.}\ \bibnamefont
  {Jiang}}, \bibinfo {author} {\bibfnamefont {S.}~\bibnamefont {Chen}},\ and\
  \bibinfo {author} {\bibfnamefont {Z.-Y.}\ \bibnamefont {Weng}},\ }\bibfield
  {title} {\bibinfo {title} {Critical role of the sign structure in the doped
  {Mott} insulator: {Luther-Emery} versus {Fermi}-liquid-like state in
  quasi-one-dimensional ladders},\ }\href
  {https://doi.org/10.1103/PhysRevB.102.104512} {\bibfield  {journal} {\bibinfo
   {journal} {Phys. Rev. B}\ }\textbf {\bibinfo {volume} {102}},\ \bibinfo
  {pages} {104512} (\bibinfo {year} {2020})}\BibitemShut {NoStop}%
\bibitem [{\citenamefont {Ma}\ \emph {et~al.}(2022)\citenamefont {Ma},
  \citenamefont {Wang},\ and\ \citenamefont {Wu}}]{Ma2022doping}%
  \BibitemOpen
  \bibfield  {author} {\bibinfo {author} {\bibfnamefont {T.}~\bibnamefont
  {Ma}}, \bibinfo {author} {\bibfnamefont {D.}~\bibnamefont {Wang}},\ and\
  \bibinfo {author} {\bibfnamefont {C.}~\bibnamefont {Wu}},\ }\bibfield
  {title} {\bibinfo {title} {Doping-driven antiferromagnetic
  insulator-superconductor transition: {A quantum Monte Carlo study}},\ }\href
  {https://doi.org/10.1103/PhysRevB.106.054510} {\bibfield  {journal} {\bibinfo
   {journal} {Phys. Rev. B}\ }\textbf {\bibinfo {volume} {106}},\ \bibinfo
  {pages} {054510} (\bibinfo {year} {2022})}\BibitemShut {NoStop}%
\bibitem [{\citenamefont {White}(1992)}]{White1992}%
  \BibitemOpen
  \bibfield  {author} {\bibinfo {author} {\bibfnamefont {S.~R.}\ \bibnamefont
  {White}},\ }\bibfield  {title} {\bibinfo {title} {Density matrix formulation
  for quantum renormalization groups},\ }\href
  {https://doi.org/10.1103/PhysRevLett.69.2863} {\bibfield  {journal} {\bibinfo
   {journal} {Phys. Rev. Lett.}\ }\textbf {\bibinfo {volume} {69}},\ \bibinfo
  {pages} {2863} (\bibinfo {year} {1992})}\BibitemShut {NoStop}%
\bibitem [{\citenamefont {{Schollw{\"o}ck}}(2011)}]{Schollwock2011MPS}%
  \BibitemOpen
  \bibfield  {author} {\bibinfo {author} {\bibfnamefont {U.}~\bibnamefont
  {{Schollw{\"o}ck}}},\ }\bibfield  {title} {\bibinfo {title} {The
  density-matrix renormalization group in the age of matrix product states},\
  }\href {https://doi.org/https://doi.org/10.1016/j.aop.2010.09.012} {\bibfield
   {journal} {\bibinfo  {journal} {Ann. Phys.}\ }\textbf {\bibinfo {volume}
  {326}},\ \bibinfo {pages} {96} (\bibinfo {year} {2011})}\BibitemShut
  {NoStop}%
\bibitem [{\citenamefont {Gleis}\ \emph {et~al.}(2023)\citenamefont {Gleis},
  \citenamefont {Li},\ and\ \citenamefont {von Delft}}]{Andreas2023CBE}%
  \BibitemOpen
  \bibfield  {author} {\bibinfo {author} {\bibfnamefont {A.}~\bibnamefont
  {Gleis}}, \bibinfo {author} {\bibfnamefont {J.-W.}\ \bibnamefont {Li}},\ and\
  \bibinfo {author} {\bibfnamefont {J.}~\bibnamefont {von Delft}},\ }\bibfield
  {title} {\bibinfo {title} {{Controlled Bond Expansion for Density Matrix
  Renormalization Group Ground State Search at Single-Site Costs}},\ }\href
  {https://doi.org/10.1103/PhysRevLett.130.246402} {\bibfield  {journal}
  {\bibinfo  {journal} {Phys. Rev. Lett.}\ }\textbf {\bibinfo {volume} {130}},\
  \bibinfo {pages} {246402} (\bibinfo {year} {2023})}\BibitemShut {NoStop}%
\bibitem [{\citenamefont {Verstraete}\ and\ \citenamefont
  {Cirac}(2004)}]{Verstraete2004renorm}%
  \BibitemOpen
  \bibfield  {author} {\bibinfo {author} {\bibfnamefont {F.}~\bibnamefont
  {Verstraete}}\ and\ \bibinfo {author} {\bibfnamefont {J.~I.}\ \bibnamefont
  {Cirac}},\ }\href@noop {} {\bibinfo {title} {Renormalization algorithms for
  quantum-many body systems in two and higher dimensions}} (\bibinfo {year}
  {2004}),\ \Eprint {https://arxiv.org/abs/cond-mat/0407066}
  {arXiv:cond-mat/0407066 [cond-mat.str-el]} \BibitemShut {NoStop}%
\bibitem [{\citenamefont {Jordan}\ \emph {et~al.}(2008)\citenamefont {Jordan},
  \citenamefont {Or\'us}, \citenamefont {Vidal}, \citenamefont {Verstraete},\
  and\ \citenamefont {Cirac}}]{Jordan2008Classical}%
  \BibitemOpen
  \bibfield  {author} {\bibinfo {author} {\bibfnamefont {J.}~\bibnamefont
  {Jordan}}, \bibinfo {author} {\bibfnamefont {R.}~\bibnamefont {Or\'us}},
  \bibinfo {author} {\bibfnamefont {G.}~\bibnamefont {Vidal}}, \bibinfo
  {author} {\bibfnamefont {F.}~\bibnamefont {Verstraete}},\ and\ \bibinfo
  {author} {\bibfnamefont {J.~I.}\ \bibnamefont {Cirac}},\ }\bibfield  {title}
  {\bibinfo {title} {{Classical Simulation of Infinite-Size Quantum Lattice
  Systems in Two Spatial Dimensions}},\ }\href
  {https://doi.org/10.1103/PhysRevLett.101.250602} {\bibfield  {journal}
  {\bibinfo  {journal} {Phys. Rev. Lett.}\ }\textbf {\bibinfo {volume} {101}},\
  \bibinfo {pages} {250602} (\bibinfo {year} {2008})}\BibitemShut {NoStop}%
\bibitem [{\citenamefont {Cirac}\ \emph {et~al.}(2021)\citenamefont {Cirac},
  \citenamefont {P\'erez-Garc\'{\i}a}, \citenamefont {Schuch},\ and\
  \citenamefont {Verstraete}}]{Cirac2021RMP}%
  \BibitemOpen
  \bibfield  {author} {\bibinfo {author} {\bibfnamefont {J.~I.}\ \bibnamefont
  {Cirac}}, \bibinfo {author} {\bibfnamefont {D.}~\bibnamefont
  {P\'erez-Garc\'{\i}a}}, \bibinfo {author} {\bibfnamefont {N.}~\bibnamefont
  {Schuch}},\ and\ \bibinfo {author} {\bibfnamefont {F.}~\bibnamefont
  {Verstraete}},\ }\bibfield  {title} {\bibinfo {title} {{Matrix Product States
  and Projected Entangled Pair States: Concepts, Symmetries, Theorems}},\
  }\href {https://doi.org/10.1103/RevModPhys.93.045003} {\bibfield  {journal}
  {\bibinfo  {journal} {Rev. Mod. Phys.}\ }\textbf {\bibinfo {volume} {93}},\
  \bibinfo {pages} {045003} (\bibinfo {year} {2021})}\BibitemShut {NoStop}%
\bibitem [{\citenamefont {Corboz}\ \emph {et~al.}(2010)\citenamefont {Corboz},
  \citenamefont {Or\'us}, \citenamefont {Bauer},\ and\ \citenamefont
  {Vidal}}]{Corboz2010Simulation}%
  \BibitemOpen
  \bibfield  {author} {\bibinfo {author} {\bibfnamefont {P.}~\bibnamefont
  {Corboz}}, \bibinfo {author} {\bibfnamefont {R.}~\bibnamefont {Or\'us}},
  \bibinfo {author} {\bibfnamefont {B.}~\bibnamefont {Bauer}},\ and\ \bibinfo
  {author} {\bibfnamefont {G.}~\bibnamefont {Vidal}},\ }\bibfield  {title}
  {\bibinfo {title} {Simulation of strongly correlated fermions in two spatial
  dimensions with fermionic projected entangled-pair states},\ }\href
  {https://doi.org/10.1103/PhysRevB.81.165104} {\bibfield  {journal} {\bibinfo
  {journal} {Phys. Rev. B}\ }\textbf {\bibinfo {volume} {81}},\ \bibinfo
  {pages} {165104} (\bibinfo {year} {2010})}\BibitemShut {NoStop}%
\bibitem [{\citenamefont {Li}\ \emph {et~al.}(2011)\citenamefont {Li},
  \citenamefont {Ran}, \citenamefont {Gong}, \citenamefont {Zhao},
  \citenamefont {Xi}, \citenamefont {Ye},\ and\ \citenamefont {Su}}]{Li2011a}%
  \BibitemOpen
  \bibfield  {author} {\bibinfo {author} {\bibfnamefont {W.}~\bibnamefont
  {Li}}, \bibinfo {author} {\bibfnamefont {S.-J.}\ \bibnamefont {Ran}},
  \bibinfo {author} {\bibfnamefont {S.-S.}\ \bibnamefont {Gong}}, \bibinfo
  {author} {\bibfnamefont {Y.}~\bibnamefont {Zhao}}, \bibinfo {author}
  {\bibfnamefont {B.}~\bibnamefont {Xi}}, \bibinfo {author} {\bibfnamefont
  {F.}~\bibnamefont {Ye}},\ and\ \bibinfo {author} {\bibfnamefont
  {G.}~\bibnamefont {Su}},\ }\bibfield  {title} {\bibinfo {title} {{Linearized
  Tensor Renormalization Group Algorithm for the Calculation of Thermodynamic
  Properties of Quantum Lattice Models}},\ }\href
  {https://doi.org/10.1103/PhysRevLett.106.127202} {\bibfield  {journal}
  {\bibinfo  {journal} {Phys. Rev. Lett.}\ }\textbf {\bibinfo {volume} {106}},\
  \bibinfo {pages} {127202} (\bibinfo {year} {2011})}\BibitemShut {NoStop}%
\bibitem [{\citenamefont {Chen}\ \emph {et~al.}(2018)\citenamefont {Chen},
  \citenamefont {Chen}, \citenamefont {Chen}, \citenamefont {Li},\ and\
  \citenamefont {Weichselbaum}}]{Chen2018XTRG}%
  \BibitemOpen
  \bibfield  {author} {\bibinfo {author} {\bibfnamefont {B.-B.}\ \bibnamefont
  {Chen}}, \bibinfo {author} {\bibfnamefont {L.}~\bibnamefont {Chen}}, \bibinfo
  {author} {\bibfnamefont {Z.}~\bibnamefont {Chen}}, \bibinfo {author}
  {\bibfnamefont {W.}~\bibnamefont {Li}},\ and\ \bibinfo {author}
  {\bibfnamefont {A.}~\bibnamefont {Weichselbaum}},\ }\bibfield  {title}
  {\bibinfo {title} {{Exponential Thermal Tensor Network Approach for Quantum
  Lattice Models}},\ }\href {https://doi.org/10.1103/PhysRevX.8.031082}
  {\bibfield  {journal} {\bibinfo  {journal} {Phys. Rev. X}\ }\textbf {\bibinfo
  {volume} {8}},\ \bibinfo {pages} {031082} (\bibinfo {year}
  {2018})}\BibitemShut {NoStop}%
\bibitem [{\citenamefont {Li}\ \emph {et~al.}(2023)\citenamefont {Li},
  \citenamefont {Gao}, \citenamefont {He}, \citenamefont {Qi}, \citenamefont
  {Chen},\ and\ \citenamefont {Li}}]{tanTRG2023}%
  \BibitemOpen
  \bibfield  {author} {\bibinfo {author} {\bibfnamefont {Q.}~\bibnamefont
  {Li}}, \bibinfo {author} {\bibfnamefont {Y.}~\bibnamefont {Gao}}, \bibinfo
  {author} {\bibfnamefont {Y.-Y.}\ \bibnamefont {He}}, \bibinfo {author}
  {\bibfnamefont {Y.}~\bibnamefont {Qi}}, \bibinfo {author} {\bibfnamefont
  {B.-B.}\ \bibnamefont {Chen}},\ and\ \bibinfo {author} {\bibfnamefont
  {W.}~\bibnamefont {Li}},\ }\bibfield  {title} {\bibinfo {title} {{Tangent
  Space Approach for Thermal Tensor Network Simulations of the {2D} {Hubbard}
  Model}},\ }\href {https://doi.org/10.1103/PhysRevLett.130.226502} {\bibfield
  {journal} {\bibinfo  {journal} {Phys. Rev. Lett.}\ }\textbf {\bibinfo
  {volume} {130}},\ \bibinfo {pages} {226502} (\bibinfo {year}
  {2023})}\BibitemShut {NoStop}%
\bibitem [{\citenamefont {Yi}\ \emph {et~al.}(2024)\citenamefont {Yi},
  \citenamefont {Meng}, \citenamefont {Li}, \citenamefont {Liao}, \citenamefont
  {Li}, \citenamefont {You}, \citenamefont {Gu},\ and\ \citenamefont
  {Su}}]{yi2024antiferromagnetic}%
  \BibitemOpen
  \bibfield  {author} {\bibinfo {author} {\bibfnamefont {X.-W.}\ \bibnamefont
  {Yi}}, \bibinfo {author} {\bibfnamefont {Y.}~\bibnamefont {Meng}}, \bibinfo
  {author} {\bibfnamefont {J.-W.}\ \bibnamefont {Li}}, \bibinfo {author}
  {\bibfnamefont {Z.-W.}\ \bibnamefont {Liao}}, \bibinfo {author}
  {\bibfnamefont {W.}~\bibnamefont {Li}}, \bibinfo {author} {\bibfnamefont
  {J.-Y.}\ \bibnamefont {You}}, \bibinfo {author} {\bibfnamefont
  {B.}~\bibnamefont {Gu}},\ and\ \bibinfo {author} {\bibfnamefont
  {G.}~\bibnamefont {Su}},\ }\bibfield  {title} {\bibinfo {title} {Nature of
  charge density waves and metal-insulator transition in pressurized
  {${\mathrm{La}}_{3}{\mathrm{Ni}}_{2}{\mathrm{O}}_{7}$}},\ }\href
  {https://doi.org/10.1103/PhysRevB.110.L140508} {\bibfield  {journal}
  {\bibinfo  {journal} {Phys. Rev. B}\ }\textbf {\bibinfo {volume} {110}},\
  \bibinfo {pages} {L140508} (\bibinfo {year} {2024})}\BibitemShut {NoStop}%
\bibitem [{\citenamefont {Zhou}\ \emph {et~al.}(2025)\citenamefont {Zhou},
  \citenamefont {Lv}, \citenamefont {Wang}, \citenamefont {Nie}, \citenamefont
  {Chen}, \citenamefont {Li}, \citenamefont {Huang}, \citenamefont {Chen},
  \citenamefont {Sun}, \citenamefont {Xue},\ and\ \citenamefont
  {Chen}}]{zhou2024ambient}%
  \BibitemOpen
  \bibfield  {author} {\bibinfo {author} {\bibfnamefont {G.}~\bibnamefont
  {Zhou}}, \bibinfo {author} {\bibfnamefont {W.}~\bibnamefont {Lv}}, \bibinfo
  {author} {\bibfnamefont {H.}~\bibnamefont {Wang}}, \bibinfo {author}
  {\bibfnamefont {Z.}~\bibnamefont {Nie}}, \bibinfo {author} {\bibfnamefont
  {Y.}~\bibnamefont {Chen}}, \bibinfo {author} {\bibfnamefont {Y.}~\bibnamefont
  {Li}}, \bibinfo {author} {\bibfnamefont {H.}~\bibnamefont {Huang}}, \bibinfo
  {author} {\bibfnamefont {W.-Q.}\ \bibnamefont {Chen}}, \bibinfo {author}
  {\bibfnamefont {Y.-J.}\ \bibnamefont {Sun}}, \bibinfo {author} {\bibfnamefont
  {Q.-K.}\ \bibnamefont {Xue}},\ and\ \bibinfo {author} {\bibfnamefont
  {Z.}~\bibnamefont {Chen}},\ }\bibfield  {title} {\bibinfo {title}
  {{Ambient-pressure superconductivity onset above 40{\thinspace}K in
  (La,Pr)$_3$Ni$_2$O$_7$ films}},\ }\href
  {https://doi.org/10.1038/s41586-025-08755-z} {\bibfield  {journal} {\bibinfo
  {journal} {Nature}\ }\textbf {\bibinfo {volume} {640}},\ \bibinfo {pages}
  {641} (\bibinfo {year} {2025})}\BibitemShut {NoStop}%
\bibitem [{\citenamefont {Weichselbaum}(2012)}]{Weichselbaum2012}%
  \BibitemOpen
  \bibfield  {author} {\bibinfo {author} {\bibfnamefont {A.}~\bibnamefont
  {Weichselbaum}},\ }\bibfield  {title} {\bibinfo {title} {Non-abelian
  symmetries in tensor networks : {A} quantum symmetry space approach},\ }\href
  {https://doi.org/10.1016/j.aop.2012.07.009} {\bibfield  {journal} {\bibinfo
  {journal} {Ann. Phys.}\ }\textbf {\bibinfo {volume} {327}},\ \bibinfo {pages}
  {2972} (\bibinfo {year} {2012})}\BibitemShut {NoStop}%
\bibitem [{\citenamefont {Weichselbaum}(2020)}]{Weichselbaum2020}%
  \BibitemOpen
  \bibfield  {author} {\bibinfo {author} {\bibfnamefont {A.}~\bibnamefont
  {Weichselbaum}},\ }\bibfield  {title} {\bibinfo {title} {X-symbols for
  {non-Abelian} symmetries in tensor networks},\ }\href
  {https://doi.org/10.1103/PhysRevResearch.2.023385} {\bibfield  {journal}
  {\bibinfo  {journal} {Phys. Rev. Research}\ }\textbf {\bibinfo {volume}
  {2}},\ \bibinfo {pages} {023385} (\bibinfo {year} {2020})}\BibitemShut
  {NoStop}%
\bibitem [{\citenamefont {Haegeman}\ \emph {et~al.}(2025)\citenamefont
  {Haegeman}, \citenamefont {Devos}, \citenamefont {Hauru}, \citenamefont
  {Nakano}, \citenamefont {Damme}, \citenamefont {Roose}, \citenamefont
  {Carlstr{\"o}m},\ and\ \citenamefont {Dong}}]{haegeman2024jutho}%
  \BibitemOpen
  \bibfield  {author} {\bibinfo {author} {\bibfnamefont {J.}~\bibnamefont
  {Haegeman}}, \bibinfo {author} {\bibfnamefont {L.}~\bibnamefont {Devos}},
  \bibinfo {author} {\bibfnamefont {M.}~\bibnamefont {Hauru}}, \bibinfo
  {author} {\bibfnamefont {H.}~\bibnamefont {Nakano}}, \bibinfo {author}
  {\bibfnamefont {M.}~\bibnamefont {Damme}}, \bibinfo {author} {\bibfnamefont
  {G.}~\bibnamefont {Roose}}, \bibinfo {author} {\bibfnamefont
  {S.}~\bibnamefont {Carlstr{\"o}m}},\ and\ \bibinfo {author} {\bibfnamefont
  {X.}~\bibnamefont {Dong}},\ }\href {https://github.com/Jutho/TensorKit.jl}
  {\bibinfo {title} {{Jutho/TensorKit. jl: v0. 14.6}}} (\bibinfo {year}
  {2025})\BibitemShut {NoStop}%
\bibitem [{\citenamefont {Fishman}\ \emph
  {et~al.}(2022{\natexlab{a}})\citenamefont {Fishman}, \citenamefont {White},\
  and\ \citenamefont {Stoudenmire}}]{ITensor}%
  \BibitemOpen
  \bibfield  {author} {\bibinfo {author} {\bibfnamefont {M.}~\bibnamefont
  {Fishman}}, \bibinfo {author} {\bibfnamefont {S.~R.}\ \bibnamefont {White}},\
  and\ \bibinfo {author} {\bibfnamefont {E.~M.}\ \bibnamefont {Stoudenmire}},\
  }\bibfield  {title} {\bibinfo {title} {{The ITensor Software Library for
  Tensor Network Calculations}},\ }\href
  {https://doi.org/10.21468/SciPostPhysCodeb.4} {\bibfield  {journal} {\bibinfo
   {journal} {SciPost Phys. Codebases}\ ,\ \bibinfo {pages} {4}} (\bibinfo
  {year} {2022}{\natexlab{a}})}\BibitemShut {NoStop}%
\bibitem [{\citenamefont {Fishman}\ \emph
  {et~al.}(2022{\natexlab{b}})\citenamefont {Fishman}, \citenamefont {White},\
  and\ \citenamefont {Stoudenmire}}]{ITensor-r0.3}%
  \BibitemOpen
  \bibfield  {author} {\bibinfo {author} {\bibfnamefont {M.}~\bibnamefont
  {Fishman}}, \bibinfo {author} {\bibfnamefont {S.~R.}\ \bibnamefont {White}},\
  and\ \bibinfo {author} {\bibfnamefont {E.~M.}\ \bibnamefont {Stoudenmire}},\
  }\bibfield  {title} {\bibinfo {title} {{Codebase release 0.3 for ITensor}},\
  }\href {https://doi.org/10.21468/SciPostPhysCodeb.4-r0.3} {\bibfield
  {journal} {\bibinfo  {journal} {SciPost Phys. Codebases}\ ,\ \bibinfo {pages}
  {4}} (\bibinfo {year} {2022}{\natexlab{b}})}\BibitemShut {NoStop}%
\bibitem [{\citenamefont {Jiang}\ \emph {et~al.}(2008)\citenamefont {Jiang},
  \citenamefont {Weng},\ and\ \citenamefont {Xiang}}]{Xiang2008SU}%
  \BibitemOpen
  \bibfield  {author} {\bibinfo {author} {\bibfnamefont {H.~C.}\ \bibnamefont
  {Jiang}}, \bibinfo {author} {\bibfnamefont {Z.~Y.}\ \bibnamefont {Weng}},\
  and\ \bibinfo {author} {\bibfnamefont {T.}~\bibnamefont {Xiang}},\ }\bibfield
   {title} {\bibinfo {title} {{Accurate Determination of Tensor Network State
  of Quantum Lattice Models in Two Dimensions}},\ }\href
  {https://doi.org/10.1103/PhysRevLett.101.090603} {\bibfield  {journal}
  {\bibinfo  {journal} {Phys. Rev. Lett.}\ }\textbf {\bibinfo {volume} {101}},\
  \bibinfo {pages} {090603} (\bibinfo {year} {2008})}\BibitemShut {NoStop}%
\bibitem [{\citenamefont {Li}\ \emph {et~al.}(2012)\citenamefont {Li},
  \citenamefont {von Delft},\ and\ \citenamefont {Xiang}}]{Li2012SU}%
  \BibitemOpen
  \bibfield  {author} {\bibinfo {author} {\bibfnamefont {W.}~\bibnamefont
  {Li}}, \bibinfo {author} {\bibfnamefont {J.}~\bibnamefont {von Delft}},\ and\
  \bibinfo {author} {\bibfnamefont {T.}~\bibnamefont {Xiang}},\ }\bibfield
  {title} {\bibinfo {title} {Efficient simulation of infinite tree tensor
  network states on the {Bethe} lattice},\ }\href
  {https://doi.org/10.1103/PhysRevB.86.195137} {\bibfield  {journal} {\bibinfo
  {journal} {Phys. Rev. B}\ }\textbf {\bibinfo {volume} {86}},\ \bibinfo
  {pages} {195137} (\bibinfo {year} {2012})}\BibitemShut {NoStop}%
\bibitem [{\citenamefont {Nishino}\ and\ \citenamefont
  {Okunishi}(1996)}]{NishinoCTMRG}%
  \BibitemOpen
  \bibfield  {author} {\bibinfo {author} {\bibfnamefont {T.}~\bibnamefont
  {Nishino}}\ and\ \bibinfo {author} {\bibfnamefont {K.}~\bibnamefont
  {Okunishi}},\ }\bibfield  {title} {\bibinfo {title} {{Corner Transfer Matrix
  Renormalization Group Method}},\ }\href {https://doi.org/10.1143/JPSJ.65.891}
  {\bibfield  {journal} {\bibinfo  {journal} {Journal of the Physical Society
  of Japan}\ }\textbf {\bibinfo {volume} {65}},\ \bibinfo {pages} {891}
  (\bibinfo {year} {1996})},\ \Eprint
  {https://arxiv.org/abs/https://doi.org/10.1143/JPSJ.65.891}
  {https://doi.org/10.1143/JPSJ.65.891} \BibitemShut {NoStop}%
\bibitem [{\citenamefont {Or\'us}\ and\ \citenamefont
  {Vidal}(2009)}]{Orus2009Simulation}%
  \BibitemOpen
  \bibfield  {author} {\bibinfo {author} {\bibfnamefont {R.}~\bibnamefont
  {Or\'us}}\ and\ \bibinfo {author} {\bibfnamefont {G.}~\bibnamefont {Vidal}},\
  }\bibfield  {title} {\bibinfo {title} {Simulation of two-dimensional quantum
  systems on an infinite lattice revisited: Corner transfer matrix for tensor
  contraction},\ }\href {https://doi.org/10.1103/PhysRevB.80.094403} {\bibfield
   {journal} {\bibinfo  {journal} {Phys. Rev. B}\ }\textbf {\bibinfo {volume}
  {80}},\ \bibinfo {pages} {094403} (\bibinfo {year} {2009})}\BibitemShut
  {NoStop}%
\bibitem [{\citenamefont {Qu}(2025)}]{Qu2025PEPSTensorKit}%
  \BibitemOpen
  \bibfield  {author} {\bibinfo {author} {\bibfnamefont {X.-Z.}\ \bibnamefont
  {Qu}},\ }\href {https://github.com/XingzhouQu/PEPS_TensorKit} {\bibinfo
  {title} {{XingzhouQu/PEPS\_TensorKit}}} (\bibinfo {year} {2025})\BibitemShut
  {NoStop}%
\bibitem [{\citenamefont {Kresse}\ and\ \citenamefont
  {Furthmuller}(1996)}]{RN142}%
  \BibitemOpen
  \bibfield  {author} {\bibinfo {author} {\bibfnamefont {G.}~\bibnamefont
  {Kresse}}\ and\ \bibinfo {author} {\bibfnamefont {J.}~\bibnamefont
  {Furthmuller}},\ }\bibfield  {title} {\bibinfo {title} {Efficient iterative
  schemes for ab initio total-energy calculations using a plane-wave basis
  set},\ }\href {https://doi.org/10.1103/physrevb.54.11169} {\bibfield
  {journal} {\bibinfo  {journal} {Phys. Rev., B Condens. Matter}\ }\textbf
  {\bibinfo {volume} {54}},\ \bibinfo {pages} {11169} (\bibinfo {year}
  {1996})}\BibitemShut {NoStop}%
\bibitem [{\citenamefont {Mostofi}\ \emph {et~al.}(2014)\citenamefont
  {Mostofi}, \citenamefont {Yates}, \citenamefont {Pizzi}, \citenamefont {Lee},
  \citenamefont {Souza}, \citenamefont {Vanderbilt},\ and\ \citenamefont
  {Marzari}}]{RN144}%
  \BibitemOpen
  \bibfield  {author} {\bibinfo {author} {\bibfnamefont {A.~A.}\ \bibnamefont
  {Mostofi}}, \bibinfo {author} {\bibfnamefont {J.~R.}\ \bibnamefont {Yates}},
  \bibinfo {author} {\bibfnamefont {G.}~\bibnamefont {Pizzi}}, \bibinfo
  {author} {\bibfnamefont {Y.-S.}\ \bibnamefont {Lee}}, \bibinfo {author}
  {\bibfnamefont {I.}~\bibnamefont {Souza}}, \bibinfo {author} {\bibfnamefont
  {D.}~\bibnamefont {Vanderbilt}},\ and\ \bibinfo {author} {\bibfnamefont
  {N.}~\bibnamefont {Marzari}},\ }\bibfield  {title} {\bibinfo {title} {An
  updated version of wannier90: {A} tool for obtaining maximally-localised
  {W}annier functions},\ }\href
  {https://doi.org/https://doi.org/10.1016/j.cpc.2014.05.003} {\bibfield
  {journal} {\bibinfo  {journal} {Comput. Phys. Commun.}\ }\textbf {\bibinfo
  {volume} {185}},\ \bibinfo {pages} {2309} (\bibinfo {year}
  {2014})}\BibitemShut {NoStop}%
\bibitem [{\citenamefont {Lu}\ \emph {et~al.}(2024{\natexlab{c}})\citenamefont
  {Lu}, \citenamefont {Chen}, \citenamefont {Zhu}, \citenamefont {Sheng},\ and\
  \citenamefont {Gong}}]{Lu2024tJ8leg}%
  \BibitemOpen
  \bibfield  {author} {\bibinfo {author} {\bibfnamefont {X.}~\bibnamefont
  {Lu}}, \bibinfo {author} {\bibfnamefont {F.}~\bibnamefont {Chen}}, \bibinfo
  {author} {\bibfnamefont {W.}~\bibnamefont {Zhu}}, \bibinfo {author}
  {\bibfnamefont {D.~N.}\ \bibnamefont {Sheng}},\ and\ \bibinfo {author}
  {\bibfnamefont {S.-S.}\ \bibnamefont {Gong}},\ }\bibfield  {title} {\bibinfo
  {title} {{Emergent Superconductivity and Competing Charge Orders in
  Hole-Doped Square-Lattice $t\text{\ensuremath{-}}J$ Model}},\ }\href
  {https://doi.org/10.1103/PhysRevLett.132.066002} {\bibfield  {journal}
  {\bibinfo  {journal} {Phys. Rev. Lett.}\ }\textbf {\bibinfo {volume} {132}},\
  \bibinfo {pages} {066002} (\bibinfo {year} {2024}{\natexlab{c}})}\BibitemShut
  {NoStop}%
\bibitem [{\citenamefont {Chen}\ \emph
  {et~al.}(2025{\natexlab{b}})\citenamefont {Chen}, \citenamefont {Haldane},\
  and\ \citenamefont {Sheng}}]{Chen2025tJ8leg}%
  \BibitemOpen
  \bibfield  {author} {\bibinfo {author} {\bibfnamefont {F.}~\bibnamefont
  {Chen}}, \bibinfo {author} {\bibfnamefont {F.~D.~M.}\ \bibnamefont
  {Haldane}},\ and\ \bibinfo {author} {\bibfnamefont {D.~N.}\ \bibnamefont
  {Sheng}},\ }\bibfield  {title} {\bibinfo {title} {{Global phase diagram of
  D-wave superconductivity in the square-lattice $t$-$J$ model}},\ }\href
  {https://doi.org/10.1073/pnas.2420963122} {\bibfield  {journal} {\bibinfo
  {journal} {Proceedings of the National Academy of Sciences}\ }\textbf
  {\bibinfo {volume} {122}},\ \bibinfo {pages} {e2420963122} (\bibinfo {year}
  {2025}{\natexlab{b}})}\BibitemShut {NoStop}%
\end{thebibliography}%

\clearpage
\newpage
\clearpage
\appendix*
\onecolumngrid
\begin{center}
\section{\textbf{\large{Appendix}}}

\end{center}



\renewcommand{\theequation}{\Alph{section}\arabic{equation}}
\setcounter{section}{0}
\setcounter{figure}{0}
\setcounter{equation}{0}
\renewcommand{\theequation}{A\arabic{equation}}
\renewcommand{\thefigure}{A\arabic{figure}}
\setcounter{secnumdepth}{3}

\section*{Hybridization versus Hund's rule coupling scenarios}
\begin{figure}[hb]
\centering
\includegraphics[width=0.5\linewidth]{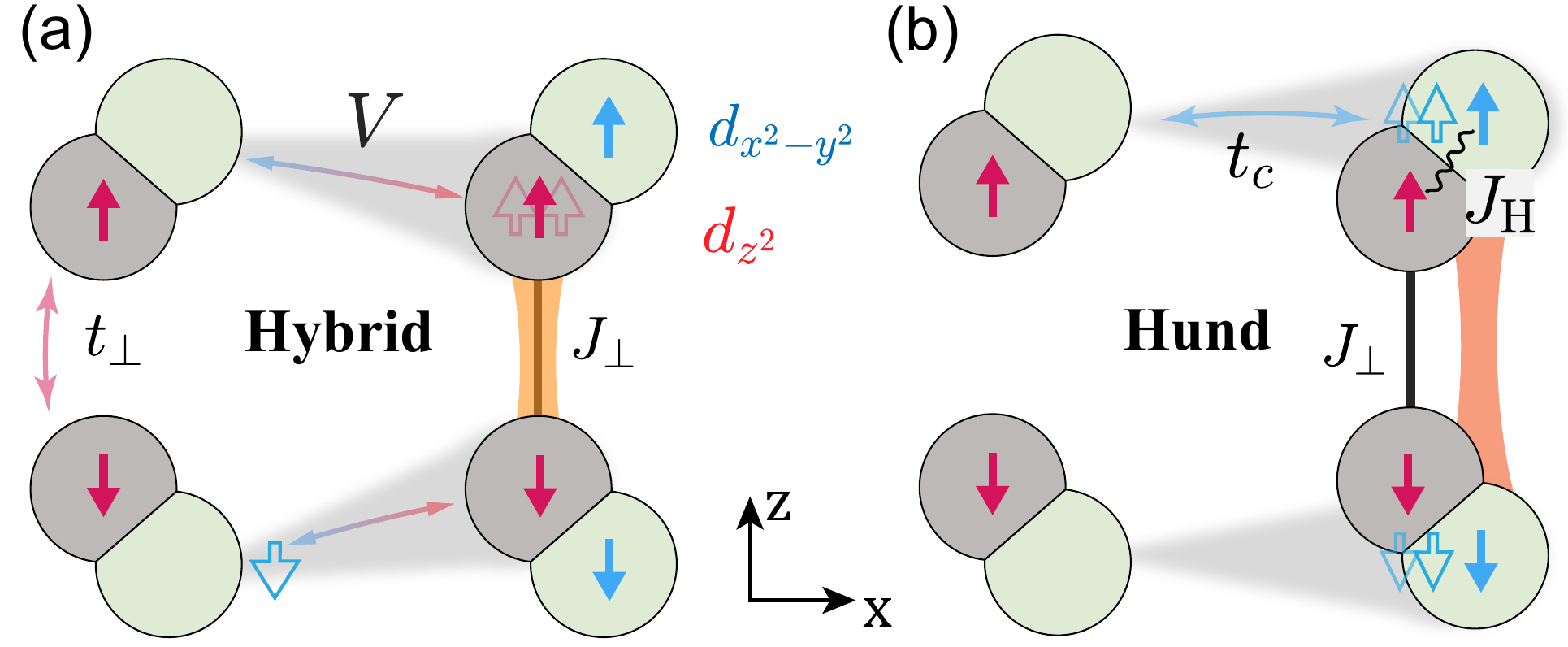}
\caption{Illustration of (a) hybridization dominated and (b) Hund's rule coupling dominated scenarios, which emphasizes respectively the hybridization $V$ and Hund's rule coupling $J_{\rm H}$ plays a primary role in forming the SC order.
}
\label{FigE1}
\end{figure}

In Fig.~\ref{FigE1}, we illustrate the hybridization and Hund SC scenarios. In the hybridization picture [Fig.~\ref{FigE1}(a)], there is a strong interlayer pairing between \ZO orbitals due to the significant AF coupling $J_\perp$. The preformed \ZO pairs gain phase coherence through hybridization with itinerant \XO orbitals. On the other hand, a different pairing scenario [Fig.~\ref{FigE1}(b)] considers that the Hund's rule coupling plays the primary role. The itinerant \XO band gains interlayer AF correlation through the strong on-site Hund's rule coupling that tends to symmetrize the spins of \XO and \ZO orbitals.

\section*{Ground-state and finite-temperature tensor network methods}
We exploit the state-of-the-art density matrix renormalization group (DMRG)~\cite{White1992, Schollwock2011MPS}, infinite projected entangled pair state (iPEPS)~\cite{Verstraete2004renorm, Jordan2008Classical, Cirac2021RMP, Corboz2010Simulation}, and thermal tensor network methods~\cite{Chen2018XTRG, tanTRG2023} to compute the zero- and finite-temperature properties. Non-Abelian and Abelian symmetries are implemented with the tensor library QSpace~\cite{Weichselbaum2012, Weichselbaum2020}, TensorKit~\cite{haegeman2024jutho} and ITensor~\cite{ITensor,ITensor-r0.3}. In DMRG calculations, we consider the $2 \times W \times L$ (with length $L$ up to 64 and width $W$ up to 2) two-orbital ladders, and retain up to $D^*=4 500$ U(1)$_{\rm charge} \times$ SU(2)$_{\rm spin}$ multiplets or $D=9 000$ U(1)$_{\rm charge} \times$ U(1)$_{\rm spin}$ individual states. The results are well converged with typical truncation error $\epsilon \sim 10^{-6}$. In tanTRG calculations, we consider system size $2\times 1 \times 24$, and a small pairing field is applied to compute the pairing susceptibility. We use $\mathbb{Z}_{2, \textrm{charge}}$ $\times$ SU(2)$_\textrm{spin}$ symmetry and retain up to $D^*=2 000$ multiplets (equivalent to $\sim 5 200$ individual states), rendering well converged results with truncation error $\epsilon \sim 10^{-4}$. The electron density can be controlled by adjusting the chemical potential term $-\mu N_\text{tot}$, where $N_\text{tot}$ represents the total electron number operator. By fine-tuning the parameter $\mu$, we ensure that the two-orbital system remains approximately at $n_e \simeq 1.5$ filling at low temperatures.

\begin{figure}[]
\centering
\includegraphics[width=0.6\linewidth]{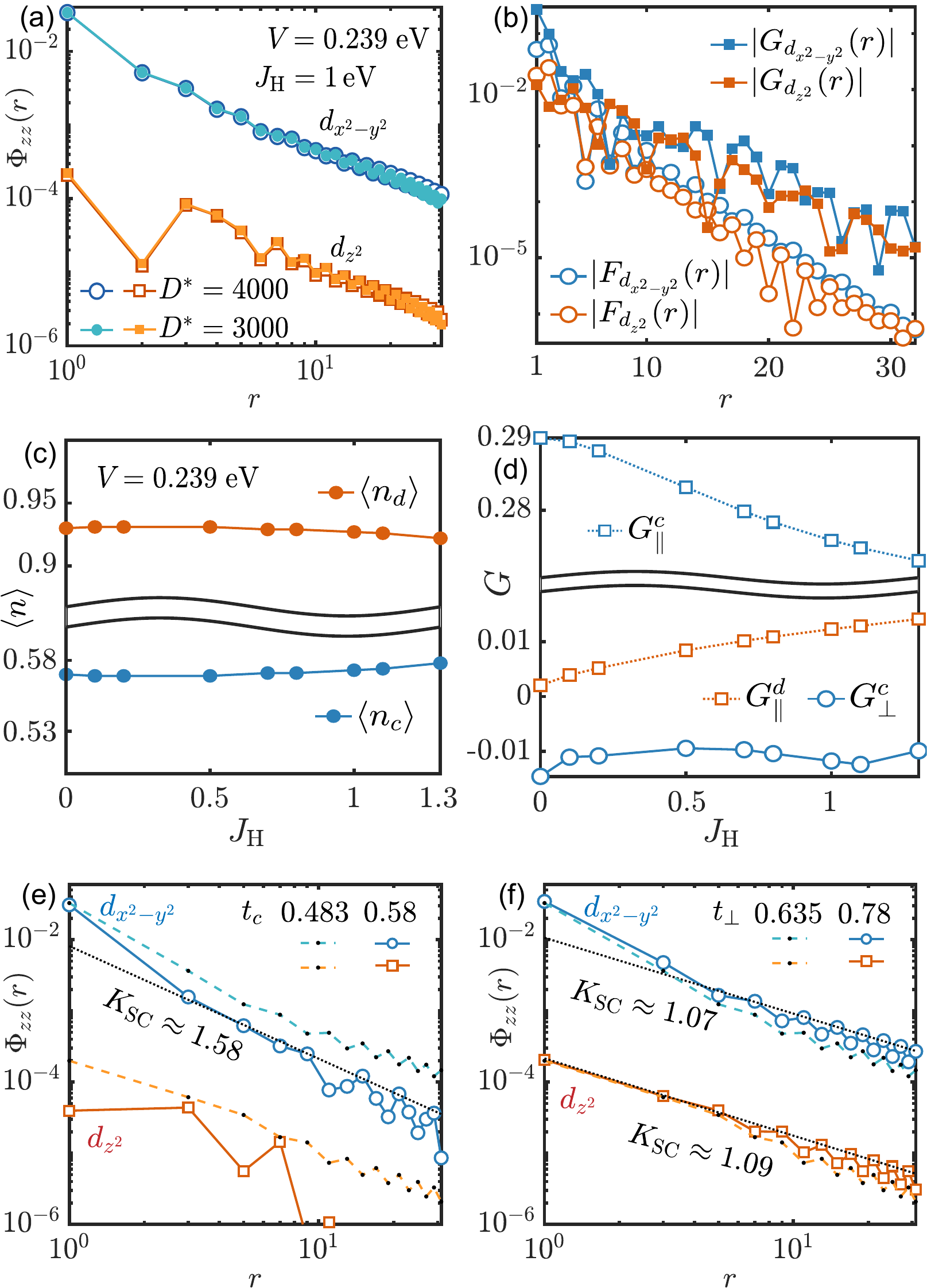}
\caption{(a) Interlayer pairing correlations $\Phi_{zz}$ and (b) spin correlation and single-particle 
Green's function are shown, where the results are well converged versus different bond multiplets $D^*$. 
(c) Electron density and (d) Green's function are calculated in the two $e_g$ orbitals.
To simulate the pressure effects, we increase (e) intralayer 
hopping $t_c$ and coupling $J_c$, and (f) interlayer hopping $t_\perp$ and coupling $J_\perp$, 
and find the results change only quantitatively with $t_c, J_c$ and $t_\perp, J_\perp$. Here we assume 
$J = 4 t^2 / U$ with $U=4$~eV, and fixed other model parameters at their pristine values.
}
\label{FigE2}
\end{figure}

\section*{DMRG convergence, orbital selectivity and impact of hopping $t_c$, $t_\perp$} 
Figure~\ref{FigE2}(a) illustrates the convergence of DMRG results, with interlayer pairing correlations from $D^* = 3000$ and $4000$ U(1)$_{\text{charge}} \times$ SU(2)$_{\text{spin}}$ multiplets showing excellent consistency up to distance $r \sim 30$. Figure~\ref{FigE2}(b) shows the exponential decay of spin correlations and single-particle Green's functions, indicating the presence of Luther-Emery SC phase.

Figures~\ref{FigE2}(c,d) show the electron densities $\langle n_{c,d} \rangle$ and the single-particle Green's function $G_{\parallel, \perp}$ in two $e_g$ orbitals. The intralayer Green's function is defined as $G_\parallel^\alpha \equiv \frac{1}{2(L-1)} \sum_{i, \mu, \sigma} \langle \alpha_{i, \mu, \sigma}^\dagger \alpha_{i+1, \mu, \sigma}\rangle$ between nearest-neighboring sites within each layer, with $\alpha = \{c, d\}$ denoting the two orbitals. The interlayer Green's function is defined as $G_\perp^c \equiv \frac{1}{L} \sum_{i, \sigma}\langle c^\dagger_{i,\mu=1,\sigma} c_{i,\mu=-1,\sigma} \rangle$, reflecting the interlayer electron hopping. The results in Fig.~\ref{FigE2}(c) indicate that the \ZO orbitals are nearly half-filled with only few holes, in distinction with the approximately quarter-filled \XO orbitals with $\langle n_c \rangle \simeq 0.58$. In Fig.~\ref{FigE2}(d), we further point out that the \ZO electrons are rather localized as $G_{\parallel}^d \ll G_{\parallel}^c$, while the \XO orbitals are itinerant and can move coherently within each layer (large $G^c_{\parallel}$), but not across two layers (negligible $G^c_{\perp}$).

In Figs.~\ref{FigE2}(e,f) we adjust the intralayer \XO hopping $t_c$ and interlayer \ZO hopping $t_\perp$ to their realistic values in over-pressured phase (70 GPa), respectively. In both cases we find algebraically decaying $\Phi_{zz}$ which accounts for robust SC order in the ground state. Nevertheless, the enhancement of $t_c$ as well as $J_c$ signifies the interorbital frustration, leading to a weaker SC with $K_{\rm SC} \simeq 1.58$ for \XO orbital and even absence of SC for \ZO orbital. On the contrary, SC slightly benefits from the enhancement of $t_\perp$, which strengthens the interlayer AF coupling. However, the impact of these parameters on the SC order is not significant. In the main text, we find that the interorbital hybridization $V$ plays an essential role in the over-pressurized regime and have chosen to show related results in Fig.~\ref{Fig4}.

\begin{figure}[]
\centering
\includegraphics[width=0.6\linewidth]{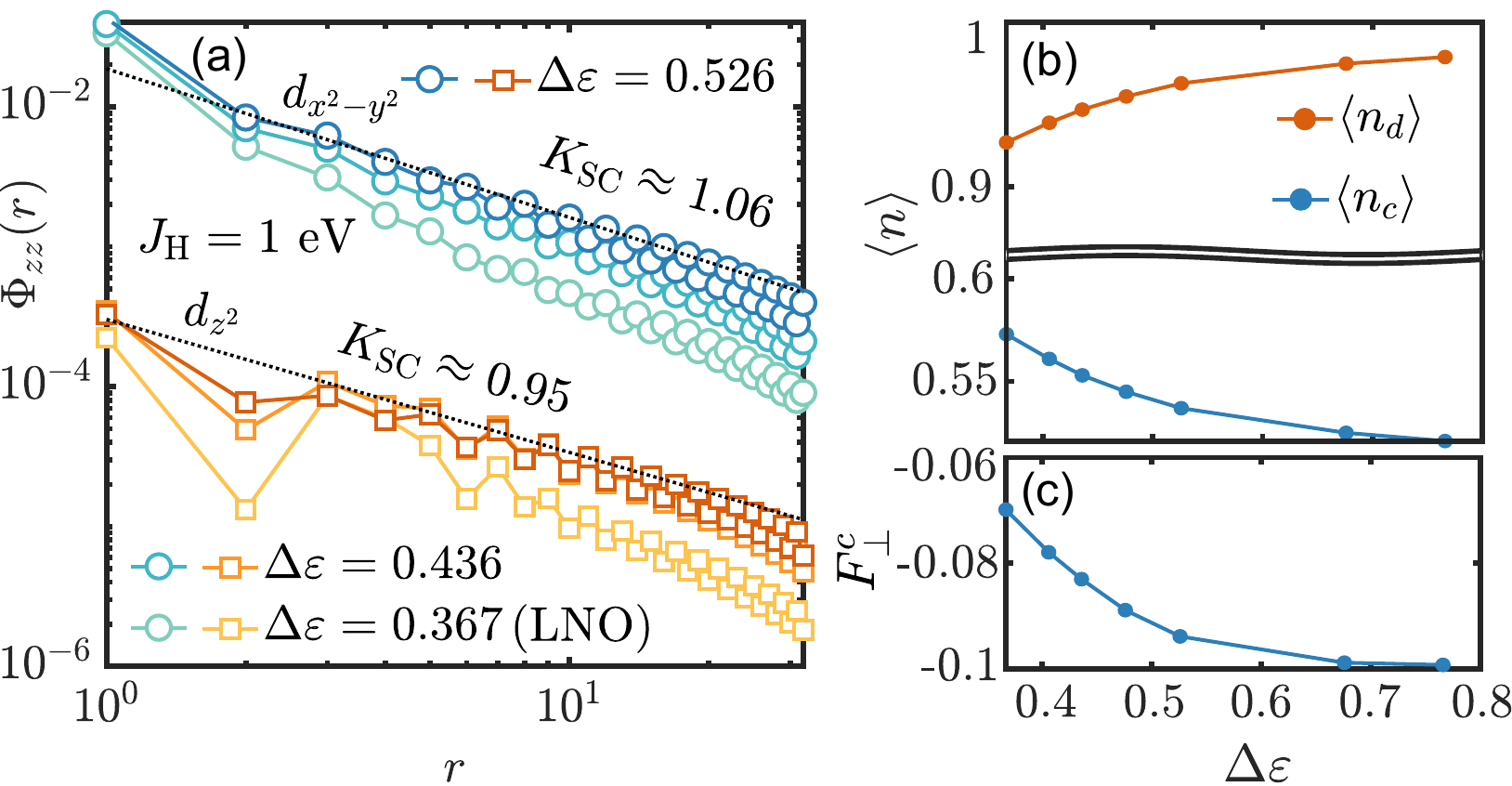}
 \caption{(a) Interlayer pairing correlations of both the $d_{x^2-y^2}$ and $d_{z^2}$ orbitals with different site energy offsets
$\Delta \varepsilon \equiv \varepsilon_c - \varepsilon_d$. $\Delta \varepsilon = 0.367$~eV is estimated from DFT calculations 
on \LNO~under about 30 GPa pressure~\cite{Luo2023Model}. We artificially increase $\Delta \varepsilon$ between the two $e_g$ 
orbitals, and compute (b) the electron densities and (c) interlayer spin correlation $F_\perp^c$.
}
\label{FigE3}
\end{figure}

\section*{Site energy difference and pairing correlation}
In Fig.~\ref{FigE3} we increase the difference in the site energies between two orbitals. As $\Delta \varepsilon$ increases, more electrons are transferred to the \ZO orbital while keeping total $n_e = 1.5$. Given all other realistic parameters remain unchanged, Fig.~\ref{FigE3}(a) demonstrates a significant enhancement in the pairing correlations for both orbitals as $\Delta \varepsilon$ increases. The resulting electron densities, $\langle n_c \rangle$ for the \XO orbital and $\langle n_d \rangle$ for the \ZO orbital, are shown in Fig.~\ref{FigE3}(b). As the \ZO orbital approaches half-filling (thus containing fewer holes), a stronger interlayer AF correlation $F_\perp^c$ can be observed between the \XO orbitals, as illustrated in Fig.~\ref{FigE3}(c). This provides insights into potential experimental strategies for achieving higher critical temperature $T_c$ or stabilizing the SC order in ambient conditions.

\section*{iPEPS calculations of two-orbital bilayer $t$-$J$ model}
\begin{figure}[H]
\centering
\includegraphics[width=0.6\linewidth]{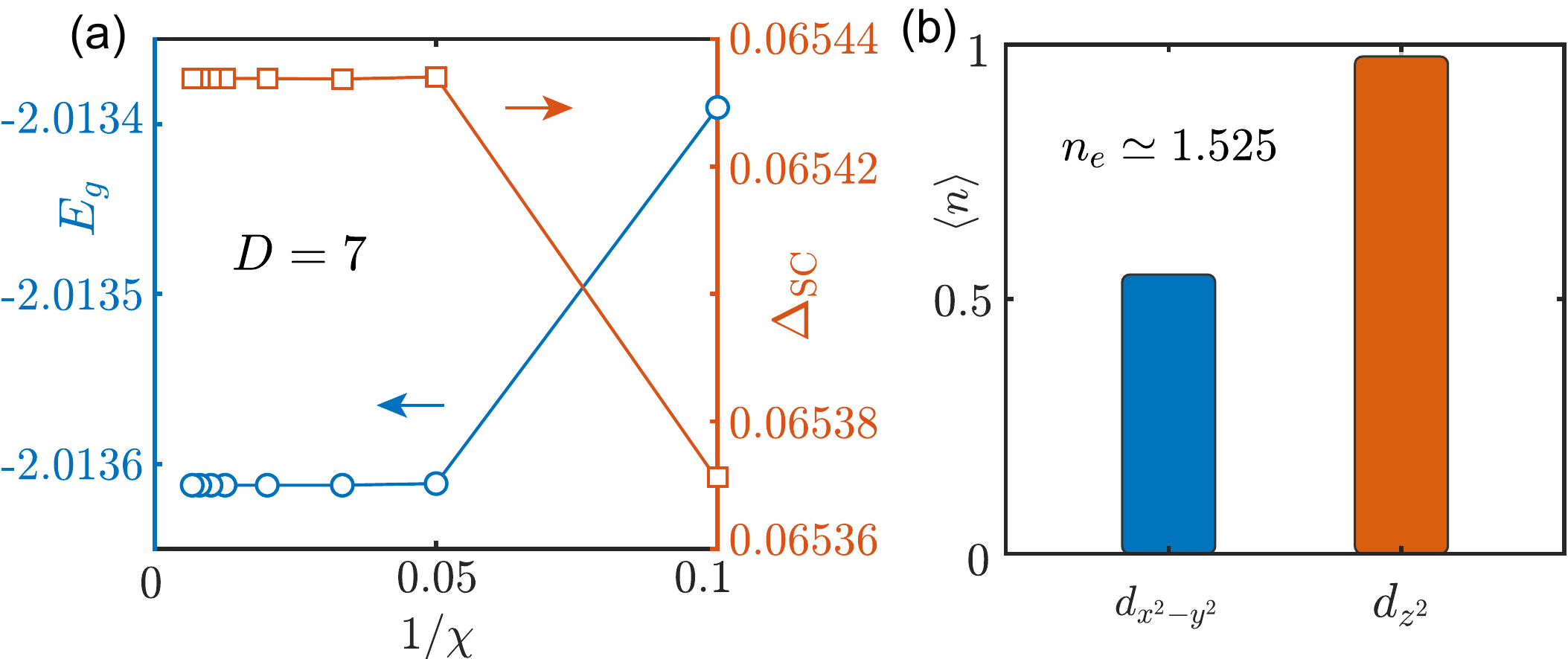}
\caption{(a) Data convergence of the iPEPS results. We fix the bond dimension $D = 7$ and vary the environment bond dimension $\chi$ in CTMRG. We find ground-state energy per site (left axis) and the SC order parameter (\XO orbital, right axis) are well converged for $\chi \gtrsim 20$. (b) Electron density distribution at $n_e \simeq 1.525$, 
with approximately $0.55$ in \XO orbital and $0.98$ in \ZO orbital.
}
\label{FigE4}
\end{figure}

Here we present the iPEPS calculation details and demonstrate the data convergence. To deal with the two-orbital bilayer model in the thermodynamic limit, we combine the four $t$-$J$ sites --- with local Hilbert space of dimension $d=3^4=81$ --- which is highly challenging within the framework of iPEPS calculations. To reduce computational complexity and make the calculations feasible, we implement $\mathbb{Z}_{2,\mathrm{charge}} \times$ SU(2)$_{\mathrm{spin}}$ symmetry~\cite{haegeman2024jutho} and fold the $d=81$ individual states into $d^*=34$ multiplets. We adopt simple update~\cite{Xiang2008SU, Li2012SU} with equivalently $D = 7$ individual bond states and perform the corner transfer-matrix renormalization group~\cite{NishinoCTMRG, Orus2009Simulation} with environment bond dimension up to $\chi = 150$. The ground-state energy $E_g$ and SC order parameter (of \XO orbital) versus $\chi$ are show in Fig.~\ref{FigE4}(a), where the results are found well converged with $\chi \gtrsim 20$. We show in Fig.~\ref{FigE4}(b) that \XO orbital is quarter-filled and \ZO orbital half-filled, consistent with the DMRG results in Fig.~\ref{FigE2}(c).
Our iPEPS code with non-abelian symmetries implemented is publicly available at the Github repository~\cite{Qu2025PEPSTensorKit}.

\section*{DFT calculations}
We use the Vienna ab initio simulation package (VASP) to optimize the lattice structure and obtain the electronic band structure \cite{RN142}. 
The convergence criterion for atomic forces during structural optimization was set to 1 meV/\AA. 
The total energy convergence threshold for electronic self-consistent processes was set to 10$^{-8}$~eV/atom. 
A plane-wave cutoff energy of 520~eV is employed. 
The $\Gamma$-centered 12×12×12 Monkhorst-Pack k-points grid in reciprocal space was utilized in the self-consistent cycle. 
The Wannier90 package is employed to downfold the band structure and construct the two-orbital model comprising Ni-$e_g$ orbitals~\cite{RN144}. 
Results obtained with various configurations are summarized in Table.~\ref{TabS1}. 
For the CDW-AF phase, the oxygen octahedra exhibit in-plane distortions, and Ni atoms can occupy two inequivalent sites. 
Consequently, the parameters like $V$ and $t_c$ exhibit spatial distributions, represented by the error bars in Fig.~\ref{Fig4}(a) of the main text. 
There are two inequivalent Wyckoff positions for La atoms in \LPNO. 
It is found that Pr doping at the La2 position (inside the NiO bilayer) is energetically preferred over the La1 site (outside the NiO bilayer), leading to energy reductions of 593.3 meV/f.u. for the Amam phase and 437.9 meV/f.u for the I4/mmm phase.
Note that in all calculations, low-pressure data (0 and 10~GPa) are obtained from \textit{Amam} orthorhombic 
phase while high-pressure results are from \textit{I4/mmm} tetragonal phase.

\begin{table}[h]
\centering
\begin{minipage}[l]{0.48\textwidth}
\begin{tabular}{ccccccc}
  \toprule
  \textbf{NM} (\LNO)& $t_c$ & $t_d$ & $V$   & $t_\perp$& $\varepsilon_c$ & $\varepsilon_d$ [eV]\\
  \cmidrule{1-7}
  0 GPa      & 0.397 & 0.073 & 0.189 & 0.618    & 0.765           & 0 \\
  \cmidrule{1-7}
  10         & 0.475 & 0.103 & 0.230 & 0.664    & 0.742           & 0 \\
  \cmidrule{1-7}
  20         & 0.502 & 0.105 & 0.252 & 0.690    & 0.752           & 0 \\
  \cmidrule{1-7}
  30         & 0.522 & 0.112 & 0.265 & 0.714    & 0.762           & 0 \\
  \cmidrule{1-7}
  40         & 0.534 & 0.125 & 0.279 & 0.724    & 0.638           & 0 \\
  \cmidrule{1-7}
  50         & 0.556 & 0.125 & 0.286 & 0.750    & 0.772           & 0 \\
  \cmidrule{1-7}
  60         & 0.570 & 0.131 & 0.295 & 0.765    & 0.773           & 0 \\
  \cmidrule{1-7}
  70         & 0.582 & 0.137 & 0.303 & 0.780    & 0.767           & 0 \\
  \cmidrule{1-7}
  80         & 0.593 & 0.143 & 0.310 & 0.794    & 0.759           & 0 \\
  \cmidrule{1-7}
  90         & 0.604 & 0.147 & 0.317 & 0.807    & 0.759           & 0 \\
  \cmidrule{1-7}
  100        & 0.614 & 0.152 & 0.324 & 0.819    & 0.750           & 0 \\
  \cmidrule{1-7}
  125        & 0.634 & 0.162 & 0.340 & 0.844    & 0.733           & 0 \\
  \cmidrule{1-7}
  150        & 0.645 & 0.167 & 0.351 & 0.861    & 0.742           & 0 \\
  \bottomrule
  \vspace{0.1cm}
\end{tabular}
\begin{tabular}{ccccccc}
  \toprule
  \makecell{\textbf{CDW-AF} \\ (\LNO)} & $t_c$                           & $t_d$                           & $V$                             & $t_\perp$  & $\varepsilon_c$                  & $\varepsilon_d$ [eV]             \\
  \cmidrule{1-7}
  0 GPa           & \makecell{0.124 \\$\sim$ 0.178} & \makecell{0.025 \\$\sim$ 0.068} & \makecell{0.088 \\$\sim$ 0.192} & 0.549      & \makecell{-1.182 \\$\sim$ 1.651} & \makecell{-1.518 \\$\sim$ 1.626} \\
  \cmidrule{1-7}
  30              & \makecell{0.069 \\$\sim$ 0.239} & \makecell{0.077 \\$\sim$ 0.084} & \makecell{0.163 \\$\sim$ 0.242} & 0.666      & \makecell{-1.542 \\$\sim$ 1.545} & \makecell{-1.468 \\$\sim$ 1.613} \\
  \bottomrule
\end{tabular}
\end{minipage}
\begin{minipage}[r]{0.48\textwidth}
\begin{tabular}{ccccccc}
  \toprule
  \textbf{AF} (\LNO)& $t_c$ & $t_d$ & $V$   &$t_\perp$ & $\varepsilon_c$ & $\varepsilon_d$ [eV] \\
  \cmidrule{1-7}
  0 GPa       & 0.286 & 0.031 & 0.132 & 0.552    & 0.420           & 0 \\
  \cmidrule{1-7}     
  10          & 0.328 & 0.010 & 0.172 & 0.665    & 0.474           & 0 \\
  \cmidrule{1-7}   
  20          & 0.421 & 0.100 & 0.229 & 0.700    & 0.495           & 0 \\
  \cmidrule{1-7}    
  30          & 0.446 & 0.106 & 0.244 & 0.734    & 0.526           & 0 \\
  \cmidrule{1-7}   
  40          & 0.470 & 0.118 & 0.259 & 0.749    & 0.543           & 0 \\
  \cmidrule{1-7}   
  50          & 0.495 & 0.127 & 0.271 & 0.765    & 0.556           & 0 \\
  \cmidrule{1-7}   
  60          & 0.507 & 0.130 & 0.279 & 0.775    & 0.580           & 0 \\
  \cmidrule{1-7}   
  70          & 0.526 & 0.138 & 0.293 & 0.789    & 0.599           & 0 \\
  \cmidrule{1-7} 
  80          & 0.546 & 0.144 & 0.303 & 0.798    & 0.617           & 0 \\
  \cmidrule{1-7} 
  90          & 0.575 & 0.148 & 0.320 & 0.837    & 0.735           & 0 \\
  \cmidrule{1-7}
  100         & 0.594 & 0.152 & 0.325 & 0.847    & 0.747           & 0 \\
  \cmidrule{1-7}  
  125         & 0.641 & 0.158 & 0.344 & 0.836    & 1.004           & 0 \\
  \cmidrule{1-7}
  150         & 0.674 & 0.163 & 0.358 & 0.842    & 1.190           & 0 \\
  \bottomrule
  \vspace{0.11cm}
\end{tabular}
\begin{tabular}{ccccccc}
  \toprule
  \textbf{NM} (\LPNO)& $t_c$ & $t_d$ & $V$   & $t_\perp$  & $\varepsilon_c$   & $\varepsilon_d$ [eV]  \\
  \cmidrule{1-7}
  0 GPa              & 0.349 & 0.049 & 0.176 & 0.488      & 0.920             & 0                     \\
  \cmidrule{1-7}       
  30                 & 0.474 & 0.099 & 0.245 & 0.660      & 0.918             & 0                     \\
  \cmidrule{1-7}     
  80                 & 0.598 & 0.150 & 0.307 & 0.795      & 0.977             & 0                     \\
  \cmidrule{1-7}        
  150                & 0.650 & 0.144 & 0.343 & 0.844      & 0.957             & 0                     \\
  \bottomrule
\end{tabular}
\end{minipage}
\caption{
Tight-binding parameters of the two-orbital model for \LNO~and \LPNO~determined by Wannier downfolding 
from DFT calculations. Non-magnetic (NM), antiferromagnetic (AF) and charge density wave (CDW) configurations are considered. 
Calculations at low pressures (0 and 10 GPa) are based on the \textit{Amam} orthorhombic phase, 
whereas the \textit{I4/mmm} tetragonal phase is used for calculations at high pressures ($\geq$20 GPa).
} 
\label{TabS1}
\end{table}

\section*{DMRG results on larger cylindrical systems}
In this section, we extend DMRG calculations of the bilayer two-orbital model to 
$(H, W, L) = (2, 2, 32)$ lattice with $W=2$ wider cylindrical geometries, thereby significantly strengthening our conclusion obtained from ladders in the main text. 

\begin{figure}[htbp]
\centering
\includegraphics[width=0.85\linewidth]{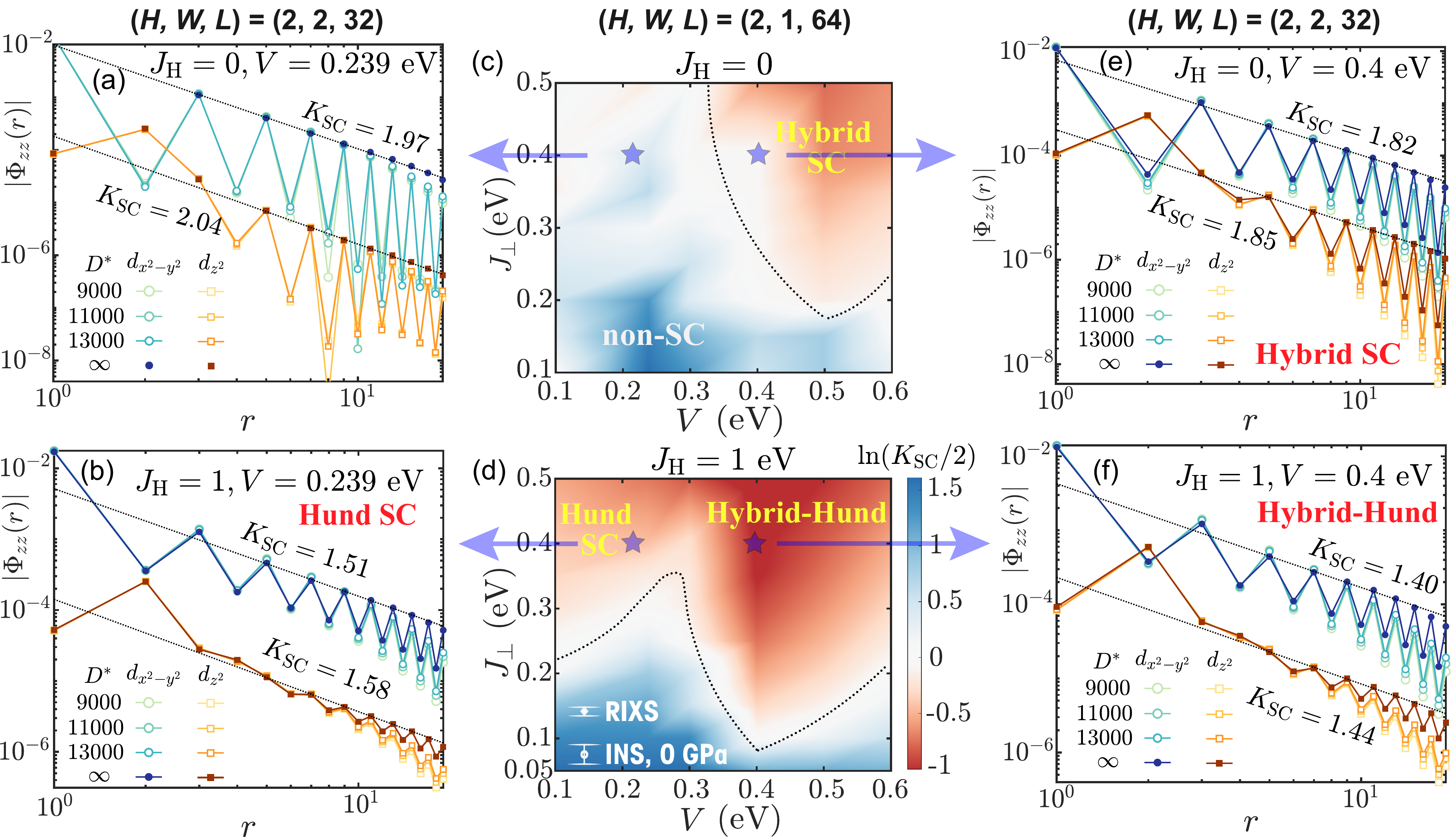}
\caption{Panels (c,d) show the ground state $J_\perp$-$V$ phase diagram, which is adapted from Figs.1(a,c) of the main text, obtained on two-orbital ladder $(H, W, L)=(2,1,64)$. With large $J_\perp \approx 0.4$~eV, there are  ``non-SC", ``Hybrid SC", ``Hund SC" and ``Hybrid-Hund" regimes, indicated by the four blue stars. In (a,b,e,f), we perform DMRG calculations on a bilayer lattice $(H, W, L)=(2,2,32)$ and show the interlayer pairing correlations $|\Phi_{zz}|$, with different retained bond dimensions. Through extrapolating the correlation functions to $D^* = \infty$ and extract the Luttinger parameters, we identify
(a) non-SC phase ($J_{\rm H} = 0$, $V = 0.239$~eV), 
(b) Hund SC regime ($J_{\rm H} = 1$~eV, $V = 0.239$~eV),
(e) Hybrid SC regime ($J_{\rm H} = 0$, $V = 0.4$~eV) and
(d) Hybrid-Hund synergistic SC regime ($J_{\rm H} = 1$~eV, $V = 0.4$~eV).
These results show excellent consistency with the phase diagram obtained from $W=1$ ladders.
}
\label{FigR1}
\end{figure}

In practical DMRG calculations, we keep $D^* = 9000, 11000, 13000$ U(1)$_{\mathrm{charge}} \times$ SU(2)$_{\mathrm{spin}}$ multiplets and perform polynomial extrapolations $C(1/D^*) = C(0) + a/D^* + b/D^{*2}$ to ensure well-controlled and reliable results. The corresponding Luttinger parameters are extracted from the correlated functions with infinite-$D^*$ extrapolation. Note the exponential growth of entanglement entropy with width $W$ severely restricts the accessible $W$ in DMRG, even with state-of-the-art implementations. While the calculations of $W=2$ cylinders are still quasi-one-dimensional, it represents the current practical limit for fermionic systems with multiple orbitals and spinful electrons --- with the computational complexity equivalent to that of width-8 extended $t$-$J$ model on a square lattice~\cite{Lu2024tJ8leg,Chen2025tJ8leg}.

Figures~\ref{FigR1}(a,b,e,f) show the interlayer pairing correlations $|\Phi_{zz}(r)|$ for both orbitals of the $W=2$ system. $|\Phi_{zz}(r)|$ is calculated between interlayer pairs $(x, y, z=1 \mid x, y, z=-1)$ and $(x+r, y, z=1 \mid x+r, y, z=-1)$, where $z, x$ mark the sites along the height and length direction and the results are averaged over width $y$. We adapt the $J_\perp - V$ phase diagram from Fig.1(a,b) of the main text, in which our analysis reveals four distinct regimes for a substantial value of $J_\perp \approx 0.4$~eV, namely ``non-SC'',``Hund SC'',``Hybrid SC'' and ``Hybrid-Hund'', each signified by a blue star. 
Figure~\ref{FigR1}(a) illustrates the ``non-SC" phase, where we get $K_{\mathrm{SC}} \approx 2$ for both orbitals, indicating the absence of (or very weak) superconductivity. The (quasi-long-range) SC orders are observed in the other three distinct SC regimes with Luttinger parameters $K_{\mathrm{SC}} < 2$, dominated by the hybridization [cf. Fig.~\ref{FigR1}(e)], the Hund's rule coupling [cf. Fig.~\ref{FigR1}(b)] and both of them [cf. Fig.~\ref{FigR1}(f)]. The most robust SC emerges in the ``Hybrid-Hund” synergistic regime. Our DMRG results from wider cylinders show very good agreement with the two-orbital ladder ($W=1$) calculations, demonstrating the robustness of interlayer pairing for various widths. This conclusion is further corroborated by our infinite-size iPEPS simulations shown in the main text.

Overall, our results show that bilayer nickelates exhibit robust interlayer pairings, well-captured even in quasi-1D simulations and barely affected by system width. This is in sharp contrast to the situation in cuprates, where the simulations of intralayer pairing show strong width dependence.

\end{document}